\pdfoutput=1
\documentclass[11pt,twoside,a4paper,cmspaper,final,collab]{cms-tdr}
\def\svnVersion{a0b020d}\def\svnDate{2024/09/04}\def\cmsCernNoTag{CERN-EP-2024-014}\def\cmsCernDate{\today}\def\cmsMessage{Published in the Journal of Instrumentation as \href{http://dx.doi.org/10.1088/1748-0221/19/09/P09004}{\doi{10.1088/1748-0221/19/09/P09004}.}}
\begin{document}\cmsNoteHeader{EGM-18-002}

\ifthenelse{\boolean{cms@external}}{\providecommand{\cmsLeft}{upper\xspace}}{\providecommand{\cmsLeft}{left\xspace}}
\ifthenelse{\boolean{cms@external}}{\providecommand{\cmsRight}{lower\xspace}}{\providecommand{\cmsRight}{right\xspace}}
\newlength\cmsTabSkip\setlength{\cmsTabSkip}{1ex}
\newcommand{\mgg}{\ensuremath{m_{\gamma\gamma}}\xspace}
\newcommand{\zee}{\ensuremath{\PZ\to\Pep\Pem}\xspace}

\cmsNoteHeader{EGM-18-002}
\title{Performance of the CMS electromagnetic calorimeter in \texorpdfstring{$\Pp\Pp$}{pp} collisions at \texorpdfstring{$\sqrt{s}=13\TeV$}{sqrt(s) = 13 TeV}}

\date{\today}

\abstract{
The operation and performance of the Compact Muon Solenoid (CMS) electromagnetic calorimeter (ECAL) are presented, based on data collected in $\Pp\Pp$ collisions at $\sqrt{s}=13\TeV$ at the CERN LHC, in the years from 2015 to 2018 (LHC Run~2), corresponding to an integrated luminosity of 151\fbinv. The CMS ECAL is a scintillating lead-tungstate crystal calorimeter, with a silicon strip preshower detector in the forward region that provides precise measurements of the energy and the time-of-arrival of electrons and photons. The successful operation of the ECAL is crucial for a broad range of physics goals, ranging from observing the Higgs boson and measuring its properties, to other standard model measurements and searches for new phenomena. Precise calibration, alignment, and monitoring of the ECAL response are important ingredients to achieve these goals. To face the challenges posed by the higher luminosity, which characterized the operation of the LHC in Run 2, the procedures established during the 2011--2012 run of the LHC have been revisited and new methods have been developed for the energy measurement and for the ECAL calibration. The energy resolution of the calorimeter, for electrons from {\PZ} boson decays reaching the ECAL without significant loss of energy by bremsstrahlung, was better than 1.8\%, 3.0\%, and 4.5\% in the $\abs{\eta}$ intervals $[0.0,0.8]$, $[0.8,1.5]$, $[1.5, 2.5]$, respectively. This resulting performance is similar to that achieved during Run~1 in 2011--2012, in spite of the more severe running conditions.
}

\hypersetup{
pdfauthor={CMS Collaboration},
pdftitle={The CMS ECAL performance in LHC Run~II},
pdfsubject={CMS},
pdfkeywords={CMS, physics, software, computing}}

\maketitle

\section{Introduction}\label{sec:introduction}
CMS is a general-purpose particle detector~\cite{Chatrchyan:2008zzk}
at the CERN LHC
that measures collision products from 
high-energy proton-proton ($\Pp\Pp$) and heavy ion collisions.
The electromagnetic calorimeter (ECAL) of the CMS detector provides a highly efficient and
accurate reconstruction of photons and electrons over a wide range of energies
from low-energy electrons (5\GeV) 
typical of multilepton events,
to electroweak-scale energies (Higgs and \PW/\PZ bosons),
up to the TeV scale typical of high-mass resonance searches. 
The ECAL also 
measures energy
deposits from hadrons and the electromagnetic component 
of jets 
contributing to jet energy and missing transverse momentum (\ptmiss) measurements.
The ECAL provides precise time-of-arrival measurements of electromagnetic showers,
used in the rejection of backgrounds with a broad
time distribution, such as electronic noise
and $\Pp\Pp$ interactions in preceding and subsequent bunch crossings.
Precise time measurements can be used to 
identify new particles with long lifetimes, typically larger than 1\unit{ns},
predicted by certain theories beyond the standard model~\cite{PhysRevD.100.112003}.

This paper describes the operation, monitoring, calibration, and performance of the ECAL during Run~2, 2015--2018.
A short description of the CMS detector and of the ECAL is provided in Section~\ref{sec:cmsdetector}, and an overview of the challenges posed by the LHC in Run~2
is given in Section~\ref{sec:introrun2}. 
The ECAL on-detector and off-detector readout and trigger systems are briefly described in
Section~\ref{sec:online}, and the details of the energy reconstruction are given in Section~\ref{sec:reconstruction}.
The methods used to monitor and to correct for the variation of the ECAL response in time are described in Section~\ref{sec:response}, and the
methods to calibrate the ECAL are briefly summarised in Section~\ref{sec:calibration}.
Quantitative assessments of the precision of the resulting calibrations are described in Section~\ref{sec:combination}.
A short description of the simulation of ECAL in CMS is presented in Section~\ref{sec:simulation}. 
Finally, the performance of the ECAL both at the trigger level and for the offline
reconstruction are described in Section~\ref{sec:performance}.
The paper is summarized in Section~\ref{sec:conclusions}.
The procedure to align ECAL with respect to the CMS tracking detector
and the details of the calibration procedure are reported in the appendices. 

\section{The CMS detector}\label{sec:cmsdetector}   
The central feature of the CMS detector is a superconducting solenoid of 6\unit{m} internal diameter, 
providing a magnetic field of 3.8\unit{T}. 
Within the solenoid volume are silicon pixel and strip trackers,
the ECAL,
and a brass and scintillator hadron calorimeter (HCAL),
each composed of a barrel and two endcap sections.
Forward calorimeters extend to $\abs{\eta}=5$ the pseudorapidity coverage provided by the barrel and endcap detectors. 
Muons are detected in gas-ionization chambers embedded in the steel flux-return yoke outside the solenoid.
Events of interest are selected using a two-tiered trigger system~\cite{TRG-17-001,CMS:2016ngn}.
The first level, known as the L1 trigger, 
composed of custom hardware processors, 
uses information from the calorimeters and muon detectors to select events at a rate of about 100\unit{kHz}
within a fixed latency period of approximately 4\mus. 
The second level, known as the high-level trigger (HLT),
consists of a farm of processors running a version of the full-event reconstruction software optimized for fast processing, 
and reduces the event rate to about 1\unit{kHz} before data storage~\cite{CMS:2016ngn}.
A more detailed description of the CMS detector,
together with a definition of the coordinate system used and the relevant kinematic variables, can be found in Ref.~\cite{Chatrchyan:2008zzk}.

The ECAL is a
homogeneous and hermetic scintillating lead tungstate crystal ($\mathrm{PbWO_4}$) calorimeter.
It is divided into a barrel region (EB) consisting of 61200 crystals, covering the pseudorapidity region $\abs{\eta} < 1.479$, 
and an endcap region (EE) consisting of two disks each with 7324 crystals, covering $1.479 < \abs{\eta} < 3.0$.
The crystals for each half-barrel are grouped in 18 supermodules each subtending $20^\circ$ in azimuth angle $\phi$.
Each supermodule comprises four modules with 500 crystals in the first module and 400 crystals in each of the remaining three modules.
The crystals in each endcap are organized in two semicircular mechanical structures, named ``Dees''.
The scintillation light is detected by avalanche photodiodes (APDs) in the EB and by vacuum phototriodes 
in the EE~\cite{CERN-LHCC-97-033, Bloch:581342}.
Preshower detectors (ES) consisting of two planes of silicon sensors interleaved with a total of 3 radiation lengths of lead are located in front of each endcap,
covering $1.653 < \abs{\eta} < 2.6$.
The calorimeter was designed to maintain excellent energy resolution for an instantaneous luminosity up to $1\times 10^{34}\percms$
and a delivered integrated luminosity of at least 500\fbinv, corresponding to
over 10 years of operation at the LHC~\cite{CERN-LHCC-97-033}.

\section{Run 2 challenges}\label{sec:introrun2}
The data-taking conditions during Run~2 (2015--2018) were significantly different from those of Run~1 (2011--2012).
During Run~2, the centre-of-mass energy was 13\TeV, compared to 7 or 8\TeV during Run~1, resulting  
not only in an increase of the cross section for interesting signal processes, 
but also in an increase of the total inelastic $\Pp\Pp$ cross section.
Moreover, the instantaneous luminosity regularly reached 
values of $2\times10^{34}\percms$ during LHC fills in Run~2,
which is a factor of three larger than typically seen in Run~1.
This resulted in a large number of overlapping interactions (pileup, PU) per bunch crossing (BX). 
The average number of pileup events was 27, 38, and 37, for 2016, 2017, and 2018, respectively,
compared to about 20 during Run~1, which was the design level of pileup for the ECAL~\cite{CERN-LHCC-97-033}.
The integrated luminosities delivered by the LHC and recorded by CMS in the different years are given in Table~\ref{tab:lumi}.

\begin{table}[htbp]
\centering
\topcaption{ 
 Delivered and recorded integrated luminosity for the Run~2 period~\cite{lumi1516, lumi17, lumi18}.
}
\begin{tabular}{ ccc } 
 Year & Delivered [\fbinv] & Recorded [\fbinv]\\ 
 \hline
 2015 &  4.3 &  3.9 \\ 
 2016 & 41.6 & 38.3 \\ 
 2017 & 49.8 & 45.0 \\ 
 2018 & 67.9 & 63.7 \\ [\cmsTabSkip]
 Total & 163.6 & 150.9
\end{tabular}
\label{tab:lumi}
\end{table}

During Run~2, the LHC collisions occurred with a minimum duration between bunch crossings of 25\unit{ns},
compared to 50\unit{ns} in Run~1,
and the LHC was filled with up to 2544 bunches per beam, almost at its design level.
Different filling schemes for the LHC were used, including ones with long trains of consecutive proton bunches (up to 144).
Isolated bunches were also included in some fills,
which were used to study pileup
subtraction and to measure the ECAL signal pulse shape.
With collisions occurring every 25\unit{ns},
the energy reconstructed for a given BX is
affected by the energy deposited in other BXs, resulting in potentially large amounts of ``out-of-time PU'' (OOT-PU).
For ECAL, this effect was mitigated by the use of a new algorithm 
for amplitude reconstruction described in Ref.~\cite{EGM18001}
and summarized in Section~\ref{sec:reconstruction}.
The larger instantaneous and integrated luminosities increased the total radiation dose and dose rate to the ECAL, 
leading to a faster variation of the convolved response of the crystals and photodetectors,
an increase of the electronic noise,
and a drift in the pedestal baseline,
all of which required frequent updates to maintain performance.

\section{Readout and trigger}\label{sec:online}
\begin{figure}[htb]
\centering
\includegraphics[width=0.6\textwidth]{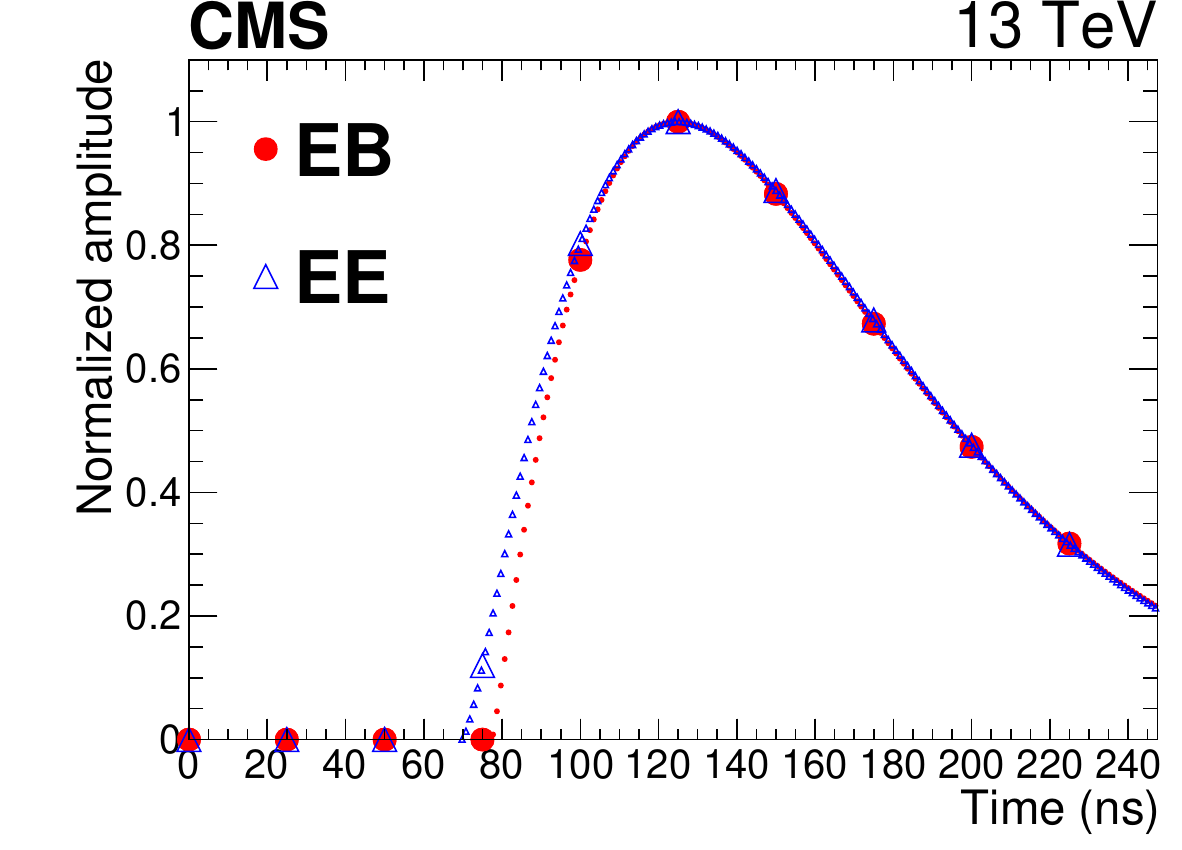}
\caption{
  Average signal pulse shape for a channel for the EB (red filled circles), and the EE (blue hollow triangles),
  after subtraction of the pedestal.
  The ten larger markers are an example of a pulse shape sampled once every 25\unit{ns}.
  The dots are the result of a granular timing scan provides a more precise measurement of the pulse.
}
\label{fig:ECALsignals:pulse_shape}
\end{figure}

The ECAL crystals and photosensors have a combined response of about 4.5 photoelectrons per MeV.
The analog electrical signal is shaped with a shaping time of about 43 ns 
and amplified by a multi-gain preamplifier (MGPA)~\cite{CERN-LHCC-97-033},
which can amplify voltages with three different gains: 12, 6, and 1. 
These three gains allow the readout to give precise measurements in the energy range from about 
40 (60)\MeV in EB (EE) (corresponding to the least significant bit of the 12 bit analog-to-digital converter (ADC) count for a signal in gain 12) to about 1.7--2.0\TeV
(corresponding to the full scale 
of the 12 bit ADC at gain 1) for energies deposited in a single channel.
The choice of the gain is automatic, with the MGPA
selecting the highest gain for which the pulse is not saturated.
The amplified signal is digitized 
with a sampling rate
of 40\unit{MHz} and stored in a buffer
in the on-detector electronics.
The acceptance of an event by the L1 trigger system 
activates
the readout to the off-detector electronics of ten consecutive samples
from the buffer
chosen such that 
the rising edge of the signal pulse occurs near to
the fourth sample    
and the pulse peak is close to the
sixth sample,
as shown in Fig.~\ref{fig:ECALsignals:pulse_shape}.

The signals measured in the ECAL are also used as inputs at a rate of 40\unit{MHz} to the L1 trigger system.
Due to bandwidth constraints these signals are used to calculate a highly compressed version of the data, 
known as a trigger primitive (TP)~\cite{Paganini_2009}, which the L1 then uses in the calorimeter trigger and global trigger systems.
For every BX, trigger primitives are formed from the energy sums 
of small groups crystals, between 5 and 25.
The TPs are computed from
the signal in strips of five crystals by the FENIX ASIC~\cite{Bloch:581342}.
Each strip is serviced by an individual FENIX chip, which performs 
amplitude estimation, applies calibration coefficients,
converts energy to transverse energy (\ET),
and identifies the BX corresponding to the energy deposit.
In the case of the EB, 
a sixth FENIX chip sums the energy of the five strips to compute the \ET of a $5\times5$ ``trigger tower'' (TT).
To this it adds a bit, known as the ``fine-grain electromagnetic bit'', which is calculated based on
the energy distribution inside the TT~\cite{Bayatyan:706847} and
computes the strip fine-grain bit
that is used to 
reject anomalous signals in the APDs~\cite{Petyt:2012sf}.
In the case of EE, 
the five-strip sums are transmitted to the off-detector trigger concentrator card (TCC) to complete the 
estimation
of \ET for the TTs. 
The TCC is responsible for transmission of the EB and EE TP’s to the L1 calorimeter trigger every BX
via a mezzanine board on the TCC motherboard, 
known as the ``optical synchronization and link board'' (oSLB).
In the L1 calorimeter trigger the ECAL TPs are combined with the TPs from the HCAL to form L1 electron, photon, {\Pgt} lepton, and jet candidates, as well as \ET sums.
The TCC also handles the storage of the TPs for subsequent reading by the ECAL data concentrator card (DCC).
Not all ECAL signals are saved for offline analysis because of limitations in the ECAL readout bandwidth that restrict the output data rate 
to about 2\unit{kbyte} per event per DCC at L1 trigger rate of 100\unit{kHz}~\cite{NAlmeida_2008}.
A selective definition of interesting regions in the detector, performed by the selective readout processor (SRP),
is used to define regions that are read out without any energy threshold (full readout), 
while the rest of the detector is read out after a zero suppression threshold is applied.
The TCC is responsible for the classification of each TT 
and its transmission to the SRP at each L1 trigger accept signal.
Details of the selective readout (SR) scheme can be found in Ref.~\cite{CMSpaperECALreadout}.

The ECAL trigger system operated with high reliability in Run~2; 
the luminosity-weighted operational efficiency of the system 
(accounting for trigger downtime and deadtime\footnote{Automatic throttling of readout decisions because of too high input rates.}) 
was greater than 99.9\%, similar to Run~1 values despite the more challenging conditions. 
The fraction of the ECAL channels 
that were incorporated as valid inputs
was larger than 99\%,
with only a few towers, strips, and individual channels permanently
excluded from use as inputs in the L1 trigger system.

Several improvements to the firmware and software of the TCCs~\cite{Thiant:2019vlg} 
were implemented to maintain the high ECAL trigger efficiencies for the Run~2 conditions:
one of these was
the automatic detection and masking of noisy or problematic signals from the front-end readout with configurable thresholds,
allowing the ECAL to eliminate spurious TPs, while continuing to deliver usable inputs to the L1 trigger without the need for manual intervention. 
The algorithm used individual thresholds per strip in the EE, which could be adapted to changing LHC conditions, 
as well as to increased radiation-induced energy-equivalent noise in the high-$\eta$ regions of the EE.
The automatic masking decisions were saved in the TP data format and were monitored continuously during operation.
Typically, at most 1-2 TTs are masked by the automatic procedure during the data taking.
An automatic single-event-upset 
recovery algorithm was also implemented to prevent masking too many strips and towers.
As a result of these improvements, which were fully implemented in both the EB and EE before the 2018 run, 
the number of incidents requiring manual intervention, as well as the deadtime and downtime associated with the ECAL trigger system, 
were reduced by about one order of magnitude compared to 2017~\cite{Thiant:2019vlg}.

\section{The ECAL signal reconstruction}\label{sec:reconstruction}
The reconstruction of high-level objects proceeds sequentially,
beginning with the raw data.
The energy deposited in each crystal of the ECAL is reconstructed according to
Eq.~(\ref{eq:uncalib_to_calib}):
\begin{equation}
E = A \, G \, LC(t) \, C(t) ,
\label{eq:uncalib_to_calib}
\end{equation}
where
\begin{itemize}
 \item $A$ is the channel signal amplitude in ADC counts.
 \item $G$ is the conversion factor between ADC counts and energy (in GeV),
  prior to any radiation damage, 
                about 40 (60)\MeV per ADC count in the EB (EE).
 \item $LC$ is a laser correction factor that takes into account crystal and photodetector response losses due to LHC irradiation. 
 It varies with time ($t$) and is measured separately for each crystal.
 An overview of the method is presented in Section~\ref{sec:response} and further details are provided in Ref.~\cite{Anfreville:2007zz}.
 \item $C$ is a combination of the different calibration constants that accounts for the intrinsic differences in individual crystal light-yield and photodetector response.
\end{itemize}

The amplitude $A$ is reconstructed from the ten digitized pulse samples (Fig.~\ref{fig:ECALsignals:pulse_shape}).
For the data collected in Run~1, a weighted sum of the signal samples
(referred to as the ``weights method''~\cite{Bruneliere:2006ra}) was used, with the weights optimized to reduce the noise
contribution for the expected pulse shape. A negative weight on the first three samples allowed a
dynamic, event-by-event pedestal subtraction. The weights method is fast and robust, but is sensitive to OOT-PU,
which can have a sizeable effect when clustering the energy in the ECAL, since it affects coherently many crystals.
To mitigate the effect of the increased pileup in Run~2, a new method called ``multifit''~\cite{EGM18001} was developed.
In this method, a template is fit
to the ten samples, with ten free parameters, corresponding to the amplitudes
of signals in different BXs. 
The signal shape for in-time signals was derived 
using collisions of isolated proton bunches
and high energy signals when there was a negligible OOT-PU contribution.
The signal shapes for the out-of-time signals were obtained from the in-time signal by shifting the time in steps of 25\unit{ns}.
The amplitudes are constrained to be non-negative and determined by a $\chi^2$ minimization procedure, where the covariance matrix 
takes into account 
the correlated components of the noise.
This method is used for 
both the offline and online event reconstruction,
using the non-negative least-squares minimization algorithm~\cite{NNLS} 
to meet the HLT computational requirements.
The multifit method is less sensitive to OOT-PU than the weights method, and it provides a better
energy resolution in Run~2 conditions, but it requires precise measurements of the pedestal, the pulse shape, 
and the covariance matrix. 
All of these
are monitored continuously, as they evolve with time.
Further details of the multifit method are reported in Ref.~\cite{EGM18001}.

Large-amplitude signals can saturate the lower gain(s) of the MGPA, 
and the highest gain, which is not saturated, is selected.
For the signals which are not read out in gain 12, a different reconstruction method is used.
For EB, the signal has a fast rise time and
samples before the switch to lower gain are usually slightly distorted
because the large $\rd{V}/\rd{t}$ saturates the output current of the amplifier.
Since the distortion induced by the OOT-PU is negligible for large pulses,
a simple ``maximum amplitude'' algorithm is used based on the difference between the 6th sample of the
pulse (as shown in Fig.~\ref{fig:ECALsignals:pulse_shape}) and the pedestal for the corresponding gain.
In EE, since the pulse has a slower rise time and a negligible slew rate, 
the multifit algorithm is applied for all gain values.

The time-of-arrival information of the energy deposit is computed with
a dedicated algorithm based on the ratio of consecutive samples, where a fixed pulse shape is assumed
and distortions due to pileup are neglected~\cite{Chatrchyan:2009aj}.
The time alignment of different channels is performed with measured data as described in Appendix~\ref{sec:sub:timingcalib}.

The amplitude and time information are then associated, 
and the energy deposition in each crystal is then calculated
applying the correction factors listed in Eq.~(\ref{eq:uncalib_to_calib}). 
 
Large anomalous signals (``spikes'') in isolated crystals are observed in EB during $\Pp\Pp$ collisions~\cite{Petyt:2012sf}
attributed to direct ionization of the APD silicon by particles.
To reject spikes, 
selections are applied
based on the ratio of the energy deposition 
in the central crystal and its four nearest neighbours,
and, for energy deposits greater than 2\GeV, based also on the time information.
Additional details on coping with the spikes are discussed in Section~\ref{subsec:performance:trigger}.

\begin{figure}[htbp]
\centering
    \includegraphics[width=0.45\textwidth]{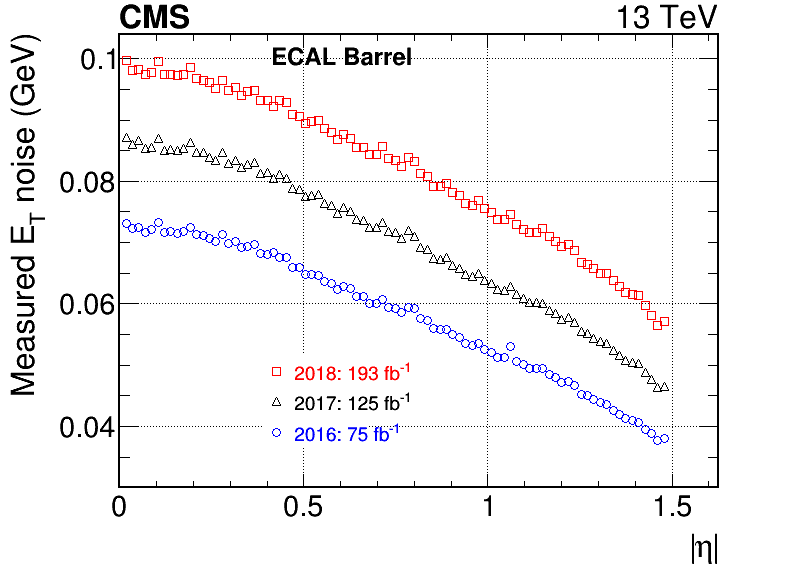}
    \includegraphics[width=0.45\textwidth]{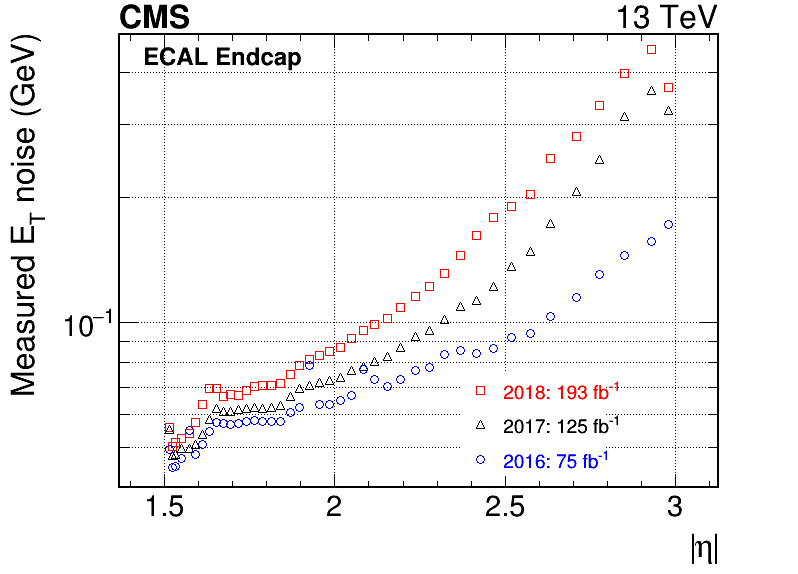}
    \includegraphics[width=0.45\textwidth]{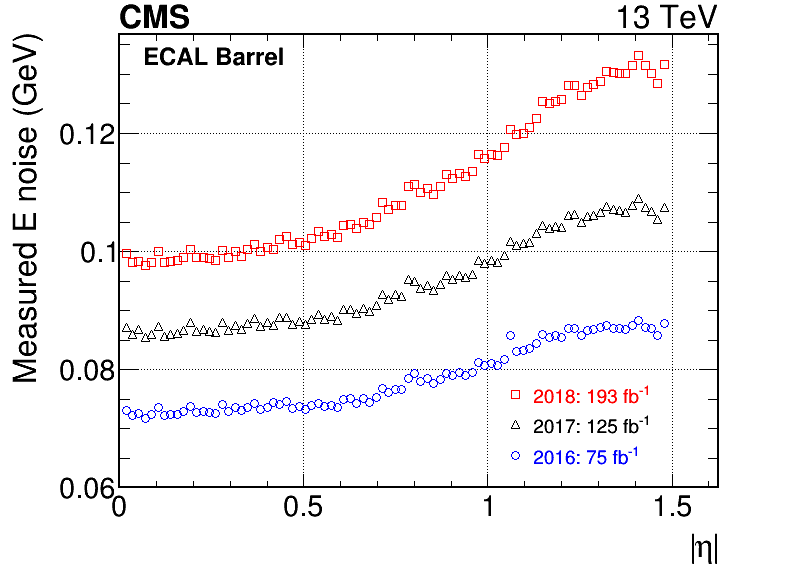}
    \includegraphics[width=0.45\textwidth]{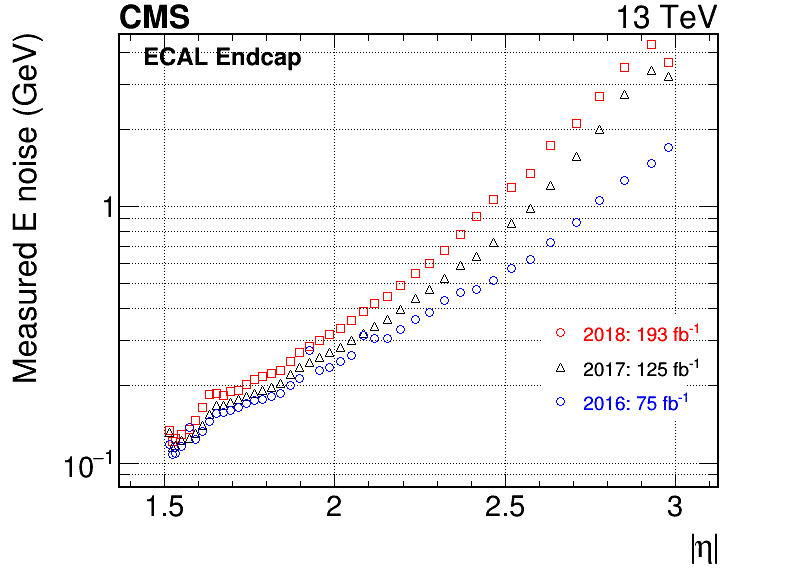} 
\caption{Average energy-equivalent noise for the EB (left column) and the EE (right column) 
   at the end of 2016, 2017, and 2018,
   measured 
   as the 1 sigma variation of the pedestal baseline values
   for each channel
   and converted into energy using Eq.~(\ref{eq:uncalib_to_calib}).
   Both transverse energy (upper row) and energy (lower row) are shown.
   The integrated luminosity for the different years refers to the cumulated delivered luminosity 
   since the beginning of Run~1.
}
\label{fig:NoiseAndThresholds}
\end{figure}

\begin{figure}[htbp]
\centering
    \includegraphics[width=0.49\textwidth]{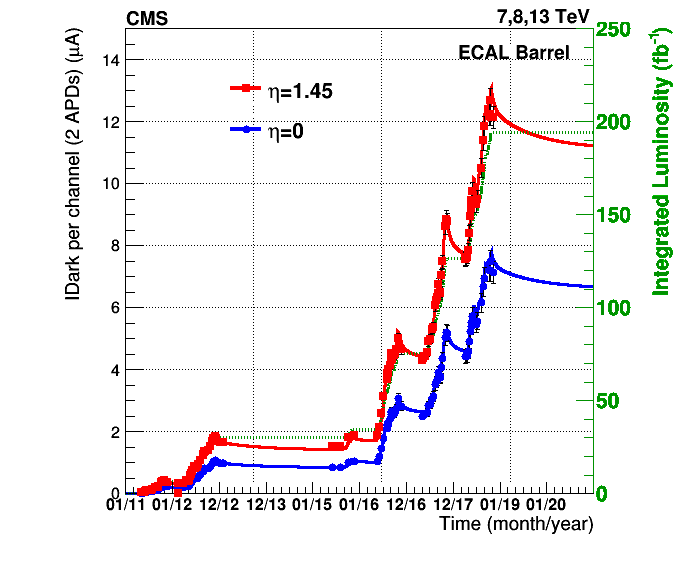}
    \includegraphics[width=0.49\textwidth]{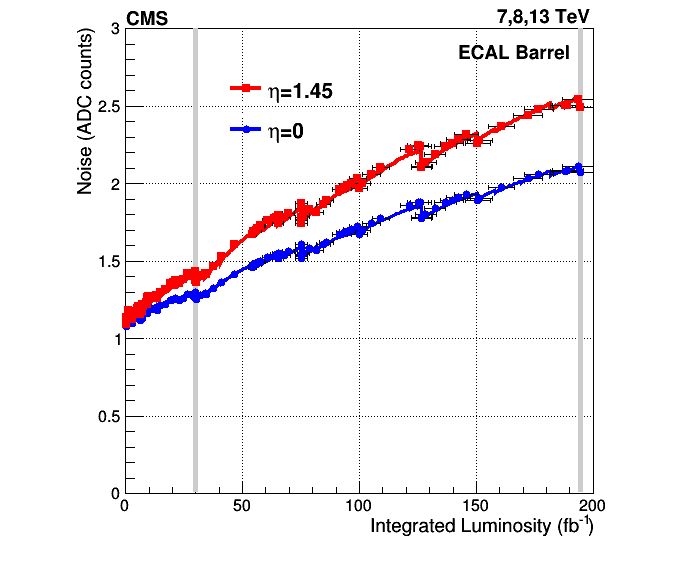}
\caption{
   The APD dark current (IDark) evolution versus time in EB (left), in red for the channels at $\abs{\eta}=1.45$ and in
   blue for the channels at $\eta=0$, with the continuous lines representing the prediction of the model.
   The delivered integrated luminosity since the beginning of Run~1 is also shown in green.
   The scale of the integrated luminosity axis is chosen such that it tends to overlap with
   the dark current at $\abs{\eta}=1.45$, in order to emphasize visually the strong correlation between 
   dark current and integrated luminosity. The vertical bars on the points represent the uncertainty in the dark current measurement.
   On the right, the APD noise estimated from the dark current (continuous line) is compared 
   with the direct noise measurement (data points) 
   for the two $\eta$ regions. The vertical grey lines in the right plot indicate the region corresponding to
   Run~2. The horizontal bars on the points represent the uncertainty on the luminosity.
}
\label{fig:NoiseAPD}
\end{figure}

The average noise levels in the EB and EE are shown in Fig.~\ref{fig:NoiseAndThresholds}.
The electronic noise is uniform in EE, and the large variation in the energy equivalent noise along $\eta$ and in time
in EE is due
to the amplification of the electronic noise due to large laser correction values.
The root-mean-square (RMS) of the noise along $\phi$ for channels at a given $\eta$ over its mean value is typically about 30\%, but it can increase by up to 100\%  
for $\abs{\eta}>2.5$.
The electronic noise in the EB increases with time due to the increase of the radiation-induced APD dark
current, shown in Fig.~\ref{fig:NoiseAPD}.
The continuous line in the figure shows the prediction of an APD ageing model,
which is in good agreement with the measurements.

After the removal of anomalous signals, the remaining hits, designated ``PF rechits'',
are input to the 
global event reconstruction that uses the 
particle-flow (PF) algorithm~\cite{Sirunyan:2017ulk} 
to reconstruct and identify the high-level objects (electrons, photons, jets, and {\Pgt} leptons), and to estimate the
missing transverse momentum
for the event.
An energy threshold is applied to the PF rechits to reduce the noise contamination shown in Fig.~\ref{fig:NoiseAndThresholds}.
In this step of the reconstruction, 
energy deposits in crystals are built around the crystal with the highest energy within its $3\times3$ matrix of neighbouring crystals,
known as a ``seed crystal''.
The seed crystal is required to have a reconstructed amplitude greater than 230 (600)\MeV in the EB (EE) 
and a transverse energy $\ET>150\MeV$ in the EE.
A topological nearest-neighbour clustering around seed crystals is then performed.
 
The crystals' energy can be shared between clusters 
with a weight calculated based on the distance between the crystal and the centres of the clusters
assuming a Gaussian profile of the energy deposition.
In addition, the energy from the ES is assigned to the
clusters 
in the EE using a geometrical matching procedure.
The ES pulse is sampled every 25\unit{ns},
and three samples are used to reconstruct the deposited energy in the silicon strips
by means of weighted sum of the samples~\cite{Chatrchyan:2013dga}.
The clusters in the EB and EE are aggregated to form ``SuperClusters'', or SCs,
which correspond to the energy deposition from electrons and photons. 
Details of the formation of electron and photon clusters,
including algorithms and corrections,
are described in Ref.~\cite{EGM17001}.
Corrections to these clusters are applied to take into account the effects of boundaries between detector modules
and to obtain a uniform response across the detector.
Additional corrections based on multivariate techniques are applied to electrons and photons.
These techniques use shower shape variables, energy, and pileup multiplicity 
to improve the energy resolution~\cite{EGM17001}.

For the electron and photon reconstruction, the EB, EE, and ES need to be aligned
with the tracker detector. The alignment procedure for EB and EE uses isolated electrons produced in \PW/\PZ boson decays,
while for ES all charged-particle tracks from minimum ionizing particles are used. 
The resulting precision in the alignment has a negligible impact on the performance
in terms of the reconstruction efficiency of electrons and photons, and their energy resolution.
A detailed description of the alignment procedures can be found in Appendix~\ref{sec:alignment}.

\section{Response, monitoring, and corrections}\label{sec:response}
The energy response of the ECAL changes continuously with time.
Different methods are adopted to constantly monitor this time-dependent drift in situ.
Corrections are computed and applied to maintain the stability of the reconstructed energy scale and resolution.
The main source of these changes is the ageing of the crystals and of the photodetectors caused by the high radiation levels at the LHC.
The effect of radiation on the crystals is the creation of colour centres that decrease the transparency and therefore reduce the light collected by the photodetector. 
This damage is both electromagnetic-induced and hadron-induced. 
At room temperature, 
only colour centres from electromagnetic-induced damage experience thermal annealing.
This results in recovery of the crystal transparency that is noticeable between LHC fills,
and, in some cases,
also toward the end of the fill, when the luminosity drops to a level where the recovery outweighs the creation of new colour centres.

A dedicated light-monitoring system~\cite{Anfreville:2007zz, Zhang:2005ip}, using lasers,
is used to measure the transparency of each crystal and the photodetector response.
During the data taking, the laser light is injected exploiting the LHC abort gap and a full cycle of the monitoring system takes 40 minutes.
A blue laser, with a wavelength of 447\unit{nm}, which is close to the peak of the lead tungstate scintillation light spectrum, 
is used to measure and correct for changes in crystal transparency and photodetector response. 
Additional lasers with different wavelengths provide complementary transparency measurements 
that are used as a cross-check of the blue laser ones,
but they are not used in the derivation of the laser correction.
Light from a light-emitting diode is also injected in the EE to keep the photodetector active
even in the absence of collisions thus stabilizing the photodetector response (Section 4.3.2 of \cite{Chatrchyan:2008zzk}).
The laser light is injected into each crystal by optical fibres 
via
a series of splitters.
The last splitter divides the light into 200 fibres, each connected to a crystal,  
and two fibres that inject the light into PN diodes used to measure the amplitude of the laser light pulse.
This group of 202 fibres is referred to as a "harness".
The variation in transparency is obtained by measuring the ratio of the light measured by the 
photodetectors and by the reference PN diodes, 
so that the transparency variation estimation is not affected by pulse-to-pulse variations in the laser amplitudes.

\begin{figure}[htb]
\centering
\includegraphics[width=0.8\textwidth]{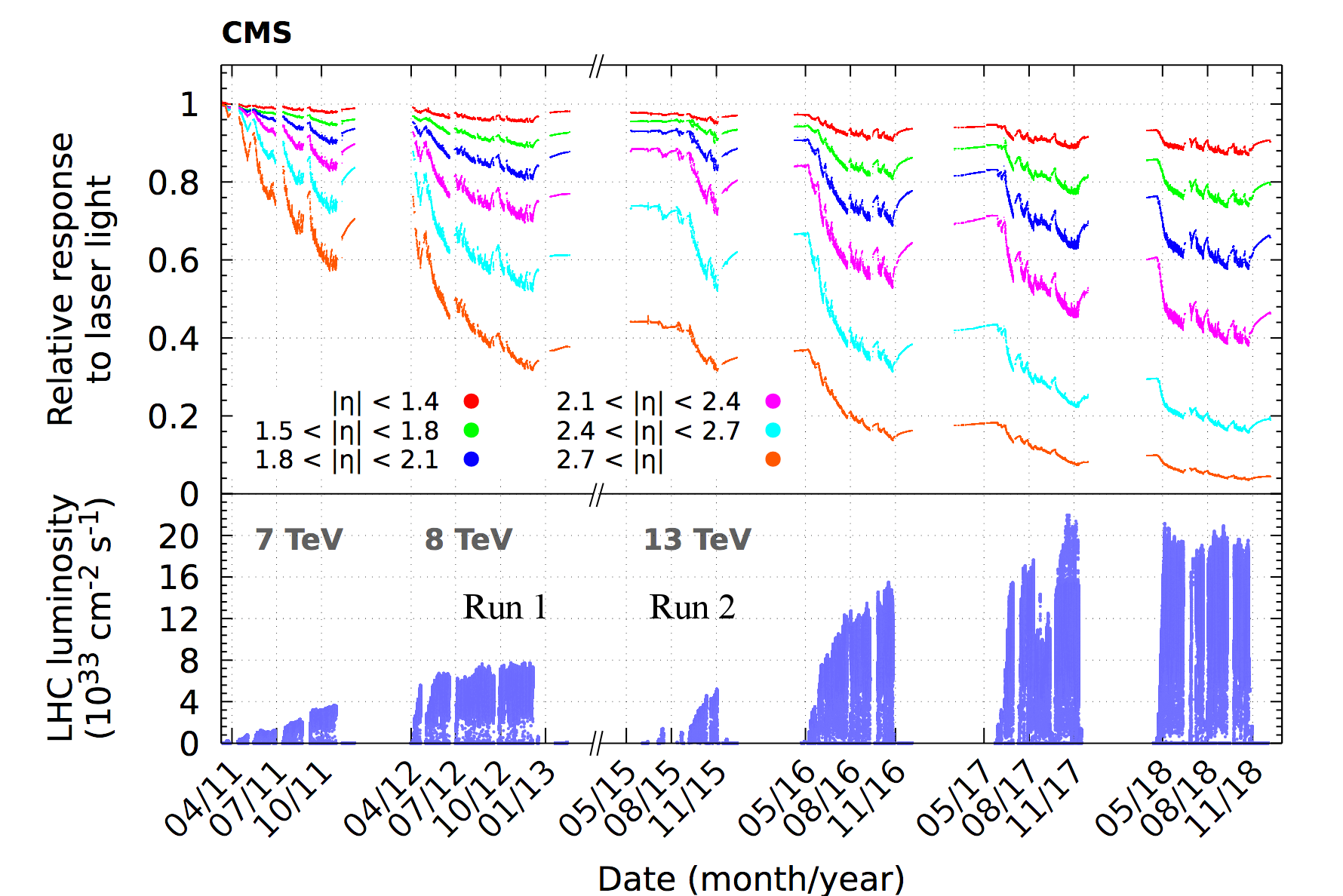}
\caption{
    Relative response to laser light (440\unit{nm} in 2011 and 447\unit{nm} from 2012 onwards) injected into the ECAL crystals,
    measured by the ECAL light monitoring system, averaged over all crystals in bins of pseudorapidity ($\abs{\eta}$).
    The lower panel shows the instantaneous LHC luminosity as a function of time.
}
\label{fig:TransparenncyEvolution}
\end{figure}

The evolution of the ECAL response to laser light between 2011 and 2018 is shown in Fig.~\ref{fig:TransparenncyEvolution}.
By construction the first point is set to 1, and points with different colours correspond to different $\eta$ regions.
The response change observed in the ECAL channels is up to 10\% in the EB and reaches up to 50\% at $\abs{\eta}=2.5$,
the limit of the tracker 
coverage.
The response change is up to
98\% in the highest $\eta$ region, above $\abs{\eta}=2.7$.
This loss in transparency, together with the pileup, has a significant impact in the resolution in this region as reported in Section~\ref{subsec:performance:energy}.
The recovery of the crystal response during periods without collisions is visible.

As mentioned in Section~\ref{sec:reconstruction}, a time-dependent correction factor $LC_{i}(t)$,
applied to the measurements of the energy deposited in the $i$th crystal at time $t$,
is derived from the
ratio of the initial response to laser light of the crystal at the start of 
the first year of 
data taking ($R_{i}(0)$)
to the response ($R_{i}(t)$)
at the time $t$ as follows: 
\begin{equation}
LC_{i}(t) = \left[\frac{R_{i}(0)}{R_{i}(t)}\right]^{\alpha}
\label{eq:laser_and_alpha}
\end{equation}
The parameter $\alpha$ accounts for the difference
between the optical paths of the laser and scintillation light.
The validity of Eq.~(\ref{eq:laser_and_alpha}) is limited to small values of transparency loss;
for larger values a more complex equation is needed~\cite{Bonamy:1998rha}.
The effect can be accounted for by changing
the parameter $\alpha$, which becomes an evolving 
parameter.
The values of $\alpha$ 
are determined to obtain the maximum stability in time of the energy response
with \zee events and 
the maximum stability in time of the ratio of the measured energy fluxes at different $\eta$ regions.
The evolution of $\alpha$ is particularly significant 
in the EE region, where the response losses are larger.
The $\alpha$ values measured in situ are about 1.5 in EB, and between 0.6 and 1.1 in EE.
The transparency of each crystal is measured every $\approx$40 minutes.
$R(t)$ at a given time $t$ is estimated with a linear interpolation between the two measurements 
before and after time $t$. Thus, smoothing of the $R(t)$ is achieved.
The $LC_{i}(t)$ corrections are provided and validated within 48 hours from the data taking,
in time for the CMS prompt offline reconstruction.

For energy measurements at the trigger level,
correction factors 
for the change in transparency
are derived using measurements 
from the light monitoring system
recorded in the preceding days.
These trigger-level corrections were first applied in 2012 and were updated
weekly
for 22 individual $\eta$ 
intervals
in EE.  
For Run~2, because of the higher beam intensities and correspondingly larger response changes,
these corrections were applied per crystal and extended to the EB. From 2017 onwards, 
an automated procedure was employed to validate the effect of the updated conditions on the L1 and HLT trigger rates, 
and the frequency of the updates was increased to twice per week to better track the response changes versus time.
These corrections are particularly important to maintain stable trigger rates and efficiencies,
and to provide the best achievable energy resolution for
electrons, photons, and jets at the L1 trigger.
In Run~3, the corrections are updated once per LHC fill, 
which is the highest frequency compatible with the CMS data acquisition system.

\subsection{Monitoring of the energy response}\label{subsec:energymonitoring}
Three independent methods are used to provide prompt feedback on the energy response stability during data taking:
\begin{itemize}
\item the \Pgpz method, based on the invariant mass of photons from \Pgpz decays reconstructed from data collected in a special data stream,
\item the $E/p$ method, where the reconstructed supercluster energy for electrons is compared to the momentum measured in the tracker,
\item \zee events are selected and the invariant mass of electron-positron pairs from {\PZ} boson decays is used.
\end{itemize}
 
\begin{figure}[htb]
\centering
\includegraphics[width=0.75\textwidth]{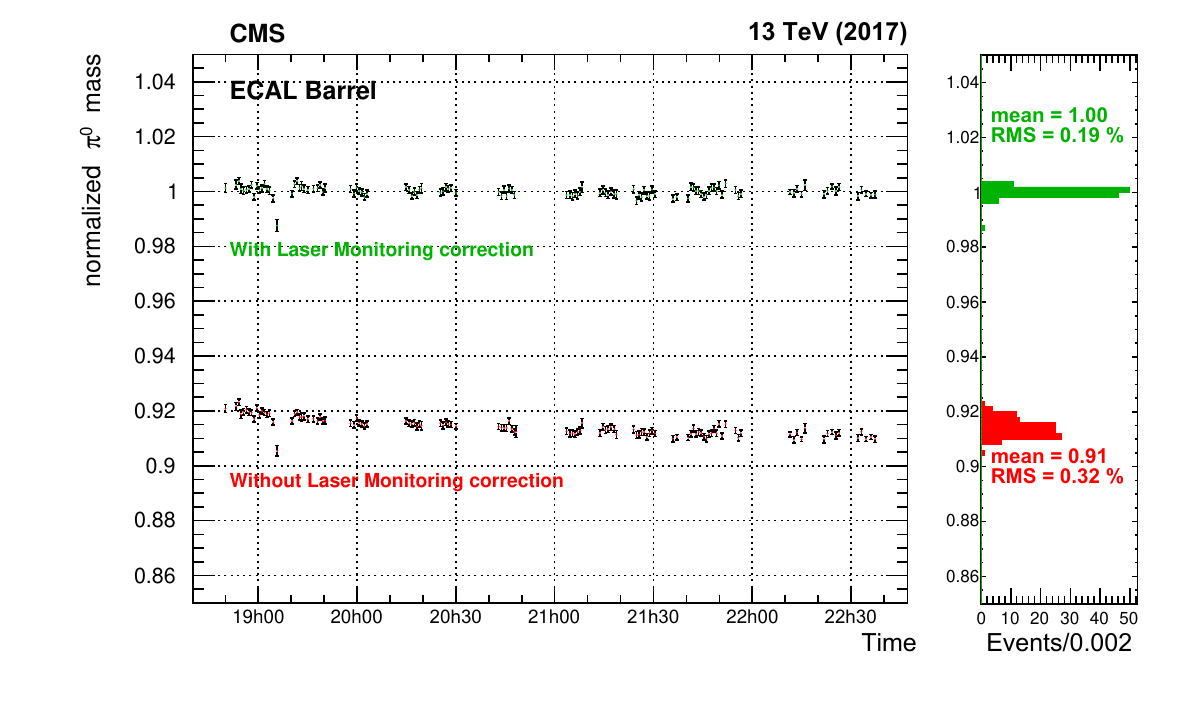}
\caption{
   The stability of the relative energy scale measured from the invariant mass distribution
    of $\Pgpz \to \Pgg \Pgg$ decays in the EB 
    as a function of time, over a period of 3 hours during an LHC fill.
    The plot shows the data with (green points)
    and without (red points) the light-monitoring corrections applied.
    The vertical bars on the points represent the statistical uncertainty.
     The right-hand panel shows 
     the 1D projections of
     the points in the left panel. The mean value and root-mean-square (RMS) are shown.
}
\label{fig:pi0monitoring}
\end{figure}

The \Pgpz monitoring uses a special high-rate data stream,
where only energy deposits in ECAL for likely \Pgpz candidates are stored.
Figure~\ref{fig:pi0monitoring} shows the stability of the energy response and 
the frequency 
of monitoring check-points
that can be reached with this method.
The reconstruction only works when the rest frame of the \Pgpz has a limited Lorentz boost in the lab frame,
so that the energy deposits of the two photons do not overlap.
This limits the energy range of the photons that can be used. 
Details of the \Pgpz stream and the reconstruction are provided in Appendix~\ref{sec:sub:pi0}.

\begin{figure}[htbp]
\centering
\includegraphics[width=0.60\textwidth]{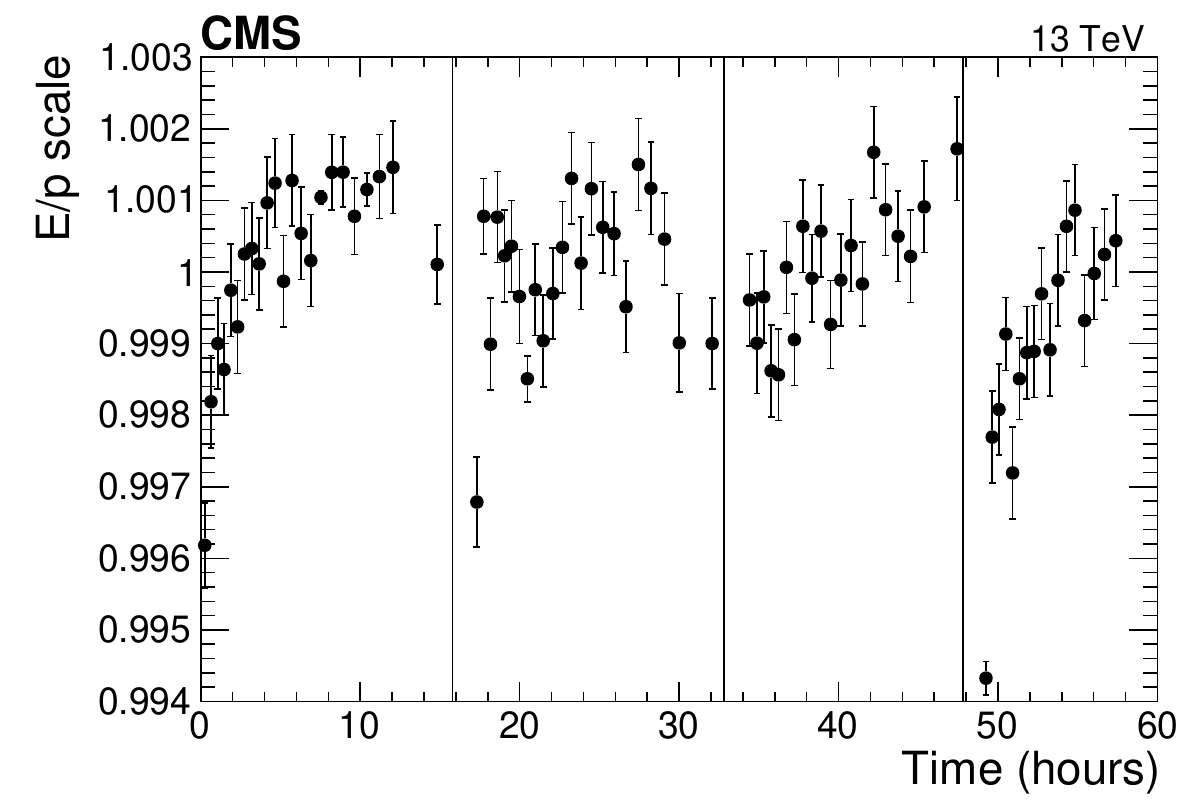}
\caption{
     The stability of the ECAL energy scaling factor,
     measured from the
     ratio of the energy of electrons, as measured by the ECAL (EB, $\abs{\eta}<0.43$), 
     and the electron  momentum,
     as measured by the tracker. 
     The stability is shown
     for a period of 4 consecutive LHC fills, with
     the limits of the fills delineated with vertical lines.
     The laser corrections were applied.
     Each point of the plot is obtained from a fit of the $E/p$ distribution 
     to approximately one hour of data taking.
     The vertical bars on the points represent the statistical uncertainty.
     The origin of the residual scale variation observed can be due to pileup variation
     and to less accurate measurement of transparency variation at the very beginning
     of the fill, when the response changes rapidly with time.
     After the beginning of a fill, the change of scale is less than 0.4\%.
 }
 \label{fig:EoPmonitoring}
 \end{figure}

The $E/p$ method uses high-energy electrons from \PW/\PZ decays and 
is based on the ratio between the supercluster energy measured in the ECAL
and the momentum of
electron tracks, measured by the tracker detector.
Compared to the \Pgpz method, the available data for the $E/p$ method is much lower,
thus the method requires more integrated luminosity to obtain a single monitoring point,
as shown in Fig.~\ref{fig:EoPmonitoring}.
However, since the average energy deposited in the ECAL is much larger than 
that
from \Pgpz decays,
the effect of 
pileup is correspondingly much smaller~\cite{EGM18001}.
The criteria to identify a sample of electrons with high purity are detailed in Appendix~\ref{sec:sub:Eop}.
The stability of the ECAL response is obtained from
template function fits
to the $E/p$ distributions in different $\eta$ regions.
The templates for each $\eta$ region are obtained from the $E/p$ distributions with all the available data.
The data are divided into time intervals with about 5000
electrons per interval in each $\eta$ region,
and a scaling factor for the reconstructed energy is determined by
the fitting procedure to match the template distribution
with an accuracy of 0.05\%.
The RMS of the results from the fitting procedure from all time intervals
is used as an estimate of the stability of the energy measurement in the ECAL.
More details on the $E/p$ method can be found in Appendix~\ref{sec:sub:Eop}.

Lastly, the invariant mass 
of electron-positron pairs from \PZ boson decays,
as described in Appendix~\ref{sec:sub:calibrationZee}, 
is used to measure the
stability of the ECAL response, and to correct for any observed drift.

\subsubsection{Regional energy drifts}\label{sec:sub:harness}
\begin{figure}[htbp]
\centering
 \includegraphics[width=.6\textwidth]{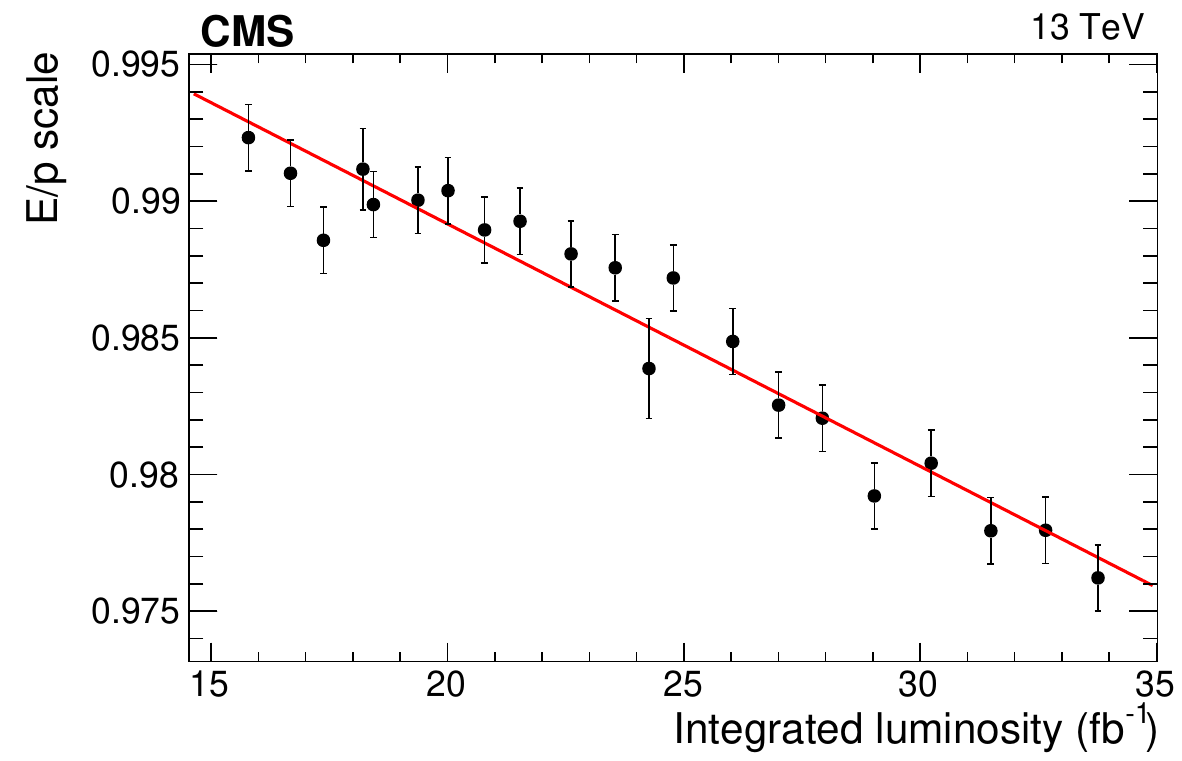}
 \caption{
        The average ratio of the ECAL energy to the track momentum for electrons from \PW boson decays 
        reconstructed in the crystals of one light-monitoring harness in EB, as a function of delivered integrated luminosity during 2018.
        The vertical bars on the points represent the statistical uncertainty.
        A linear fit to the data (red line) is superimposed.  
        For this particular light-monitoring harness a drift of 1\% every 10\fbinv is measured.
        The dimension of the intervals on the horizontal axis is chosen such that each point is obtained from a sample of about 10000 electrons.
        This corresponds to about 1--2 days of data taking for the modules in the inner EB ($\abs{\eta} < 0.8$),  
        and about 2--3 days for the modules in the outer EB ($0.8 < \abs{\eta} < 1.5$).
        }
 \label{fig:Eop_drift_fit}
 \end{figure}

Drifts in the energy scale of up to a few percent per year have been observed in the EB 
with a regional granularity corresponding to a single light-monitoring harness.
The variations depend on the radiation damage induced by ionizing radiation in the monitoring system (the PN diode and light distribution system),
which is directly proportional to the integrated luminosity.
In fact, as reported at the beginning of Section~\ref{sec:response}, the laser corrections in the EB are calculated by measuring 
the ratio of the laser light detected by the APDs to the signal in the reference PN diodes. 
An internal charge injection system, designed to monitor and correct the response of the PN readout electronics to a 0.1\% accuracy,
did not perform as designed for some modules, and the corresponding corrections were not applied during most of Run~2.
Therefore, an effective correction as a function of the integrated luminosity based on the $E/p$ method was used to stabilise the response of the crystals,
and was applied for each of the 324 harnesses.
As shown in Fig.~\ref{fig:Eop_drift_fit}, the dependency on the luminosity can be parametrised with a linear function, 
which is then used to correct the energy scale in each event.
Given the layout of the light distribution system (with the fibres behind the crystals for EE, 
and in front, closer to the interaction point, for the EB), 
the laser monitoring system for the EE is exposed to less radiation than the system for the EB,
and the drifts in the energy scale observed in the EB are not noticeable in the EE.

\section{Calibration methods}\label{sec:calibration}
The calibration of the ECAL
proceeds in three main steps:
\begin{itemize}
 \item stabilization of the energy scale for each crystal/region as a function of time,
 \item equalization of the ECAL response for different crystals at the same $\eta$ coordinate (intercalibration),
 \item equalization of the ECAL absolute scale as a function of $\eta$ ($\eta$-scale).
\end{itemize}

The procedure for the first step is described in Section~\ref{sec:response}.
It includes response corrections measured by the light monitoring system, 
corrections for the energy scale calculated with \zee events, 
and corrections for regional drifts with the $E/p$ method.

Four different methods have been developed to perform the intercalibration process
and derive for each crystal a multiplicative intercalibration coefficient (IC).
These are:
\begin{itemize}
\item the $\phi$-symmetry method, which uses the azimuthal symmetry of the energy flow in the ECAL,
\item the \Pgpz method, based on the  invariant mass of photons from \Pgpz decays,
\item the $E/p$ method, based on the ratio between the energy in the calorimeter and the momentum of electrons from \PW or \PZ boson decay,
\item the \zee method, based on the invariant mass of electron-positron pairs from \PZ boson decays.
\end{itemize}

In the absolute scale calibration the energy response for each $\eta$-ring
is equalized using \zee decays, 
and a correction factor is applied to the measured data so that
they match the 
simulation.

The calibration constant $C$ in Eq.~(\ref{eq:uncalib_to_calib}) accounts for both the intercalibration and the absolute scale calibration as a function of $\eta$.

During Run~2, the noise and pileup levels were substantially larger than in Run~1.
Changes were made to the calibration procedure, including raising
the thresholds for the $\phi$-symmetry and \Pgpz selections due to the higher noise level and pileup.
Details of these changes are given in Appendices~\ref{sec:sub:phisymm} and ~\ref{sec:sub:pi0}.
Increasing these thresholds reduced the number of events,
reducing the ability to 
track the scale variations over short intervals
and also preventing the use of
some calibration methods in certain regions of the detector, in particular the \Pgpz method in the high-$\abs{\eta}$ regions in EE.

The increased luminosity, noise, and pileup 
also required higher L1 and HLT thresholds for single-electron triggers, 
reducing the number of events available for calibration with the $E/p$ method.
This was improved by using new approaches,
such as creating a dedicated event stream with regional reconstruction of the electrons,
and the adoption of optimized identification and isolation selections.
Using these new methods a stable rate of single-electron events was achieved, sufficient to calibrate
the ECAL in narrower regional and time intervals than in Run~1.
Moreover, the $\eta$-scale corrections have also been derived more frequently, approximately every 5\fbinv
of collected integrated luminosity,
to correct for any slow drifts in detector response versus time. 
 
One consequence of the increased delivered luminosity was that it allowed for the use of new 
calibration techniques,
such as a complete calibration with \zee events.
In Run~1, \zee events were only used to perform the absolute scale measurement,
while
in Run~2, they were used for the intercalibration.
This allowed for a
precise intercalibration in the highest $\abs{\eta}$ regions in the EE, outside the tracker acceptance,
where the other methods are not viable.
With the selection of 
\zee candidates and exploiting the
invariant mass constraint with one electron reconstructed using only ECAL information in the high-$\abs{\eta}$ region,
both the per-crystal intercalibrations and the absolute energy scale were measured.

A detailed description of the calibration procedures is given in Appendix~\ref{sec:calibration-intro-appendix}, 
while the procedures used to estimate the precision of the intercalibration and their results are detailed in the following section.

\section{Precision of the intercalibration methods and combination of the results}\label{sec:combination}

\begin{figure}[htbp]
\begin{center}
\includegraphics[width=0.78\textwidth]{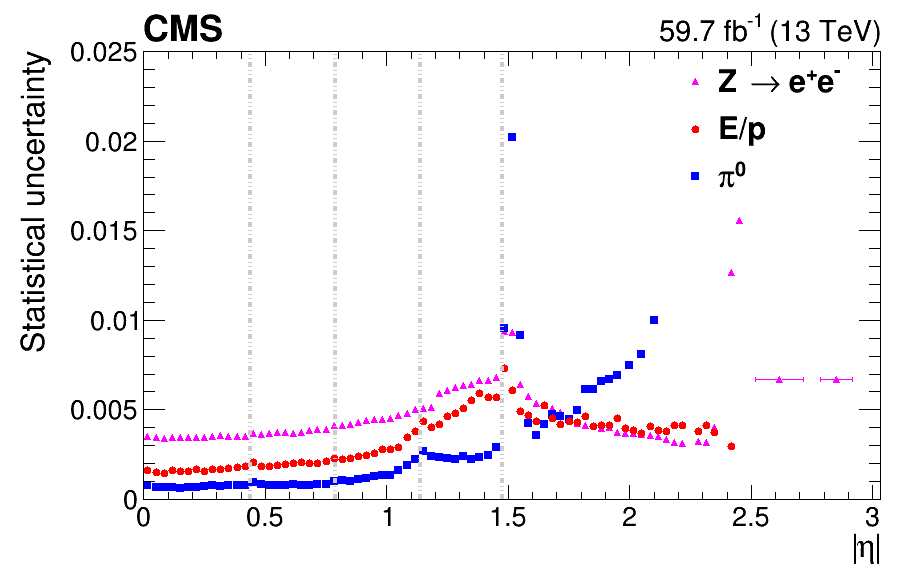}
\caption{
    The statistical uncertainties in the different intercalibration methods
    for data collected in 2018.
    The vertical dotted lines mark the boundary between the ECAL modules in the EB
    and the EB/EE transition.
    The increase in the statistical uncertainties for the last two points
    for the \zee method close to $\abs{\eta} = 2.5$ are due to 
    a reduction in the efficiency of the selection of two electrons
    since this $\abs{\eta}$ value corresponds to the end of the tracker coverage.
    A similar performance is observed for 
    data collected in 2016 and 2017.
}
\label{fig:icPrecisionStat}
\end{center}
\end{figure}

The intercalibration methods to equalize the ECAL energy response in the azimuthal coordinate $\phi$
use data sets that are largely independent, 
and therefore have different statistical and systematic uncertainties.
The statistical precision of each method, except \zee intercalibration,
is evaluated by comparing the ICs derived from two non-overlapping subsamples 
containing events with odd and even event number.
For the \zee intercalibration, the statistical uncertainty is obtained directly from the fitting procedure used in this method.
The estimated statistical precision is similar for the three years of Run~2 and is shown for 2018 in Fig.~\ref{fig:icPrecisionStat}.
All methods show an increase of the statistical uncertainty close to the transition between the EB and the EE ($\abs{\eta} = 1.479$), 
related to a reduction in the efficiency of the selections in that region.

For the $\phi$-symmetry method, a statistical uncertainty of better than 0.5\% is achieved with 
an integrated luminosity of about 0.5\fbinv.
However, the method is limited in its ultimate precision due to systematic uncertainties
arising from the presence of material between the interaction point and the ECAL.
Since in Run~2 
the uncertainty in the $\phi$-symmetry intercalibration is much
larger than in the other methods, 
this method is not used in the final determination of the ICs.
While the $\phi$-symmetry method does not yield precise ICs, 
most of the systematic uncertainties do not impact the time evolution of the $\phi$-symmetry ICs.
In particular, the variation of the $\phi$-symmetry IC versus time can be used 
to promptly measure the time evolution of the ECAL calibration for each crystal.
In the EB, the $\phi$-symmetry method has been also used to derive and check the parameter $\alpha$ of Eq.~(\ref{eq:laser_and_alpha}).

For the \Pgpz method,
a statistical uncertainty of better than 0.4\%
is reached in the EB 
with
an integrated luminosity of about 10\fbinv.
It varies from 0.1\% in the inner EB modules ($\abs{\eta}<0.8$),
to about 0.4\% in the outer EB modules ($0.8<\abs{\eta}<1.5$).
Variations in the statistical uncertainty in the EB
are correlated with the amount of material upstream of ECAL, which is larger at higher $\abs{\eta}$,
and leads to more frequent photon conversions,
thus causing more events to fail the \Pgpz selection.
A statistical uncertainty in the EE of about 1\% can be achieved with an integrated luminosity of about 10\fbinv. 
It is larger than in the EB due to the larger particle multiplicity and detector noise in the EE, which
necessitates a tighter event selection and leads to a lower attainable efficiency.
In Run~2, the \Pgpz method is used to obtain ICs in the EE for $\abs{\eta}<2.1$;
above that, the signal-to-background ratio is significantly reduced and the \Pgpz invariant mass peak
cannot be reconstructed with sufficient accuracy.
 
The $E/p$ method requires an integrated luminosity of about 50\fbinv,
which corresponds to an integrated luminosity collected during a 
calendar year in Run~2,
to reach a statistical precision
that is similar to the other methods.
A precision of about 0.2\% in the inner EB ($\abs{\eta}<0.2$) and about 0.7\% in the outer EB ($1.3<\abs{\eta}<1.5$) is achieved.
The reduction of the statistical precision with increasing $\abs{\eta}$ is related to 
the number of events collected for each crystal, 
which is 40\% higher in the inner EB than in the outer EB,
and to the narrower $E/p$ distribution in the central part of the detector.
The systematic uncertainty of this method is mainly due to 
differences in the material upstream of ECAL for different $\phi$ regions, that cause
various levels of bremsstrahlung emission, 
introducing biases in the reconstruction of the electron and positron momenta.

The \zee calibration method uses only the energy deposited in the ECAL and the invariant mass distribution of the dielectron system.
As a result this method is largely independent from the uncertainties in
the 
upstream detector components
in front of the ECAL. 
Like the $E/p$ method,
the \zee method requires an integrated luminosity of about 50\fbinv
to reach a statistical precision
that is similar to the other methods,
varying from about 0.4\% in the inner EB to about 0.9\% in the outer EB, and about 0.5\% in the EE.
In the calculation of the calibration constant, 
the electron energy is scaled by the IC of the seed crystal of the electron SC,
while it should involve all the ICs of the crystals belonging to the SC.

This is the dominant systematic bias for this method.
To estimate the 
bias from this assumption, 
the fit is performed a second time after applying the ICs from the first iteration.
The difference between the two is approximately $0.7\,\sigma_{\text{stat}}$ in the EB
and $1.3\,\sigma_{\text{stat}}$ in the EE, 
and is taken as an estimate of the systematic uncertainty.

The ICs calculated by each method ($\Lambda$) are then combined
using an estimate of their overall precision $\sigma_{\Lambda}$, including their systematic uncertainties.
Estimates of the systematic uncertainty in the \Pgpz and $E/p$ methods are 
obtained by studying the impact of the calibrations on the line shape of the \PZ boson invariant mass distribution
to the reconstructed peak width in \zee decays.
The electron energy resolution contributes to the 
width of the reconstructed \PZ peak,
and the method 
developed for the intercalibration with \zee events also  
measures the relative energy resolution per electron $\left(\sigma_E/E\right)$,
as a function of $\eta$. This is discussed further in Appendix~\ref{sec:sub:calibrationZee}.
A series of
simulation-based
studies have been performed to
analyze possible biases and uncertainties related to the use of the \zee method to 
estimate the scale and the resolution of electrons, as summarized
in Appendix~\ref{app:Zee}.
There is a small 
overestimation
in the resolution, which is taken into account in the procedure to measure the
uncertainty of the different intercalibration methods detailed below.

The energy resolution $\left(\sigma_E/E\right)_\Lambda$ 
measured from data, when using the ICs from method $\Lambda$ in the energy reconstruction,
is parametrised from:
\begin{equation}
\left(\frac{\sigma_E}{E}\right)_\Lambda^2 = \rho^2 \sigma_{\Lambda}^2 + \sigma_{0}^2,
\end{equation}
where $\sigma_{0}$ represents the contribution to the resolution not related to the IC precision,
and $\rho$ is a parameter that relates the precision of the IC to the electron energy resolution.
The parameter $\rho$ is measured by applying
a smearing of the ICs according to Gaussian distributions,
and fitting the electron energy resolution as a function of the Gaussian smearing.
The resulting value for the parameter $\rho$ is about 0.7, with a mild $\eta$ dependence. 
This value is consistent with the fraction of SC energy
contained in the central crystal, and reflects the fact that the ICs' uncertainty is diluted when 
summing over all the crystals in an SC. 
By definition, the parameter $\rho$ also absorbs the bias in the measurement of $\sigma_E/E$ 
described in Appendix~\ref{app:Zee}.

Once the uncertainty of the ICs for a reference method ($\sigma_{\text{ref}}$) is known,
the precision of the calibration constants for a particular IC method $\Lambda$ can be obtained as:
\begin{equation}
\sigma_\Lambda = \sqrt{ (\sigma_{\text{ref}})^2  +  \rho^{-2}\, \left[ \left(\frac{\sigma_E}{E}\right)_\Lambda^2 -  \left(\frac{\sigma_E}{E}\right)_{\text{ref}}^2  \right] }.
\label{eq:ic:master}
\end{equation}

The \zee method is chosen as the reference method (ref),
since it is less sensitive to biases arising from the tracker momentum calibration, pileup, and upstream material budget.

The calibration methods can be combined by weighting each method with their respective IC precision:
\begin{equation}
w_\Lambda = \frac{ \frac{1}{(\sigma_\Lambda)^2 } }{\sum_k  \frac{1}{(\sigma_k)^2}}
\end{equation}
where the index $k$ runs over the calibration methods to be combined.
Assuming all the methods are independent, 
the precision of the combined ICs ($\sigma_{\text{comb}}$) is then given by:
\begin{equation}
\sigma_{\text{comb}} = \sqrt{\frac{1}{\sum_k  \frac{1}{(\sigma_k)^2}}}.
\label{eq:sigmacombo}
\end{equation}

\begin{figure}[htbp]
\begin{center}
\includegraphics[width=0.78\textwidth]{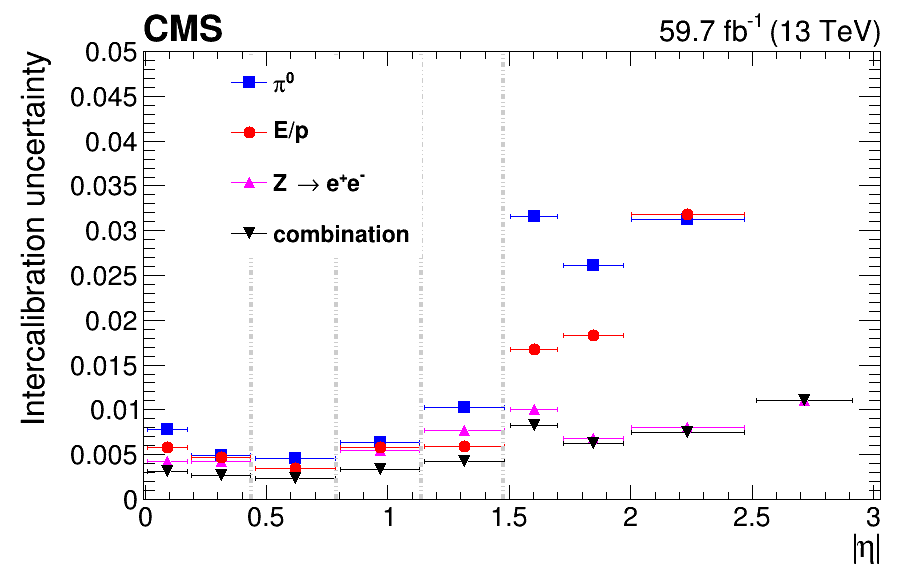}
\caption{
    The precision of the different IC measurement methods, as well as their combination, in 2018.
    The vertical dotted lines mark the boundary between the ECAL modules in the EB
    and the EB/EE transition.
    In the region outside the tracker coverage, $\abs{\eta}>2.5$, only the \zee method is available for calibration.
    Similar performance is observed in data collected in 2016 and 2017.
}
\label{fig:icPrecision}
\end{center}
\end{figure}

The measured uncertainties in the calibrations obtained with 2016, 2017, and 2018 data are similar.
Figure~\ref{fig:icPrecision} shows the estimated uncertainties for 2018.

The precision of the \Pgpz method is limited by its systematic uncertainty.
The $E/p$ method has comparable statistical and systematic uncertainties in the EB, but is dominated by the systematic uncertainty in the EE.
The \zee method is statistically limited.
The $\phi$-symmetry method is not included in the combination due to its large systematic uncertainties.
The resulting energy resolution computed using these combined methods is discussed in Section~\ref{sec:performance}.

\section{Simulation}\label{sec:simulation}
A simulation of the ECAL is included in the \GEANTfour-based~\cite{Allison:1035669,Agostinelli:2002hh} CMS 
simulation framework.
It includes a detailed description of the detector geometry and implements a simulation of 
the 
light yield,
the electromagnetic shower propagation,  
the photodetector and electronics response,
and includes an emulation of the digitisation and the 
selection thresholds applied, as described in Section~\ref{sec:online}.
The propagation of light within the crystal is not simulated, 
instead, a parametrisation of the response as a function of the 
deposited energy in the crystal is used.
The uncertainties in the per-channel ICs are included in the simulation. 
A detector ageing model including the evolution of crystal transparency and photodetector noise with time
is also included,
to ensure the agreement between data and simulation.
In this ageing model, the predicted average response loss is used to simulate the expected 
evolution of the detector
for the data-taking runs in the upcoming year.
The following aspects are included since they have an effect on the
energy scale and resolution of ECAL:
\begin{itemize}
 \item Electronic noise.      The electronic noise is measured in data from the fluctuations of the baseline
                              in the absence of a real energy deposit and is accounted for
                              by adding into the simulation a Gaussian smearing to the
                              pedestal baseline
                              in each channel taking into account the correlation matrix of the noise.
 \item Crystal transparency.  The simulation of the propagation of the optical photons
                              to the photodetector takes into account the variation of the crystal  
                              response due to the radiation damage. 
                              The reduced crystal response leads to an increase of the Poissonian contribution to the energy resolution
                              and 
                              an increase in effective electronic noise, as shown in Eq.~(\ref{eq:uncalib_to_calib}).
\end{itemize}

The noise and crystal transparency parameters
are updated in the simulation
once per year
using parameters that are representative of the expected conditions averaged over the year.
The evolution within a year is currently not included in the simulation.
While the detailed simulation provides a satisfactory description of the data, 
some discrepancies remain due to the time evolution of the detector response,
and to the imperfect modelling of the detector components and mechanical structures in front of the ECAL.
A simulation with higher time granularity of the detector conditions
could mitigate some of the discrepancies 
at the cost of a more complex procedure to handle the generation of events.
As described in Section~\ref{sec:calibration}, the energy scale for the experimental data is set to match the
simulation
for reconstructed \zee decays.
In the 
simulation
an additional term is added to the energy resolution of electrons and photons 
that is tuned to match the observed resolution
for reconstructed \zee decays~\cite{EGM17001}.

\section{Performance}\label{sec:performance}
The production of \PZ bosons provides a clean sample of electrons that is used to study and to assess
the stability of the energy scale, the energy resolution, and the timing performance.
For this,
the electron energy is reconstructed using only the ECAL information,
unlike the standard CMS reconstruction, which combines information from the ECAL and the tracking detector.
The electrons in CMS experience bremsstrahlung emissions interacting with the tracker material, 
and this process influences the spatial distribution of the energy in the calorimeter.
The electrons are subdivided into two categories corresponding to 
high and low bremsstrahlung emission.
The separation between the two categories is based on the energy distribution in the SC, 
using the $\RNINE$ variable, which is a measure of the fraction of the SC energy contained in a central $3 \times 3$  
crystal matrix, 
and its value is close to 1 for low-bremsstrahlung electrons.
Hereafter, low-bremsstrahlung electrons correspond to the selection $\RNINE>0.965$.
More details of the \zee method
are given in Appendix~\ref{sec:sub:calibrationZee}.
The resolution of low-bremsstrahlung electrons is closer to the intrinsic performance of the ECAL,
while the resolution of the high-bremsstrahlung electrons is 
significantly influenced by the clustering of the energy deposits.

\subsection{The ECAL energy scale and resolution performance}\label{subsec:performance:energy}
\begin{figure}[htbp]
\begin{center}
\includegraphics[width=0.79\textwidth]{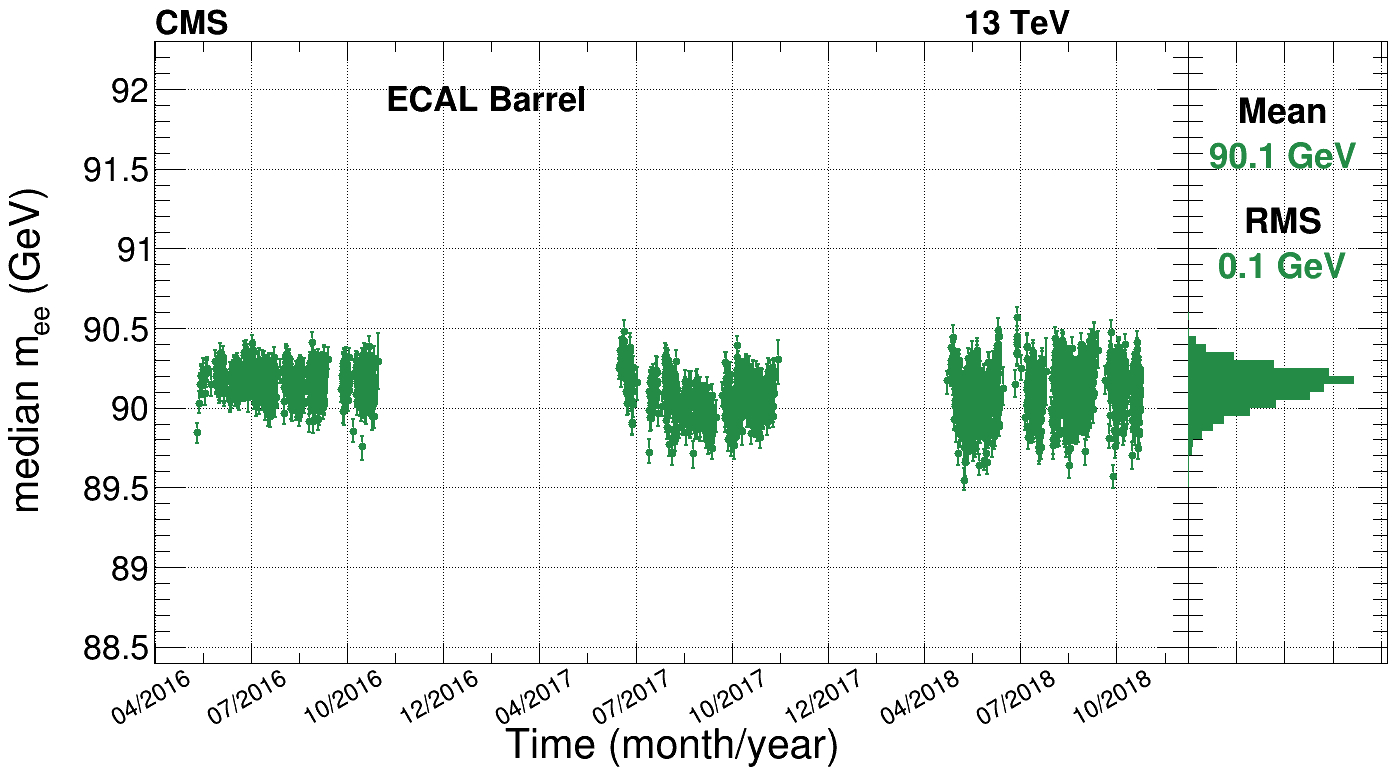}
\includegraphics[width=0.79\textwidth]{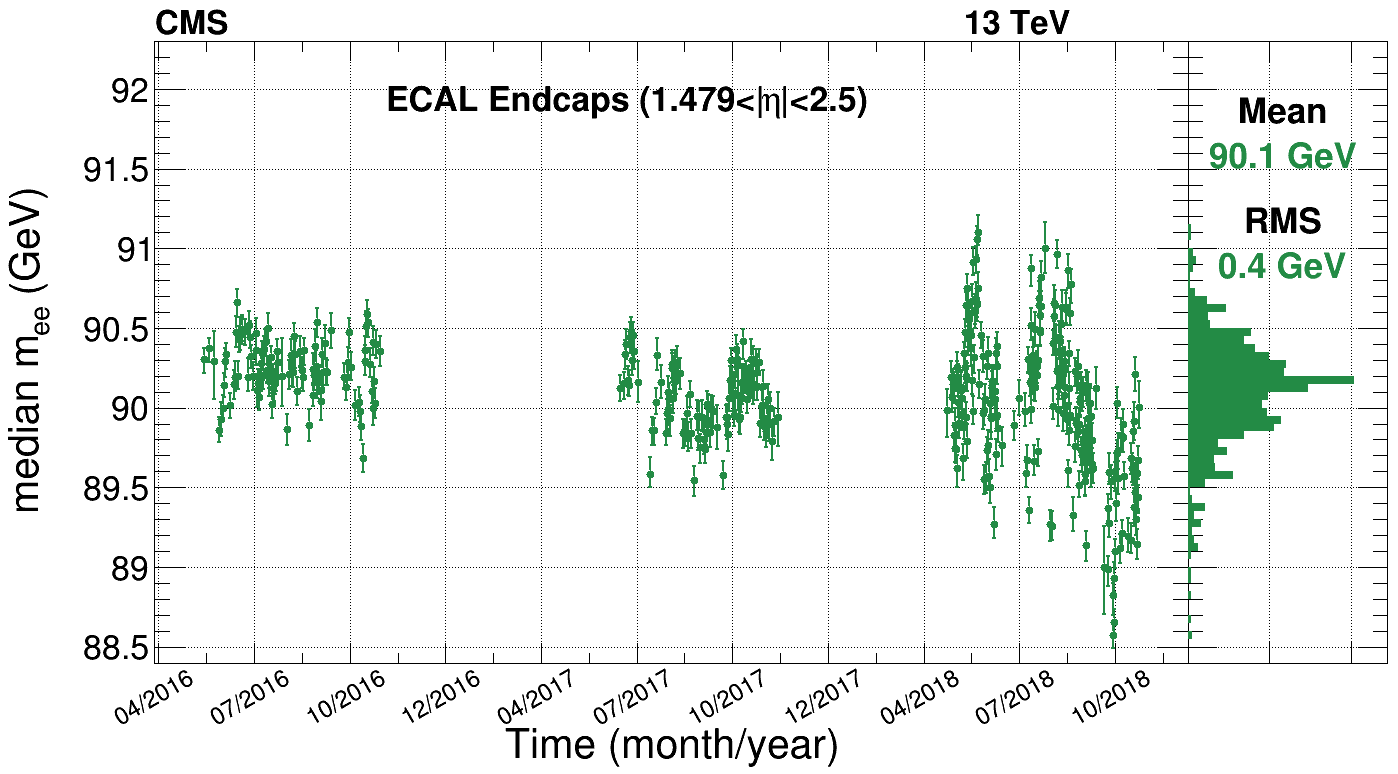}
\caption{
    Time stability of the dielectron invariant mass distribution for the Run~2 data-taking period using \zee electrons. 
    Both electrons are required to be in the EB (upper) or in the EE (lower). 
    Each time bin contains about 
    ten thousand
    events. 
    The error bar on the points denotes the statistical uncertainty (at 95\% confidence level) on the median.
    The right panel shows the distribution of the medians.
}
\label{fig:ZeeMassPerformanceStability}
\end{center}
\end{figure}

The stability of the energy scale is determined from the median value of the invariant mass distribution of \Pep\Pem pairs
from \PZ boson decays. 
For the full Run~2 data set, the stability of the scale is within 
0.1 (0.2)\% in the EB (EE), 
as shown in Fig.~\ref{fig:ZeeMassPerformanceStability}.
The spread observed, in particular in the EE, has a negligible impact on the energy resolution.
The slight degradation of the stability in 2018 in the EE is due to less frequent energy scale updates, 
as described in Section~\ref{sec:calibration},
but the stability still meets the performance requirements for physics.
The small residual drifts in the electron and photon energy scale with time shown in Fig.~\ref{fig:ZeeMassPerformanceStability}
are corrected by
using
\zee events at physics analysis level in approximately 18-hour intervals corresponding to the length of one LHC fill.

\begin{figure}[htbp]
\begin{center}
\includegraphics[width=0.49\textwidth]{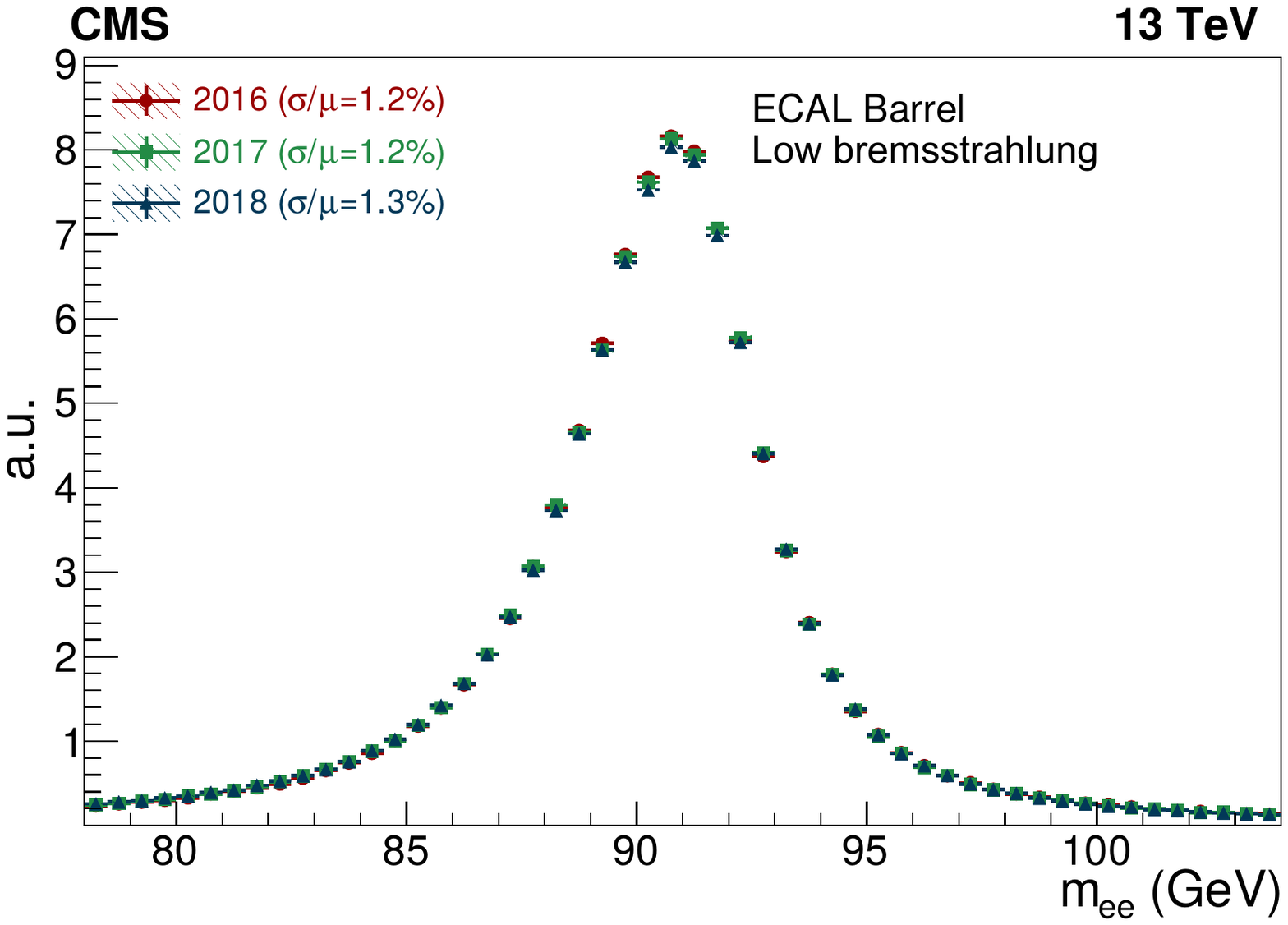}
\includegraphics[width=0.49\textwidth]{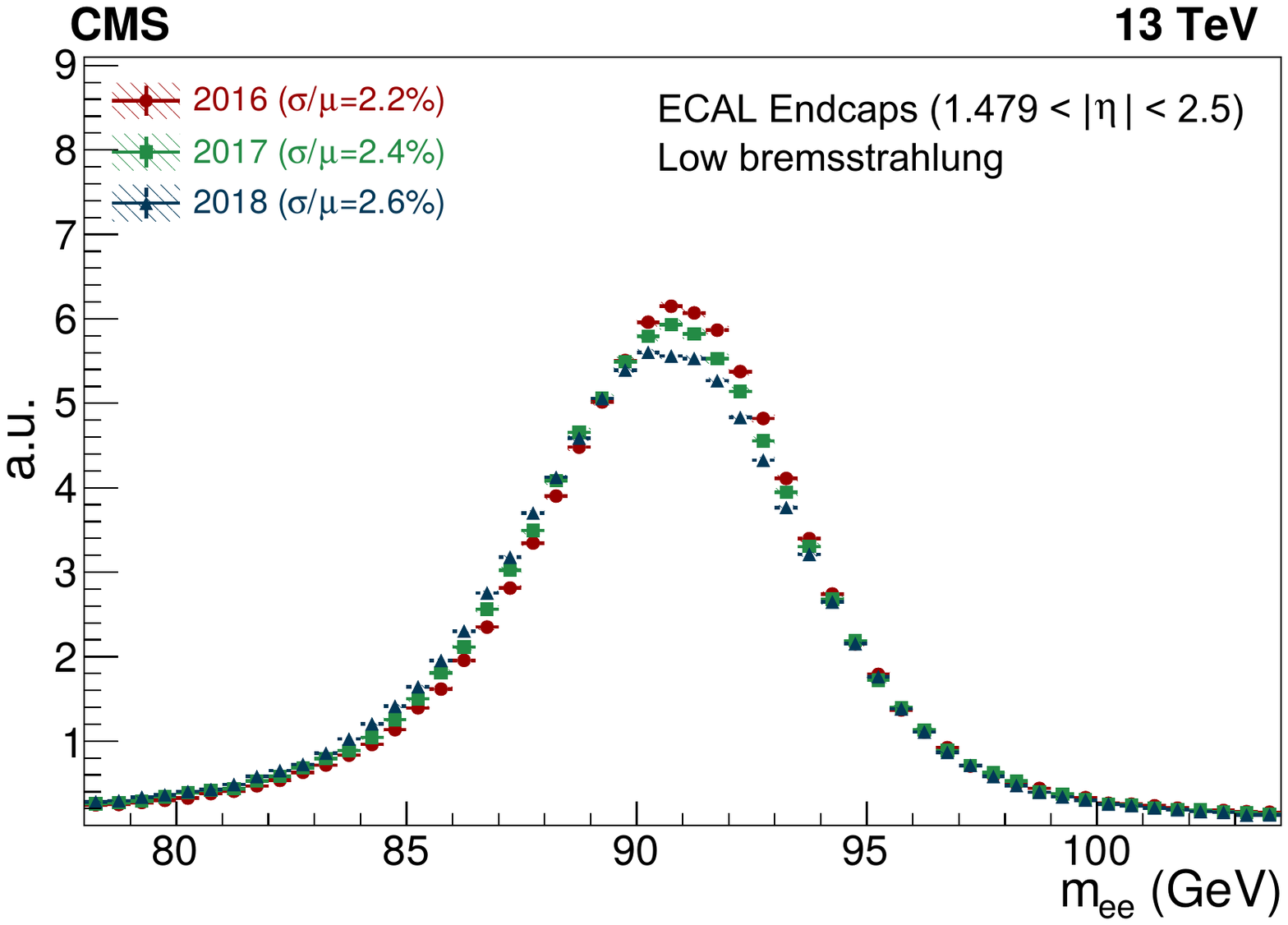}
\caption{
   Invariant mass distribution for electron-positron pairs from \PZ boson decays using low-bremsstrahlung electrons.
   The distributions from data recorded in 2016, 2017, and 2018 are shown with different colours. 
   The event selection requires two electrons to be in the EB (left) or in the EE within the tracker acceptance (right).
   The vertical bars on the points represent the statistical uncertainty.
}
\label{fig:ZeeMassPerformance}
\end{center}
\end{figure}

\begin{figure}[htbp]
\begin{center}
\includegraphics[width=0.49\textwidth]{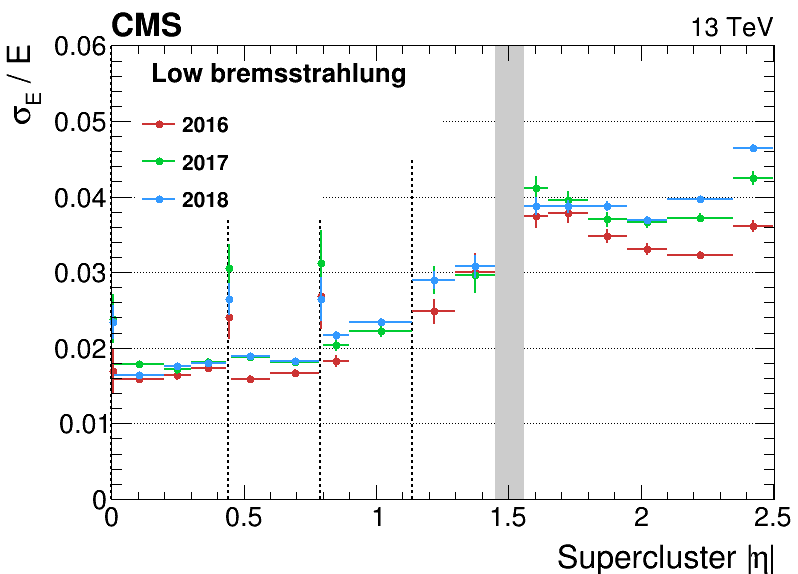}
\includegraphics[width=0.49\textwidth]{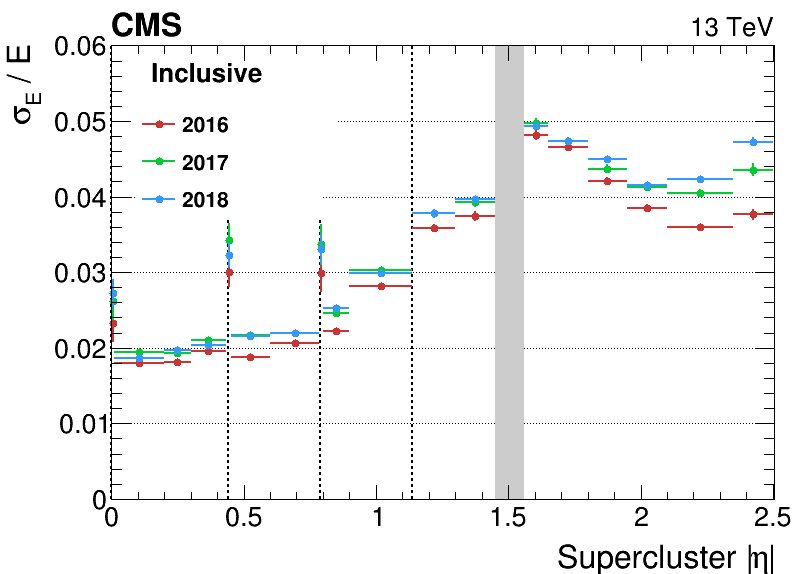}           
\caption{
   Relative energy resolution for electrons from \zee decays as a function of $\abs{\eta}$. 
   The energy resolution is measured by the method presented in Appendix~\ref{sec:sub:calibrationZee} 
   for low-bremsstrahlung electrons (left) 
   and for the inclusive sample 
   (right).
   The vertical bars on the points represent the statistical uncertainty.
   The vertical dotted lines mark the boundaries between the ECAL modules in the barrel, where a slight worsening of the resolution is observed 
   due to the material of the mechanical structures.
   The shaded grey band corresponds to the EB/EE transition.   
   }
\label{fig:ResolutionVsEta}
\end{center}
\end{figure}

The invariant mass distribution of \Pep\Pem pairs is shown in Fig.~\ref{fig:ZeeMassPerformance} for 
low-bremsstrahlung electrons, separately in EB and EE and for 2016, 2017 and 2018.
The electron energy resolution is estimated
using the \zee method as mentioned in Section~\ref{sec:calibration}.
The measured resolution for electrons is shown in Fig.~\ref{fig:ResolutionVsEta}, separately for 
low-bremsstrahlung electrons and for an inclusive sample, for the different years of Run~2.
Despite the large increase in radiation dose and pileup during Run~2,  
the resolution is only marginally degraded compared to that obtained in Run~1,
which for low bremsstrahlung electrons from \PZ boson decays was better than
1.6\% in the central barrel, 
3.0\% in the outer barrel,
4.0\% in the endcap~\cite{photonrun1}.
There are minor differences between 2016, 2017, and 2018 performance, due to larger 
response losses and pileup during 2017 and 2018.

The energy scale linearity has also been measured with \zee events.
The ratio between the energy scale in data and simulation has been computed
in \ET and $\eta$ windows, requiring both electrons to be in the same window. 
The deviations from linearity in the \ET range of 20 to 80\GeV
are less than 0.5\%,
which is corrected at the object level in physics analyses~\cite{EGM17001}.

The energy scale performance was cross-checked with a sample of photons from $\PZ\to\Pgm\Pgm\Pgg$ events both in data and simulation,
selected~\cite{EGM17001} with very high purity ($\approx$99\%), with transverse momentum greater than 20\GeV and with a narrow energy distribution in the SC 
(using the same selection, $\RNINE>0.965$, used to define the low-bremsstrahlung electrons sample).
The energy scale is extracted for data and simulation from the mean of the distribution 
of a per-event estimator~\cite{Chatrchyan:2013dga} defined as $s=(m_{\Pgm\Pgm\Pgg}^2-m_{\Pgm\Pgm}^2)/(m_\PZ^2-m_{\Pgm\Pgm}^2))-1$, 
where $m_\PZ$ denotes the nominal {\PZ} boson mass.
The scale difference between data and simulation in EB is less than 0.05\%.
This is well within the quadratic sum of the statistical and systematic uncertainties associated with the scale extraction process
from $\PZ\to\Pgm\Pgm\Pgg$ events, which is 0.09\% and includes variations in the fit function and fit range.

The ECAL extends up to $\abs{\eta}=3$, while the tracker only covers up to $\abs{\eta}<2.5$,
thus in the region $2.5<\abs{\eta}<3.0$ the ECAL's main contribution to physics analyses is for the jet energy scale and missing transverse momentum.
The intercalibration in this region is based solely on \zee events.
Despite being outside the tracker coverage,
the accuracy reached is such that
the energy resolution is dominated by the electronic noise, which reaches levels as high as few GeV
due to low crystal transparency,
as shown in Fig.~\ref{fig:TransparenncyEvolution}, and by the pileup, and not by the intercalibration precision. 

\begin{figure}[htbp]
\begin{center}
\includegraphics[width=0.49\textwidth]{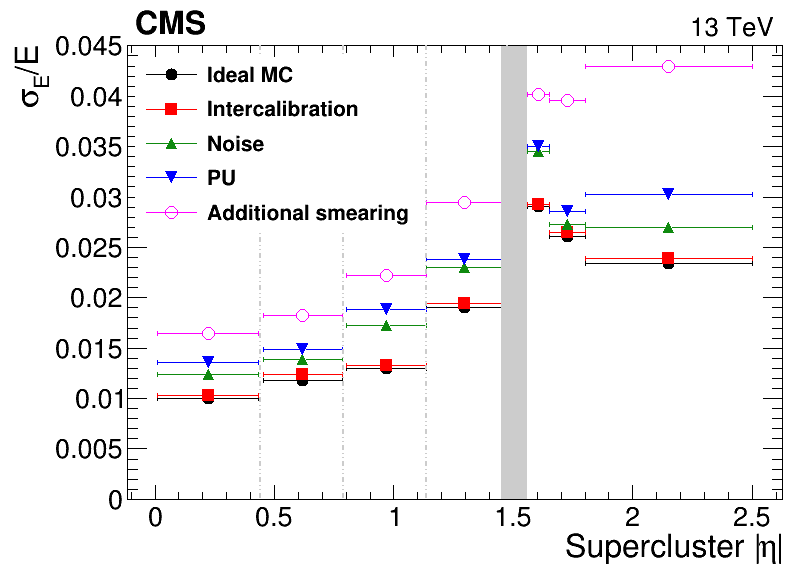}
\includegraphics[width=0.49\textwidth]{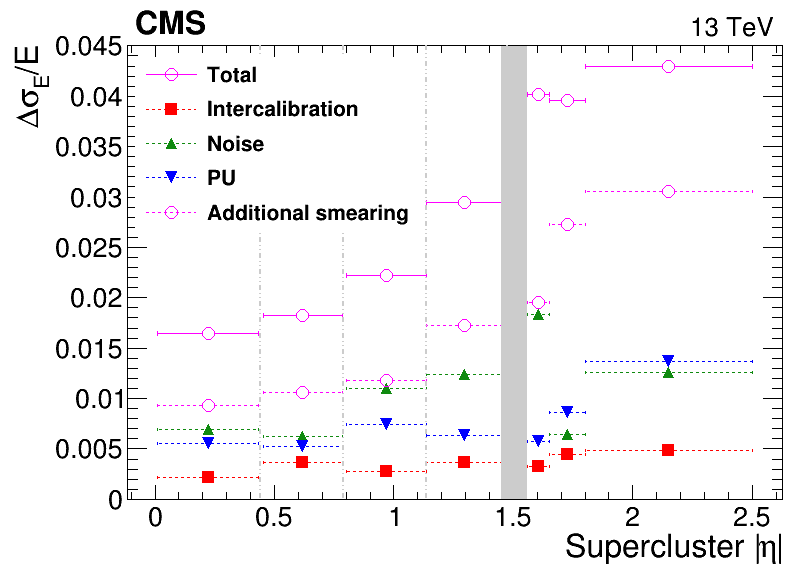}
      \caption {
    Left: the relative energy resolution for electrons from simulated \zee decays. 
    The energy resolution is measured by the method presented in Appendix~\ref{sec:sub:calibrationZee} 
    for low-bremsstrahlung electrons. 
    The different simulations correspond to the dedicated scenarios.
    The label ``Additional smearing'' is the observed resolution in data, 
    corresponding to the ``PU'' scenario with the inclusion of the additional smearing.
    Right: the contributions of different effects to the resolution (intercalibration accuracy, noise, PU) and the additional smearing.
    The total resolution,
    which in the left pane corresponds to the sample labelled ``Additional smearing''
    is also reported in the right pane, labelled as ``Total''.
    The plots are shown as functions of the $\abs{\eta}$ of the SC.
    The vertical dotted lines mark the boundaries between the ECAL modules in the EB.
    The shaded grey band corresponds to the EB/EE transition.
}
\label{fig:MCresolution}  
\end{center}
\end{figure}

The contributions to the measured energy resolution by pileup, electronic noise, and intercalibration uncertainties, have been
evaluated by comparing the resolution on
samples of \zee events simulated without 
pileup, noise and the energy thresholds applied to the PF rechits, and assuming a perfectly calibrated detector. 
The results are shown in Fig.~\ref{fig:MCresolution},
where the energy resolution is found for the simulated samples produced with the different scenarios corresponding to:

\begin{itemize}
\item  ``Ideal MC'';          ideal detector (perfect calibration, without energy thresholds, without noise, and without pileup).
\item  ``Intercalibration'';  realistic calibration, without energy thresholds, without noise, and without pileup,
\item  ``Noise'';             realistic noise, energy thresholds and calibration, without pileup,
\item  ``PU'';                realistic noise, energy thresholds, calibration, and pileup,
\end{itemize}

The largest effects are due, in the order of size, to the pileup and to the noise.
The contribution to the resolution from calibration uncertainties
was found to be negligible compared to the other effects (except for a small effect in the last $\abs{\eta}$ window).

There are several effects that are not completely quantified or simulated, 
such as the precise details of the material in front of the ECAL, and 
the accuracy of the time-dependent corrections for the ECAL.
These effects are accounted for by introducing in the simulation an additional energy smearing for electrons and photons.
The magnitude of this additional smearing is similar to the sum of all the other contributions.

\begin{figure}[htbp]
\begin{center}
\includegraphics[width=0.49\textwidth]{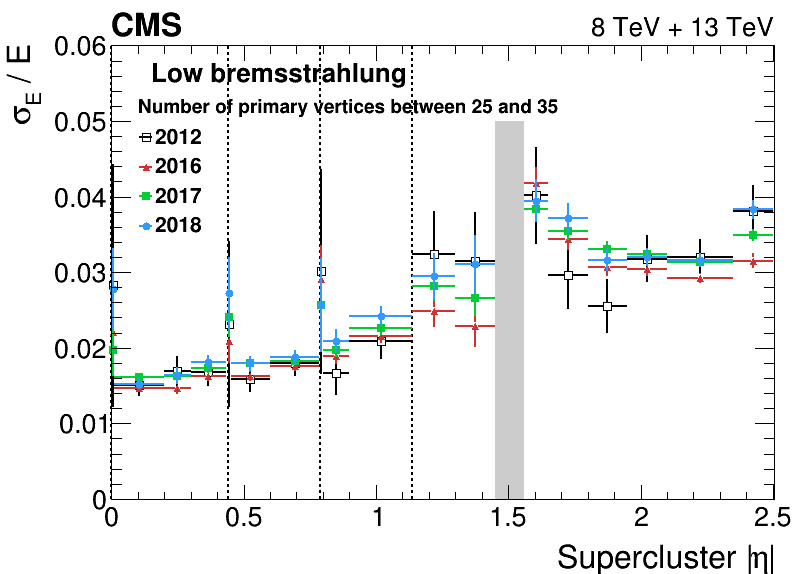}  
\includegraphics[width=0.49\textwidth]{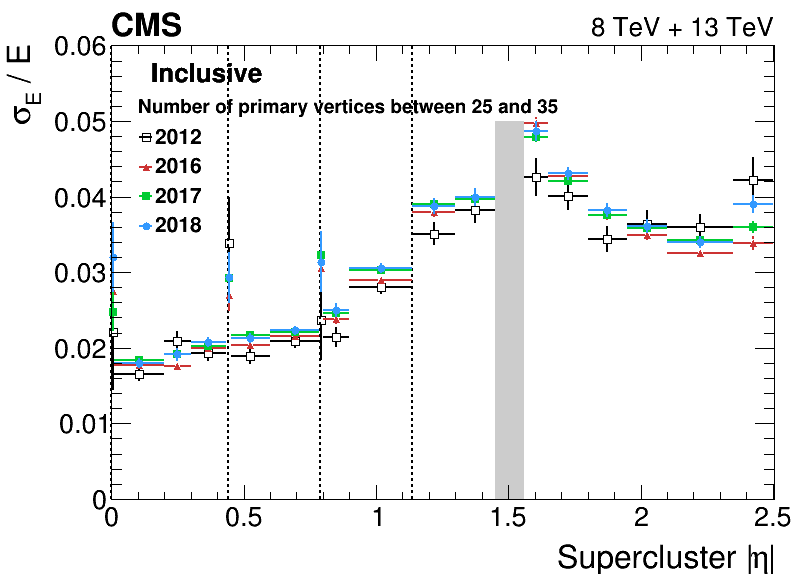} 
\caption{
   Relative energy resolution for electrons from \zee decays versus $\abs{\eta}$ for
   events with low number of vertices and weighted to the Run~1 pileup distribution.
   The energy resolution is measured by the method presented in Appendix~\ref{sec:sub:calibrationZee} 
   for low-bremsstrahlung electrons (left) 
   and for the inclusive sample 
   (right).
   The vertical bars on the points represent the statistical uncertainty.
   The vertical dotted lines mark the boundary between the ECAL modules in the barrel, where a slightly worsening of the resolution is observed 
   due to the material of the mechanical structures.
   The shaded grey band corresponds to the EB/EE transition.   
   }
\label{fig:ResolutionVsEtaPURunI}
\end{center}
\end{figure}

To evaluate with experimental data the various contributions to the energy resolution, 
events with a small number of vertices and weighted to the Run~1 pileup distribution were selected to measure the stability of the performance.
This is shown in Fig.~\ref{fig:ResolutionVsEtaPURunI}.
The similarity with the performance in Run~1 shows how,
even in the harsher environment of Run~2 with its increased radiation and OOT-PU (25 versus 50\unit{ns} BX spacing),
the resolution does not degrade significantly. 
Instead of the standard reconstruction algorithm that was used for Run~2, 
a regression algorithm like the one used in Run~1
that used both the EE and the ES information
has been trained.
The small degradation
of the EE performance in Run~2 compared to that in Run~1
is due to the effective noise increase
due to the average transparency loss,
which is partially compensated by the \zee
calibration of the Run~2 data.

\subsection{Performance of the ECAL at the trigger level}\label{subsec:performance:trigger}
Localized energy deposits in the ECAL are used both in the L1 trigger to identify electromagnetic particle candidates,
and in the HLT to reconstruct electrons and photons,
using algorithms similar to those used in the offline reconstruction. 
The performance of the L1 trigger for electrons and photons in Run~2 has been reported in a separate
paper~\cite{TRG-17-001}.
 
Since the rate of spikes, introduced in Section~\ref{sec:reconstruction}, 
is proportional to the collision rate of the proton beams~\cite{Petyt:2012sf}, 
they complicate trigger selections, particularly at high instantaneous luminosity. 
At the L1 trigger, spike-like energy deposits are suppressed by exploiting an additional functionality of the FENIX ASIC~\cite{Bloch:581342},
that can be configured, with suitable energy thresholds, to flag events with isolated energy deposits.
The parameters were updated for the more challenging beam and detector conditions of Run~2.
This reduced the contamination of spikes in ECAL trigger primitives with $\ET>30\GeV$ by a factor of two,
with a negligible impact on the triggering of electromagnetic signals with $\ET>20\GeV$.

\begin{figure}[htbp]
\centering
\includegraphics[width=0.70\textwidth]{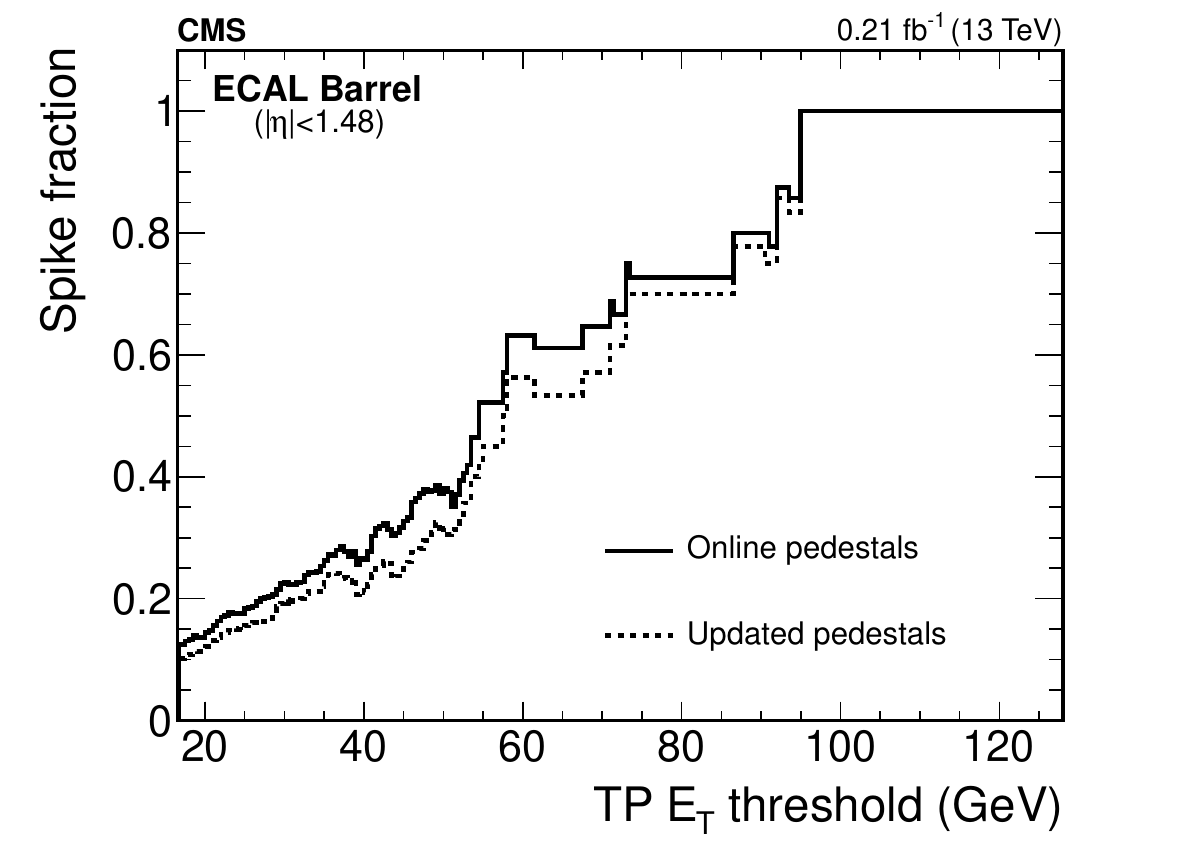} 
\caption
              {
              Fraction of ECAL TPs in the barrel region that are matched to an offline spike, 
              as a function of the trigger primitive $\ET$ threshold, 
              before and after updated pedestal values are applied. 
              \label{fig:spike1}
              }
\end{figure}

The efficiency of this selection is sensitive to drifts in the values of the pedestals. 
Periodic (up to twice yearly in 2018) updates of the pedestals used in TP formation were required to maintain stable spike identification efficiencies. 
Figure~\ref{fig:spike1} shows the improvement gained by applying more up-to-date pedestal values on the spike identification efficiency,
measured from data recorded in mid-2018. This efficiency is measured as the fraction of TPs that are matched to a spike
(identified via more sophisticated discriminating variables applied offline) versus the TP transverse energy threshold, 
which is kept as low as possible. By periodically updating the pedestal values, the spike contamination for TPs 
with  $\ET>30\GeV$ was maintained at below 20\% during the 2018 run. 

\begin{figure}[htbp]
\centering
\includegraphics[width=0.70\textwidth]{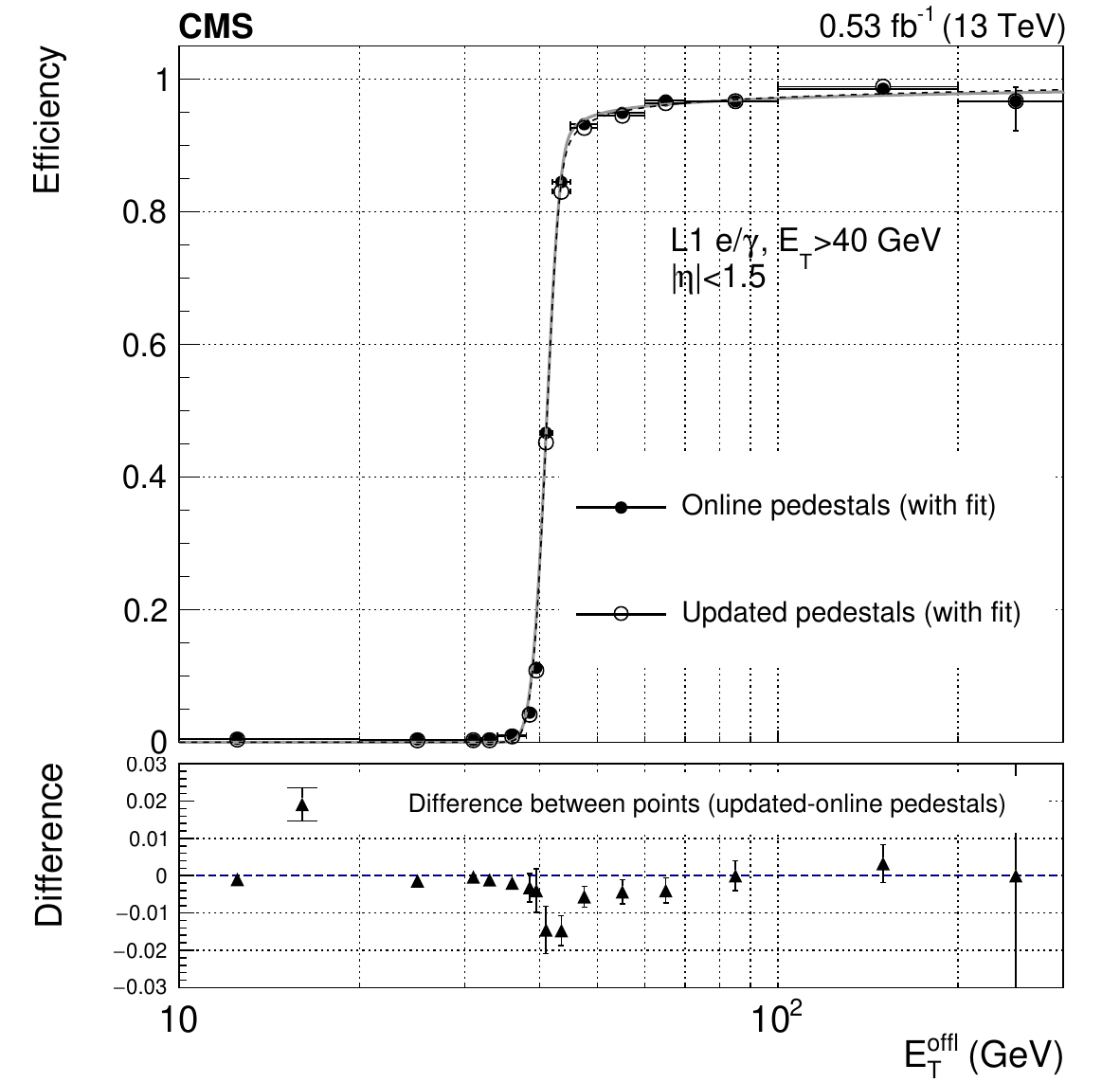}
\caption
              {
              The efficiency of the L1 electron/photon trigger with a transverse energy threshold of 40\GeV, 
              measured using the ``tag-and-probe'' method on \zee events, 
              as a function of the transverse energy of the tag electron, reconstructed offline,  
              before and after updated pedestal values are applied, and fitted with a sigmoid function.
              The vertical bars on the points represent the statistical uncertainty.
              In the lower panel the difference between the efficiencies before and after
              the update of the pedestals is shown.
              \label{fig:spike2}
              }
\end{figure}

Figure~\ref{fig:spike2} shows the effect of applying more up-to-date pedestal values on the efficiency for triggering on electron/photon candidates 
with an L1 transverse energy threshold of 40\GeV, measured using a tag-and-probe method on \zee events~\cite{TagAndProbe}.
The difference is minimal, showing that the improved spike identification efficiency does not have major effects on the efficiency for triggering on signal events.  

These improvements in TP calibration and spike rejection, together with improvements in the L1 trigger system itself, 
allowed the L1 electron/photon trigger to operate with high efficiency and the lowest possible $\ET$ thresholds throughout Run~2~\cite{TRG-17-001}.

The majority of the high-energy spikes are filtered out at L1, and they are further reduced by the HLT by means of additional quality requirements 
using methods that are also used in the offline reconstruction, as reported in Section~\ref{sec:reconstruction}.

Due to a mistiming in the ECAL readout,
which happened mostly in 2017,
an inefficiency in the L1 trigger 
occurred 
that affected
the triggering of events
with a large signal in the ECAL, 
which were assigned to the previous BX.
Since the CMS trigger does not permit triggers from two successive BXs, 
the event in the correct BX was not read out.
This effect, known as ``prefiring'',
produces a loss of efficiency
ranging from a few percent up to 80\% in a few critical regions
for high-energy (\pt $>$ 200~\GeV ) and $|\eta|>2.5$ electromagnetic deposits,
which needed to be accounted for in physics analyses.
In 2018 a frequent update of the detector timing reduced this effect to negligible levels.

\subsection{The ECAL timing performance}\label{subsec:performance:timing}
The timing precision of the ECAL was measured 
before Run~1
with 
electrons in a test beam, cosmic-ray muons, and 
muons when the LHC beam was dumped on collimators
located approximately 150\unit{m} upstream of CMS~\cite{Chatrchyan:2009aj}.
The timing precision for large energy deposits ($E>10$--20\GeV in the EB)
was estimated to be better than 100\unit{ps}.
During collisions there are additional effects that degrade this performance,
such as  
the variations in the clock distribution between different regions of the ECAL and different CMS runs,
uncertainty in the time calibration,
and crystal transparency changes, which affect also the shape of the pulse.
A first measurement of the timing performance and time intercalibration of
the CMS ECAL during collisions has been reported in Run~1 at 7\TeV~\cite{Chatrchyan:2013dga}, 
where the time precision was about
190\unit{ps} in EB and 280\unit{ps} in EE. 
The time precision of adjacent channels in an electromagnetic cluster was measured to be 70\unit{ps}
for large energies, if the channels belonged to the same electronics readout (RO) unit
(a matrix of $5\times5$ crystals)
and 130\unit{ps} for channels belonging to different readout units~\cite{DelRe:1706325}.
For Run~2, the timing precision was evaluated 
following the procedure described in Ref.~\cite{DelRe:1706325},
where the time difference 
between adjacent crystals with similar energy deposits in an SC was measured.
Pairs of crystals are considered to have a similar energy if $0.8 < E_1/E_2 < 1.25$ 
and the energy of both is between 1 and 120\GeV.

\begin{figure}[htbp]
  \begin{center}
    \includegraphics[width=0.49\textwidth]{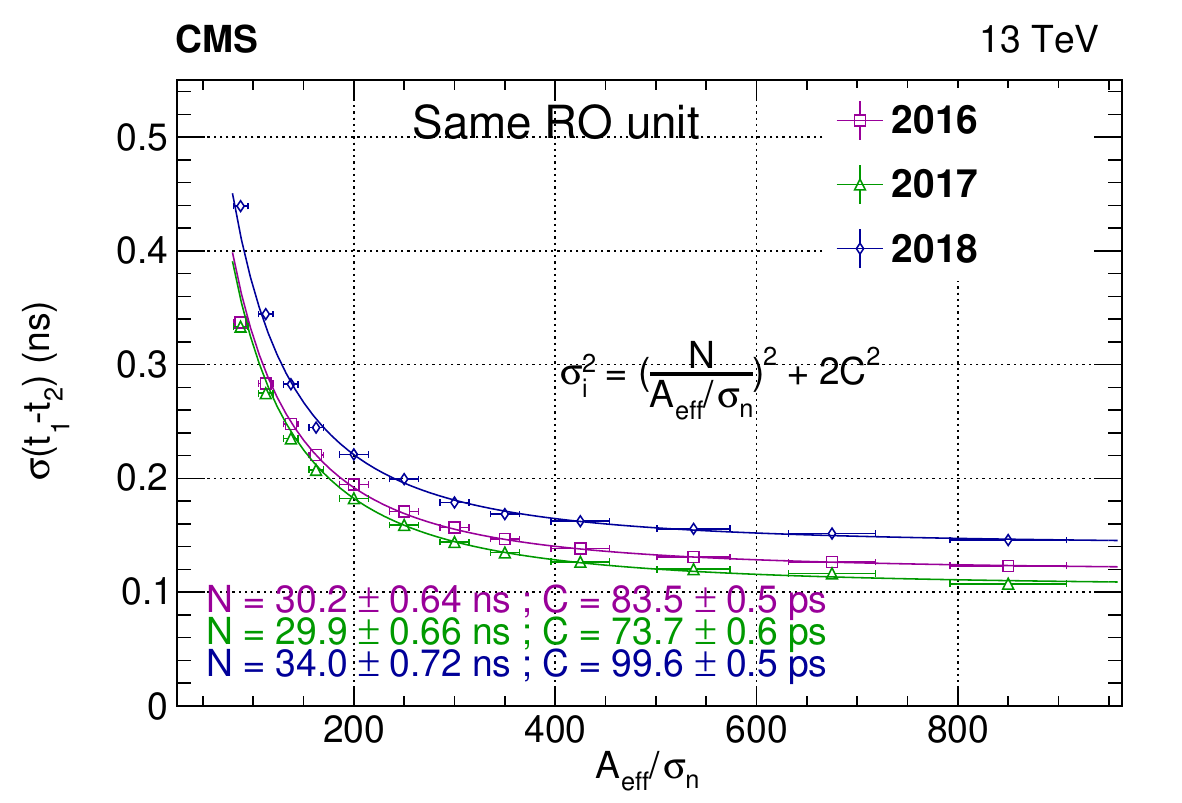}
    \includegraphics[width=0.49\textwidth]{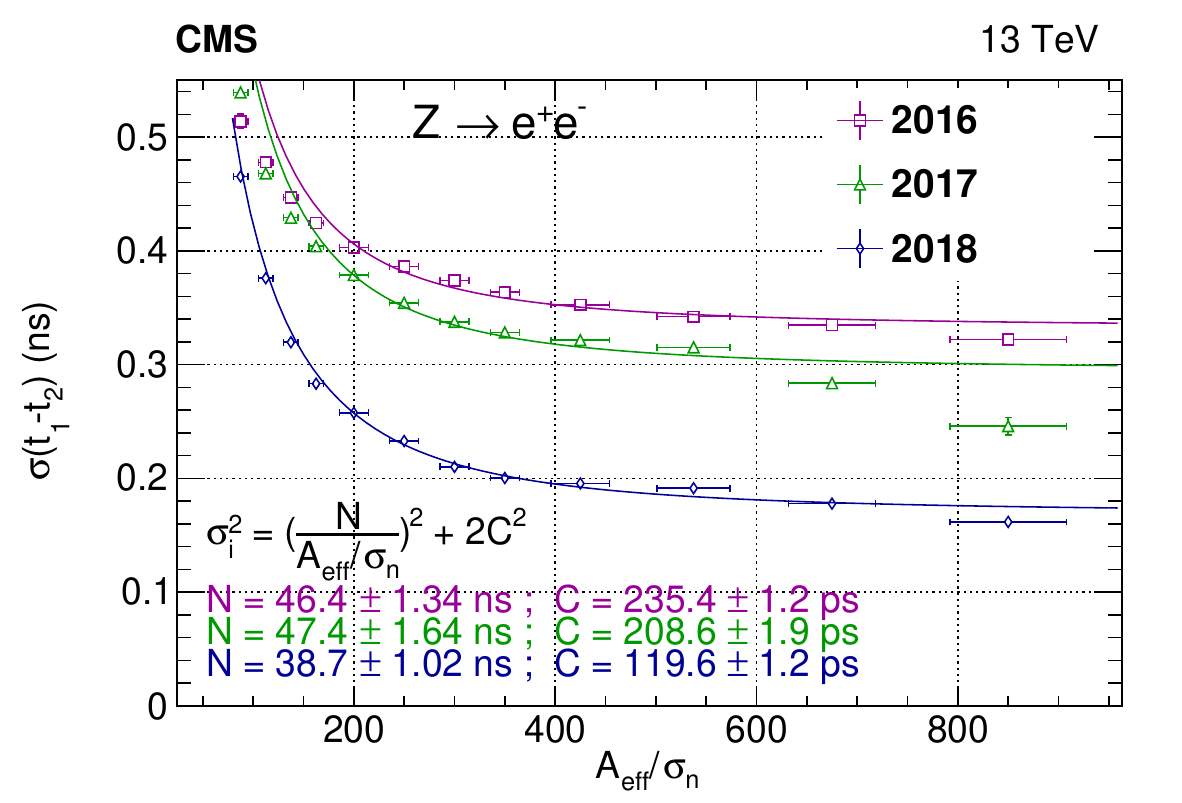}   
 \caption{
        The ECAL timing precision as measured from 
        adjacent crystals sharing the same electronic readout (left),
        in 2016, 2017, and 2018 data.
        The timing precision extracted from \zee events 
        comparing the time of flight of the {\Pep} and {\Pem} is shown as well (right).
        The vertical bars on the points represent the statistical uncertainty.
        }
    \label{fig:timeperformance:2016and2017}
  \end{center}
\end{figure}

The intrinsic timing precision obtained with crystals belonging to the same readout unit is
used to eliminate the contribution due to synchronization effects between different readout units.
The overall timing precision is found from 
the time difference between the electrons from \zee events
corrected for the vertex location.
The result is shown in 
Fig.~\ref{fig:timeperformance:2016and2017}, 
as a function of the effective amplitude,
which is defined as $A_{\text{eff}} = (A_1 A_2)/\sqrt{A_1^2 +A_2^2}$,
normalized to the electronic noise.
The noise term is very similar to that obtained at the test beam~\cite{Chatrchyan:2009aj} and the constant term
is on average 83\unit{ps} in 2016, 74\unit{ps} in 2017, and 100\unit{ps} in 2018 data.
The difference between the results for crystals in the same readout units
and those obtained with the electrons from \zee decays
are understood to be due, among other possible reasons, to the instabilities in the clock initialization in the 
different readout units in every run.
An improvement in 2017 and 2018 in the overall timing precision compared to 2016 is visible in Fig.~\ref{fig:timeperformance:2016and2017} (right).
This is due to updating the time intercalibration constants with higher frequency
in 2017 and further increasing the rate in 2018.
This approach mitigated the effects of time offsets between different readout units.
However, the performance obtained in the 2016 data is already well within the requirements for
current physics analyses.

\section{Summary}\label{sec:conclusions}
The challenge of maintaining the same excellent performance of the CMS electromagnetic calorimeter
achieved in Run~1 has been met despite
the increased levels of radiation damage and pileup.
This has included an increase in the transparency loss 
by the crystals and larger energy-equivalent electronic noise levels.
To meet this challenge,
a new algorithm has been developed to reconstruct the energy deposited in the crystals;
the detector conditions have been monitored using different and 
complementary methods and updated continuously, 
and residual time-dependent corrections to the detector 
have been derived with electrons from the decays of {\PW} and {\PZ} bosons.
Because of these changes in Run~2, the stability of the energy scale was better than 
0.1\% in the barrel and 0.4\% in the endcaps.
For electrons from {\PZ} boson decays with low bremsstrahlung (for the inclusive sample),
the energy resolution was better than
1.8 (2.0)\% in the central barrel, 
3.0 (4.0)\% in the outer barrel, 
and 4.5 (5.0)\% in the endcaps.
These techniques and methods 
will continue to be used and improved
to maintain the excellent performance of the CMS electromagnetic calorimeter
during the even more challenging operating conditions in the LHC Run~3.

\begin{acknowledgments}
  We congratulate our colleagues in the CERN accelerator departments for the excellent performance of the LHC and thank the technical and administrative staffs at CERN and at other CMS institutes for their contributions to the success of the CMS effort. In addition, we gratefully acknowledge the computing centres and personnel of the Worldwide LHC Computing Grid and other centres for delivering so effectively the computing infrastructure essential to our analyses. Finally, we acknowledge the enduring support for the construction and operation of the LHC, the CMS detector, and the supporting computing infrastructure provided by the following funding agencies: SC (Armenia), BMBWF and FWF (Austria); FNRS and FWO (Belgium); CNPq, CAPES, FAPERJ, FAPERGS, and FAPESP (Brazil); MES and BNSF (Bulgaria); CERN; CAS, MoST, and NSFC (China); MINCIENCIAS (Colombia); MSES and CSF (Croatia); RIF (Cyprus); SENESCYT (Ecuador); ERC PRG, RVTT3 and MoER TK202 (Estonia); Academy of Finland, MEC, and HIP (Finland); CEA and CNRS/IN2P3 (France); SRNSF (Georgia); BMBF, DFG, and HGF (Germany); GSRI (Greece); NKFIH (Hungary); DAE and DST (India); IPM (Iran); SFI (Ireland); INFN (Italy); MSIP and NRF (Republic of Korea); MES (Latvia); LMTLT (Lithuania); MOE and UM (Malaysia); BUAP, CINVESTAV, CONACYT, LNS, SEP, and UASLP-FAI (Mexico); MOS (Montenegro); MBIE (New Zealand); PAEC (Pakistan); MES and NSC (Poland); FCT (Portugal); MESTD (Serbia); MCIN/AEI and PCTI (Spain); MOSTR (Sri Lanka); Swiss Funding Agencies (Switzerland); MST (Taipei); MHESI and NSTDA (Thailand); TUBITAK and TENMAK (Turkey); NASU (Ukraine); STFC (United Kingdom); DOE and NSF (USA).

  \hyphenation{Rachada-pisek} Individuals have received support from the Marie-Curie programme and the European Research Council and Horizon 2020 Grant, contract Nos.\ 675440, 724704, 752730, 758316, 765710, 824093, 101115353,101002207, and COST Action CA16108 (European Union); the Leventis Foundation; the Alfred P.\ Sloan Foundation; the Alexander von Humboldt Foundation; the Science Committee, project no. 22rl-037 (Armenia); the Belgian Federal Science Policy Office; the Fonds pour la Formation \`a la Recherche dans l'Industrie et dans l'Agriculture (FRIA-Belgium); the Agentschap voor Innovatie door Wetenschap en Technologie (IWT-Belgium); the F.R.S.-FNRS and FWO (Belgium) under the ``Excellence of Science -- EOS" -- be.h project n.\ 30820817; the Beijing Municipal Science \& Technology Commission, No. Z191100007219010 and Fundamental Research Funds for the Central Universities (China); the Ministry of Education, Youth and Sports (MEYS) of the Czech Republic; the Shota Rustaveli National Science Foundation, grant FR-22-985 (Georgia); the Deutsche Forschungsgemeinschaft (DFG), under Germany's Excellence Strategy -- EXC 2121 ``Quantum Universe" -- 390833306, and under project number 400140256 - GRK2497; the Hellenic Foundation for Research and Innovation (HFRI), Project Number 2288 (Greece); the Hungarian Academy of Sciences, the New National Excellence Program - \'UNKP, the NKFIH research grants K 131991, K 133046, K 138136, K 143460, K 143477, K 146913, K 146914, K 147048, 2020-2.2.1-ED-2021-00181, and TKP2021-NKTA-64 (Hungary); the Council of Science and Industrial Research, India; ICSC -- National Research Centre for High Performance Computing, Big Data and Quantum Computing, funded by the EU NexGeneration program (Italy); the Latvian Council of Science; the Ministry of Education and Science, project no. 2022/WK/14, and the National Science Center, contracts Opus 2021/41/B/ST2/01369 and 2021/43/B/ST2/01552 (Poland); the Funda\c{c}\~ao para a Ci\^encia e a Tecnologia, grant CEECIND/01334/2018 (Portugal); the National Priorities Research Program by Qatar National Research Fund; MCIN/AEI/10.13039/501100011033, ERDF ``a way of making Europe", and the Programa Estatal de Fomento de la Investigaci{\'o}n Cient{\'i}fica y T{\'e}cnica de Excelencia Mar\'{\i}a de Maeztu, grant MDM-2017-0765 and Programa Severo Ochoa del Principado de Asturias (Spain); the Chulalongkorn Academic into Its 2nd Century Project Advancement Project, and the National Science, Research and Innovation Fund via the Program Management Unit for Human Resources \& Institutional Development, Research and Innovation, grant B37G660013 (Thailand); the Kavli Foundation; the Nvidia Corporation; the SuperMicro Corporation; the Welch Foundation, contract C-1845; and the Weston Havens Foundation (USA).
\end{acknowledgments}

\bibliography{auto_generated}

\appendix
\numberwithin{figure}{section}
\numberwithin{equation}{section}
\section{The ECAL alignment}\label{sec:alignment}
\subsection{The EB and EE alignment}\label{sec:alignment_EBEE}
The EB and EE provide precise position measurements of the impact points of electrons and 
photons based on the energy distribution among the crystals in the corresponding SCs,
which are used in the photon and electron reconstruction algorithms.
The energy deposits in the EB and EE are critical
for measuring the trajectories 
of photons that do not convert in the tracker.
Additionally, to identify and remove 
particles misidentified as electrons,
energy deposits in the EB and EE are matched with hits in the tracker.
A dedicated procedure used to align the EB and EE with respect to the tracker
increases the 
efficiency
of electron identification
and improves the  invariant mass reconstruction of photon pairs,
in particular for decays such as the Higgs boson $\PH\to\Pgg\Pgg$.

The alignment procedure for the EB and EE uses electrons from {\PW} and {\PZ} boson decays.
Since it is not statistically limited, a sample of electrons with negligible 
bremsstrahlung
emission is selected
to reduce possible uncertainties in the modelling and the clustering of bremsstrahlung energy.
Each electron has an associated track
in the tracker detector and the 
track position can be extrapolated to the SC position. 
The distances $\Delta\phi$ and $\Delta\eta$
between this extrapolated position 
and the SC position measurement, based on an energy-average position of the constituent crystals of an SC,
are used to construct a $\chi^{2}$ function.
Since the supermodules in the EB and the half-disks in the EE
are rigid elements that do not undergo internal motion, 
the $\chi^{2}$ function is minimized 
with respect to the position of each element, 
with three-dimensional translation in EB and also with a rotation parametrised with three Euler angles in EE,
to reproduce the expected $\Delta\phi$ and $\Delta\eta$ average values from simulation.

\begin{figure}
    \centering
    \includegraphics[width=0.40\textwidth]{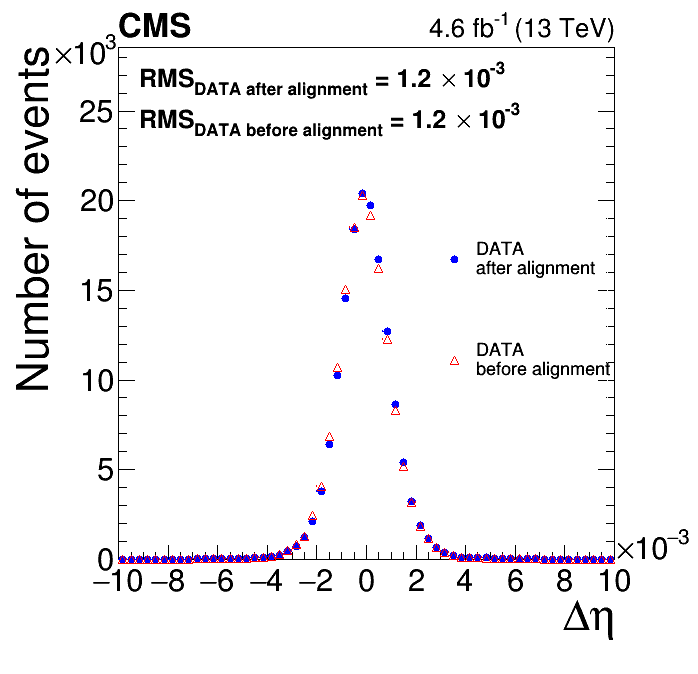}
    \includegraphics[width=0.40\textwidth]{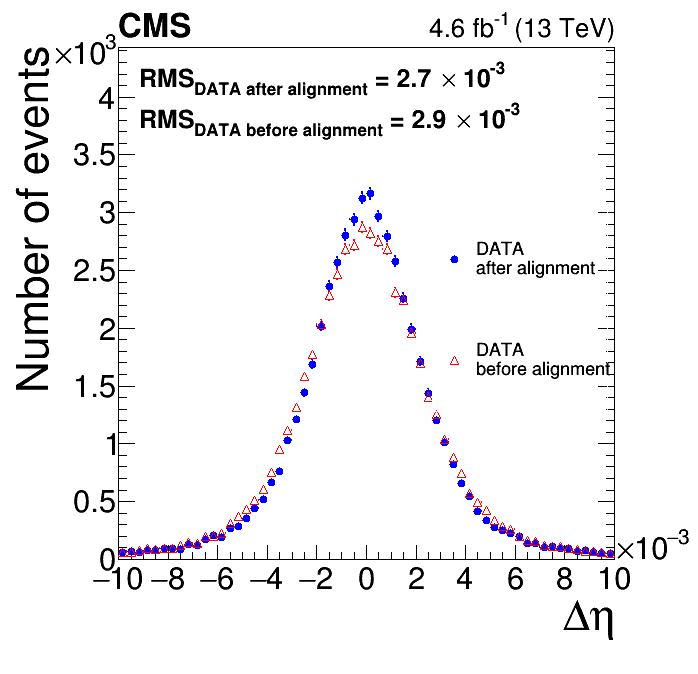}
    \caption{
        The $\Delta\eta$ between the extrapolated tracker position and the position measurement provided by the ECAL,
        before (red triangles) and after (blue circles) the alignment, for (left) the EB and
        (right) the EE, measured using electrons at the start of the 2017 data-taking period.
        The vertical bars on the points represent the statistical uncertainty.
        }
    \label{fig:ecalalignment1}
\end{figure}

An example of the $\Delta\eta$ distribution in the EB and EE,
before and after alignment, is shown in Fig.~\ref{fig:ecalalignment1}.
The alignment of the EB and EE is performed relative to the tracker; therefore, a new alignment is needed
after every tracker alignment update.
As such, frequent updates to the alignment are not needed during data-taking periods.
They are usually required after long shutdown periods, when detector repairs and/or movements of the CMS detector wheels occur.
For this reason, at the start of each year of data taking, the matching requirement (between the tracker hits and the EB or EE SC)
for electrons is loosened at the trigger level
to remove any potential bias and efficiency loss due to changes in the alignment accuracy.

Continuous monitoring of the alignment is performed during data taking.
A relative ECAL-tracker alignment accuracy of better than $3\times 10^{-3}$ 
in both $\Delta\eta$ and $\Delta\phi$ has been achieved,
which meets the required accuracy for the electron identification~\cite{EGM17001}.

\subsection{The ES alignment}\label{sec:alignment_ES}
Similarly to the EB and EE, a dedicated procedure is adopted to align the ES with respect to the tracker.

\begin{figure}[htbp]
\centering
\includegraphics[width=.7\textwidth]{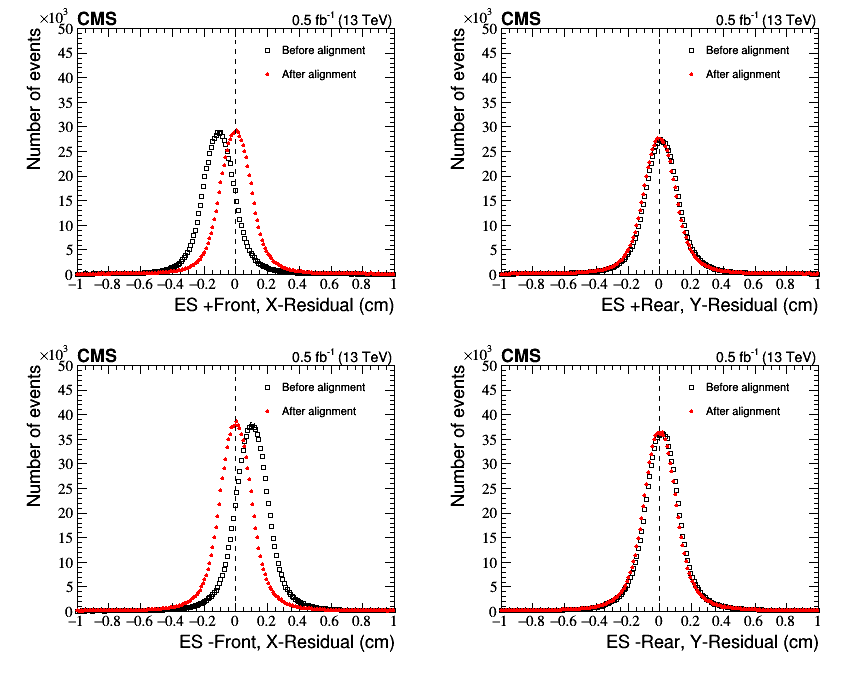}
\caption{
    Distribution of the residuals
    between the hits in the ES and the expected hit position from the reconstructed tracks,    
    before (black open squares) and after (red solid circles) the alignment of each ES plane
    was performed,
    using the first 0.5\fbinv of data recorded in 2017.
    The upper (lower) row corresponds to the ES detector at positive (negative) z, 
    and the left (right) column corresponds to the front (rear) detector plane.
    The vertical bars on the points represent the statistical uncertainty.
    }
\label{fig:AlignmentResidualBAPlot}
\end{figure}

The ES alignment algorithm calculates the residuals in x and y positions (in the plane transverse to the beams~\cite{Chatrchyan:2008zzk})
between the hits in the ES and the expected hit position extrapolated from the reconstructed tracks of minimum ionizing particles (mostly charged pions).
Each ES detector plane is a rigid element and therefore a three-dimensional translation and a rotation parametrised with three Euler angles are applied.
A method based on a $\chi^{2}$ minimization is used.
Figure~\ref{fig:AlignmentResidualBAPlot} shows the effect of the ES alignment on the distribution of the residuals for the four planes with
the first data from 2017.
A shift of about 0.1\cm was present in the front plane, which is corrected by the alignment procedure.
The RMS of the residuals is 0.055\cm, which is compatible with the width of the silicon strips.

\section{Calibration methods}\label{sec:calibration-intro-appendix}
In this appendix, the details of the different calibration methods are reported. 
The intercalibration is performed in $\eta$-rings, defined as rings along $\phi$ of one-crystal width in the EB,
ranging from 1 to 85, for the positive and negative $\eta$ sides of the EB,
and 39 rings in each of the two EE disks.

\subsection{The \texorpdfstring{$\phi$}{phi}-symmetry method}\label{sec:sub:phisymm}
\begin{figure}[!h]
  \centering
  \includegraphics[width = 0.48\textwidth]{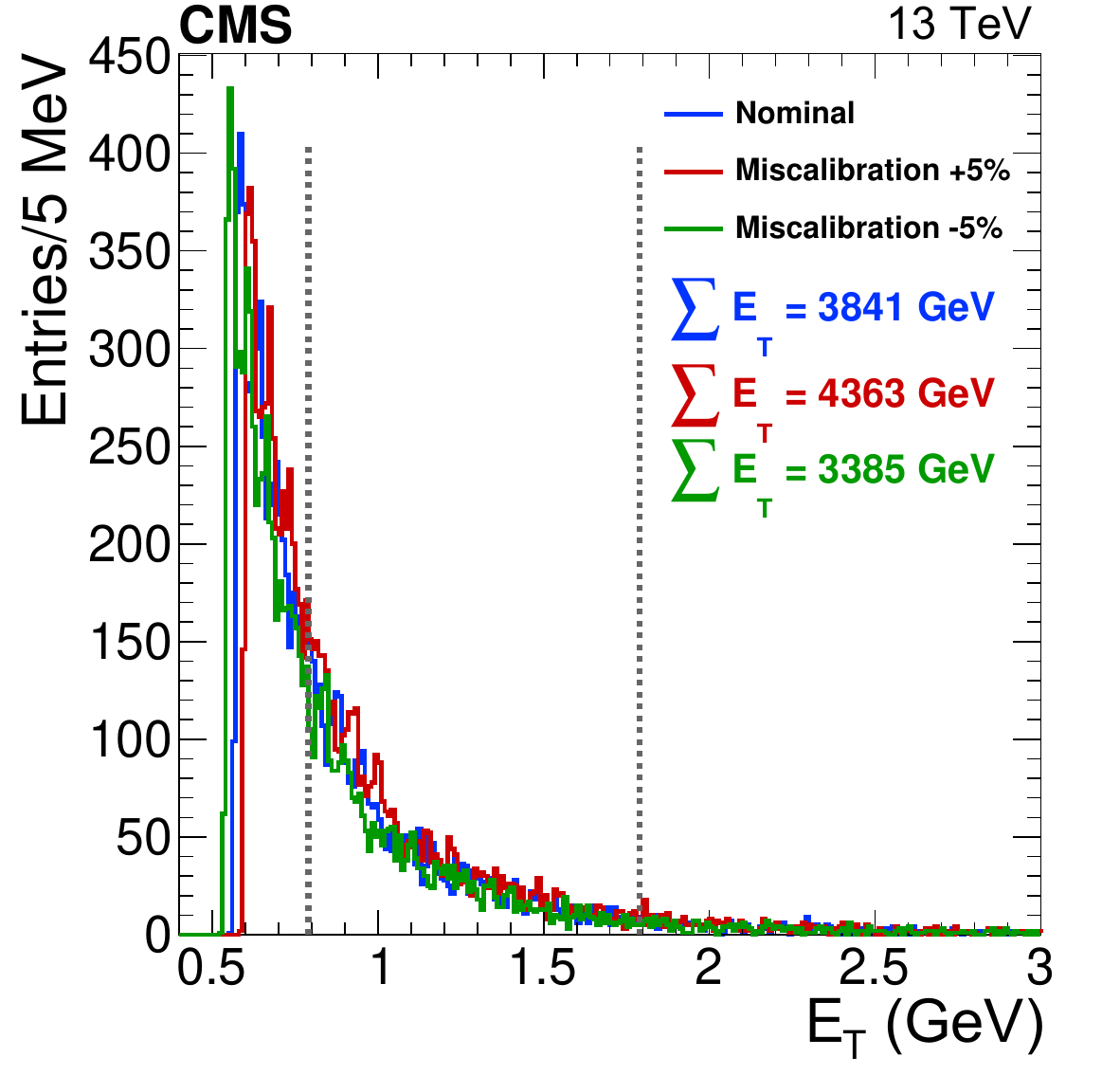}
  \includegraphics[width = 0.48\textwidth]{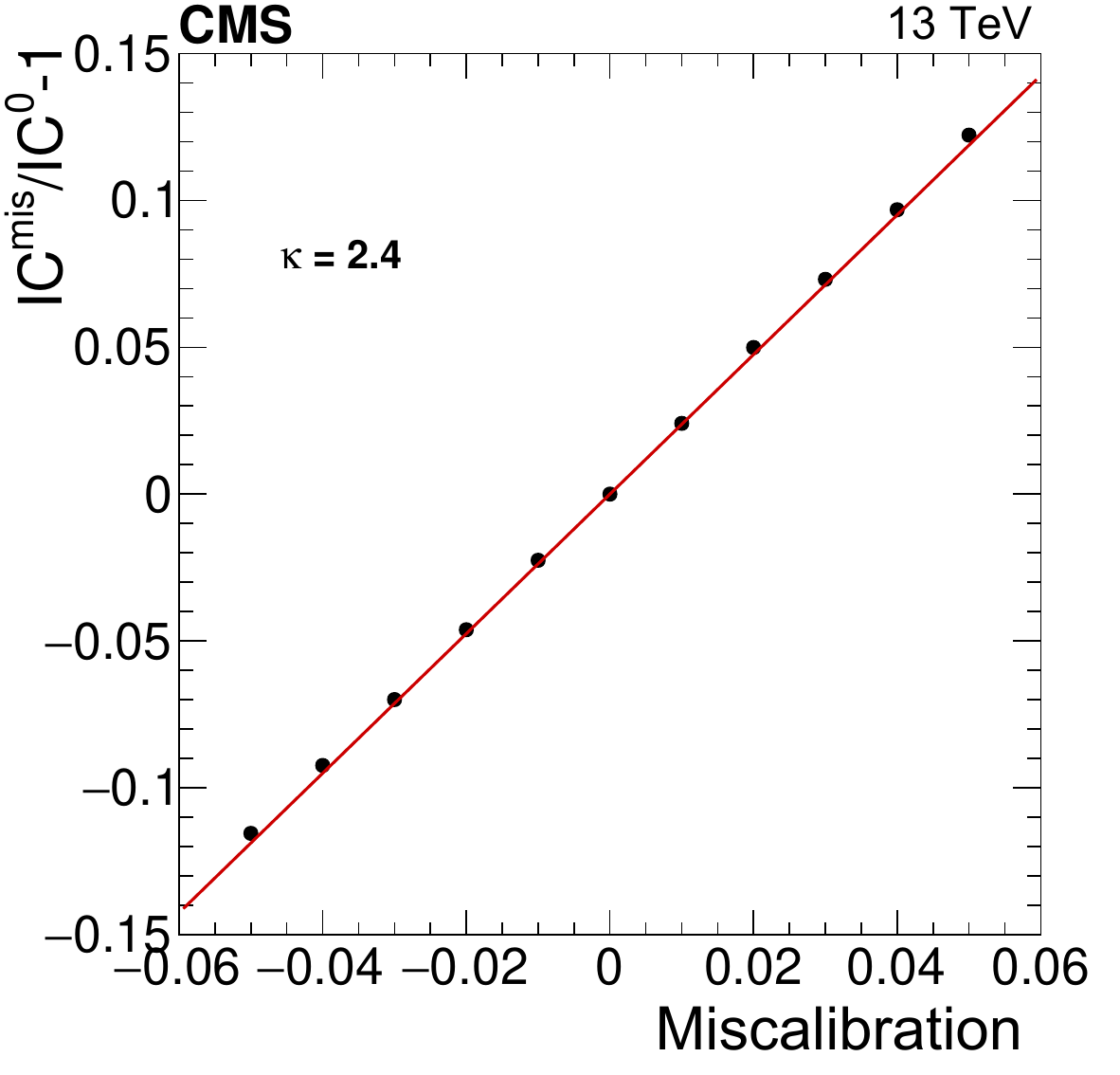}
  \caption{
  On the left, the energy spectrum of events selected by the $\phi$-symmetry HLT path in a chosen crystal in the central EB.
  The blue histogram corresponds to the measured energy deposition, while the red and green histograms are obtained from the blue histogram by
  injecting a miscalibration of $\pm5\%$. Vertical lines represent the interval of events selected in that crystal for the calibration. For each histogram,
  the sum of the energy of the selected events is also reported. 
  On the right, a typical fit 
  to extract the $\kappa$-factor.
  The $x$-axis is the injected miscalibration,
  while the $y$-axis is the variation in the IC constant with a given miscalibration
  minus one.
  A linear fit is superimposed (red line) and the $\kappa$-factor is the slope of the line.
  }
  \label{fig:PhiSymkFact}
\end{figure}

The $\phi$-symmetry intercalibration exploits the azimuthal symmetry of the energy distribution in soft $\Pp\Pp$ interactions.
The integrated energy deposition in the detector in ECAL for a large sample of collisions
is expected to be uniform along $\phi$. 
For any fixed $\eta$, the energy deposited in the crystals located at different $\phi$ will be, on average, the same;
therefore, any measured difference in the energy deposited is attributed to a
variation
in the response of the crystal itself.
Events used for $\phi$-symmetry calibration are selected by a zero-bias L1 trigger,
which accepts random bunch crossings without any requirements on detector activity,
to avoid  any possible bias in the calibration.  
The HLT selects events where the energy deposited in at least one ECAL crystal is above 7 times the expected noise.
The $\phi$-symmetry trigger is able to provide very high rates
by saving for each event only signals from the ECAL crystals above configurable thresholds. 
This trigger rate is larger than 2\unit{kHz} during data taking and up to 30\unit{kHz} during commissioning periods,
whereas the ensemble of standard physics triggers has a rate of about 1\unit{kHz}.
Events passing the online selection are further refined offline by applying an $\eta$-dependent lower energy threshold of 10 times the average RMS noise for channels at fixed $\eta$.
An upper energy threshold, such that the $\ET$ window has a range of 1\GeV, 
is applied to avoid biases from sporadic events arising from hard interactions.
The relative response among crystals in the same $\eta$-ring is measured.
It is computed from the ratio of the $\ET$
deposited in each crystal and the ring average, as illustrated in Eq.~(\ref{PhiSymCorr}):
\begin{equation}
\mathrm{IC}^{i} = \frac{\sum_{j}\ET^{j}/N}{\ET^{i}} \kappa^{i}
\label{PhiSymCorr}
\end{equation}
Here IC$^{i}$ represents the measured response variation for the $i$th ECAL crystal.
The denominator is the sum of the transverse energy of the hits in the $i$th crystal within the specified energy window,
while the numerator is the sum of the transverse energy of the hits in the $\eta$-ring to which the $i$th crystal belongs, averaged over the $N$ crystals of the ring.
For a perfectly calibrated detector the ratio would be 1 
and any deviation from unity is interpreted as the variation of the channel response. 
This measurement is performed in time intervals of a few days, long enough to provide a sufficient number of 
deposits in each crystal.
The parameter $\kappa$ is a correction factor to account for the effect of using a fixed energy window.
Events with an energy close to the window boundary are not
only shifted by the presence of a miscalibration, but might also fall outside the accepted window.
The left panel of Fig.~\ref{fig:PhiSymkFact} illustrates
how the shift of the energy due to a miscalibration influences the energy deposited in the
window.
The $\kappa$ factor, which is necessary to correct for the threshold effect,
is computed by injecting a set of known miscalibrations in the data and rederiving them from Eq.~(\ref{PhiSymCorr}).
For miscalibrations of a few percent, the derived miscalibration as a function of the true value is fitted by a linear function and
the slope is the $\kappa$ factor, as illustrated by the right panel of Fig.~\ref{fig:PhiSymkFact}. Typical values of the $\kappa$ factor are about 2.

\begin{figure}[!h]
  \centering
    \includegraphics[width = 0.49\textwidth]{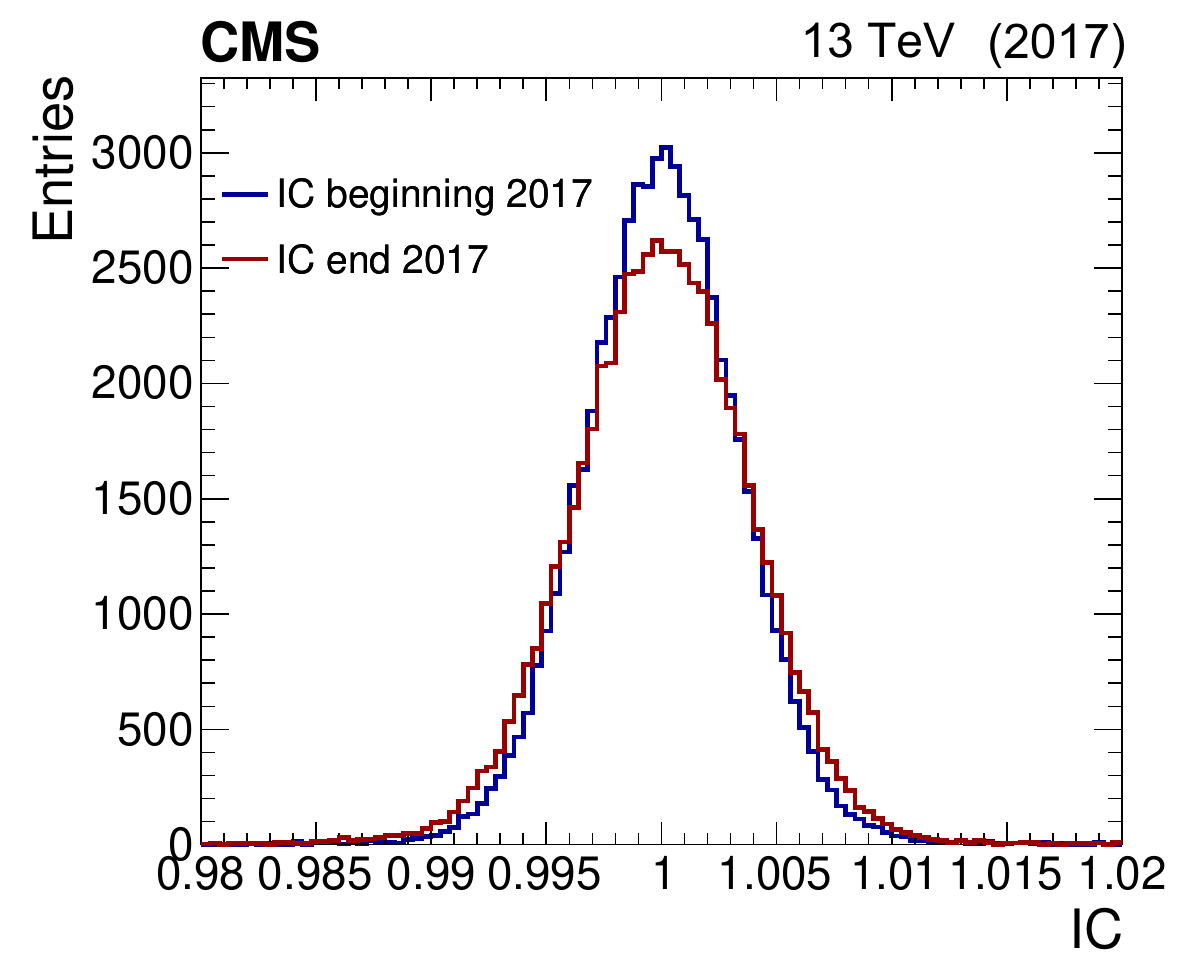}
    \includegraphics[width = 0.49\textwidth]{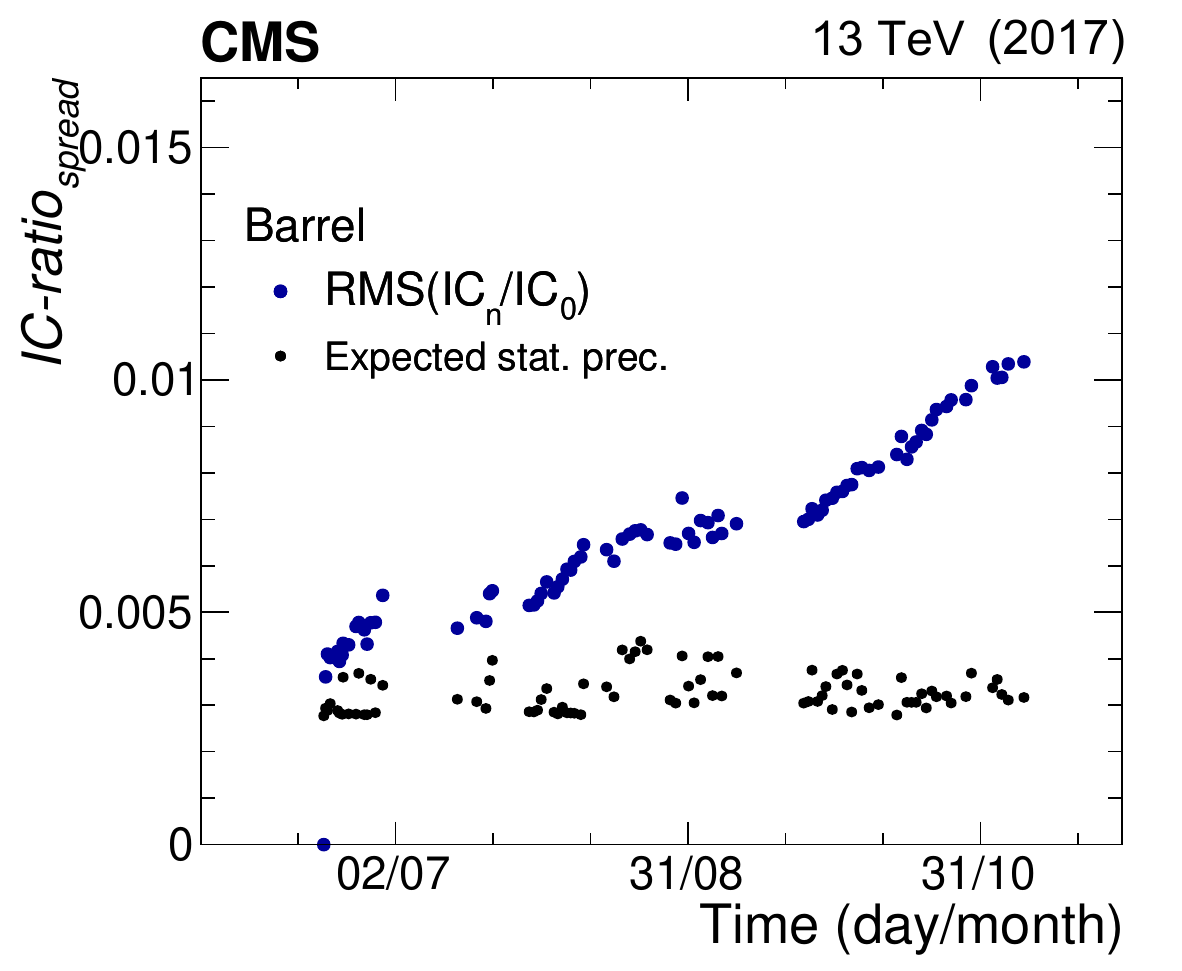}
  \caption{
  On the left, the distribution of the normalized IC constants of all EB crystals
  shown for two time bins, at the beginning (blue) and at the end (red) of the 2017 data-taking period.
  The IC value is normalised, for each crystal, to the initial value.
  On the right, the RMS of the distribution of the
  ratio of the IC constants at a given point in time 
  over the beginning of the year is shown in blue.
  The black points show the statistical precision of the method, evaluated by randomly splitting the data
  into two subsets.
  }
  \label{fig:PhiSymMonitoring}
\end{figure}

The $\phi$-symmetry method can be exploited both for the monitoring of the ECAL response during data taking
and for the derivation of the IC constants.
During data taking, the $\phi$-symmetry can provide a prompt method to assess whether the radiation damage induced
in the ECAL is properly corrected by the laser monitoring system.
IC values are computed for each crystal in the different time intervals, as in Eq.~(\ref{PhiSymCorr}). 
The ratio between the IC in a given time interval and a reference value, usually the first time interval of each year, is considered.
The full detector (or a region of it) can be monitored from the RMS of the distribution
of the IC ratios for all the crystals versus time.
With a perfectly calibrated detector the ratio would fluctuate around 1 with an RMS 
 equal to the statistical accuracy of the method. 
The left panel of Fig.~\ref{fig:PhiSymMonitoring}
shows the IC ratio distribution for two time bins, at the beginning and at the end of the 2017 data-taking period.
As expected, the two distributions peak at 1, while the RMS is wider at the end of the data-taking period.
The right panel shows the evolution of the RMS of the IC ratios versus time.
For a detector with a stable response,
the RMS would be flat versus time, at a value corresponding to the statistical accuracy of the method.
The increase of the RMS versus time is 
due to residual effects of radiation-induced damage that remain even after corrections from the other methods.
Such a drift in the RMS indicates that a time dependent calibration of the detector is required.
 
The IC constants are derived from Eq.~(\ref{PhiSymCorr}), integrated over the whole year.
In the EB, the values are corrected to account for the nonuniformity of the material budget in front of the ECAL.
The energy deposition in the tracker services and support structures causes the energy measured in the crystals behind
them to be lower than the ring average; thus the IC value would be correspondingly enhanced.
Since the material in front of the ECAL has a relative effect on the IC that is constant along $\eta$,
due to the fact that the structures run at constant $\phi$,
the correction is computed by averaging the IC along $\eta$ and
dividing the IC by these averages.
Similar effects are observed in the EE, but the different geometry of the
ECAL crystals and of the tracker structures prevents from factorising the effect of the material from the crystal response variation.
As a result the IC values from $\phi$-symmetry in EE are affected by large systematic uncertainties that prevent their use for calibration purposes.
However, since the effect of the nonuniform material does not vary with time,
it does not affect the monitoring variable, as illustrated in Fig.~\ref{fig:PhiSymMonitoring}, where only ratios of the IC values are considered.

\subsection{The \texorpdfstring{\Pgpz}{pi0} calibration method}\label{sec:sub:pi0}
The \Pgpz intercalibration method uses the value of the peak in the invariant mass distribution of unconverted photon pairs from \Pgpz meson decays.
This method benefits from the
large
production of \Pgpz particles in $\Pp\Pp$ collisions.
Events are selected by a dedicated trigger stream, which saves only limited
information from the ECAL detector in the vicinity of the selected photon candidates.
The resulting event size is about 2\unit{kbyte}, 
three orders of magnitude smaller than the typical size of a CMS physics event.
This feature minimizes the usage of the CMS readout bandwidth and storage space, allowing the stream to operate
with a rate of about 7 (2)\unit{kHz} in the EB (EE).
The lower rate in the EE comes from the tighter kinematic selections applied to reduce the
contribution from noise and guarantee a high signal purity.

The stream uses events selected by the L1 trigger
that have at least one electromagnetic object
or at least one hadronic jet is found.
It employs a simplified clustering algorithm that identifies photons as $3\times3$ 
crystal matrices centred on specific crystals,
called seeds,
with an energy deposit 
of greater than 0.5 (1.0)\GeV in the EB (EE).
Photon candidates are built
starting from the most energetic seed, 
and no energy sharing is allowed in the case of partially overlapping matrices.
Additional kinematic selection criteria are applied to photons and \Pgpz candidates to improve the signal purity:
photon candidates with $\pt>0.65\,(1.0)\GeV$ and \Pgpz candidates with $\pt>1.75\,(2.0)\GeV$ are required  
to have at least 6\,(7) crystals in each cluster in the EB (EE).
The pair of clusters must be isolated from other nearby energy deposits.
The isolation, defined as the ratio of the scalar sum of \pt from all clusters (excluding those forming the \Pgpz) 
found within a cone of radius $\sqrt{\smash[b]{\Delta\eta^2 + \Delta\phi^2}}= 0.2$ centred on the \Pgpz candidate 
and the \pt of the \Pgpz candidate itself, is required to be less than 0.5.
Finally, the  invariant mass of the diphoton system (\mgg) 
is required to be within the interval of [60, 250]\MeV.
The transverse energy distribution of selected photons 
peaks between 1 and 3\GeV
for the leading photon in the $\Pgg\Pgg$ pair, with an exponentially decreasing tail extending above 10\GeV.
Although the $3\times3$ matrix is sufficient to contain the largest fraction of the energy of these photons,
the method is highly sensitive to detector noise and pileup, 
which can potentially bias the measured energy 
or create spurious clusters. In addition, the 
selective readout
thresholds can lead to a significant loss of information in the
reconstruction of \Pgpz mesons.
The kinematic selection was optimized to minimize the
impact of these effects on the measured mass and the ICs.
A fraction of the photon energy is expected to leak outside the cluster or to be lost if the cluster is formed in the
vicinity of detector gaps or dead channels. The effect of gaps is particularly relevant in the EB due to the boundaries between modules and between supermodules.
These energy losses in the EB are corrected for using a dedicated set of containment corrections obtained from simulation.

\begin{figure}
\centering
\includegraphics[width=0.49\textwidth]{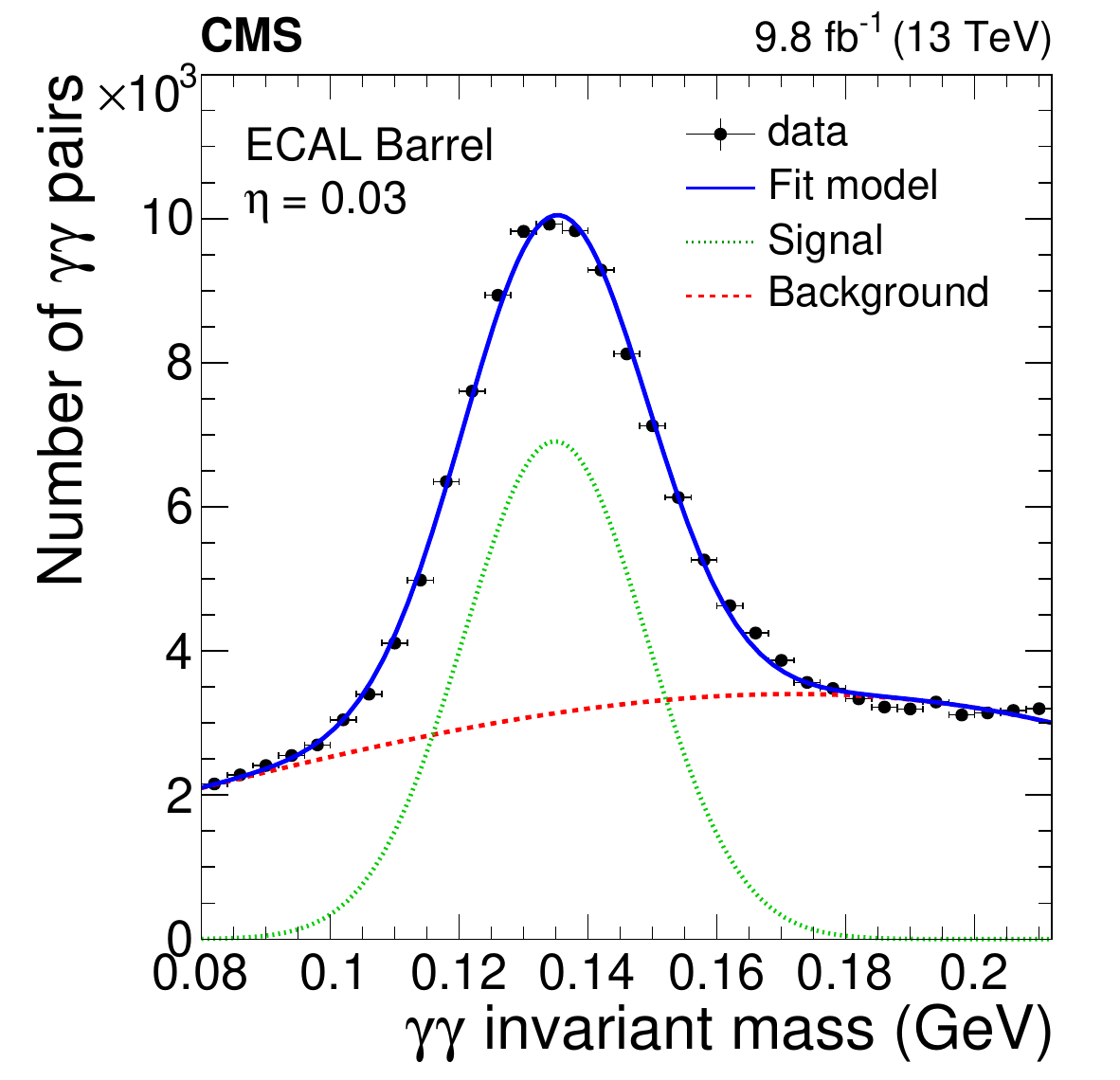}\label{fig:pi0MassEB}
\includegraphics[width=0.49\textwidth]{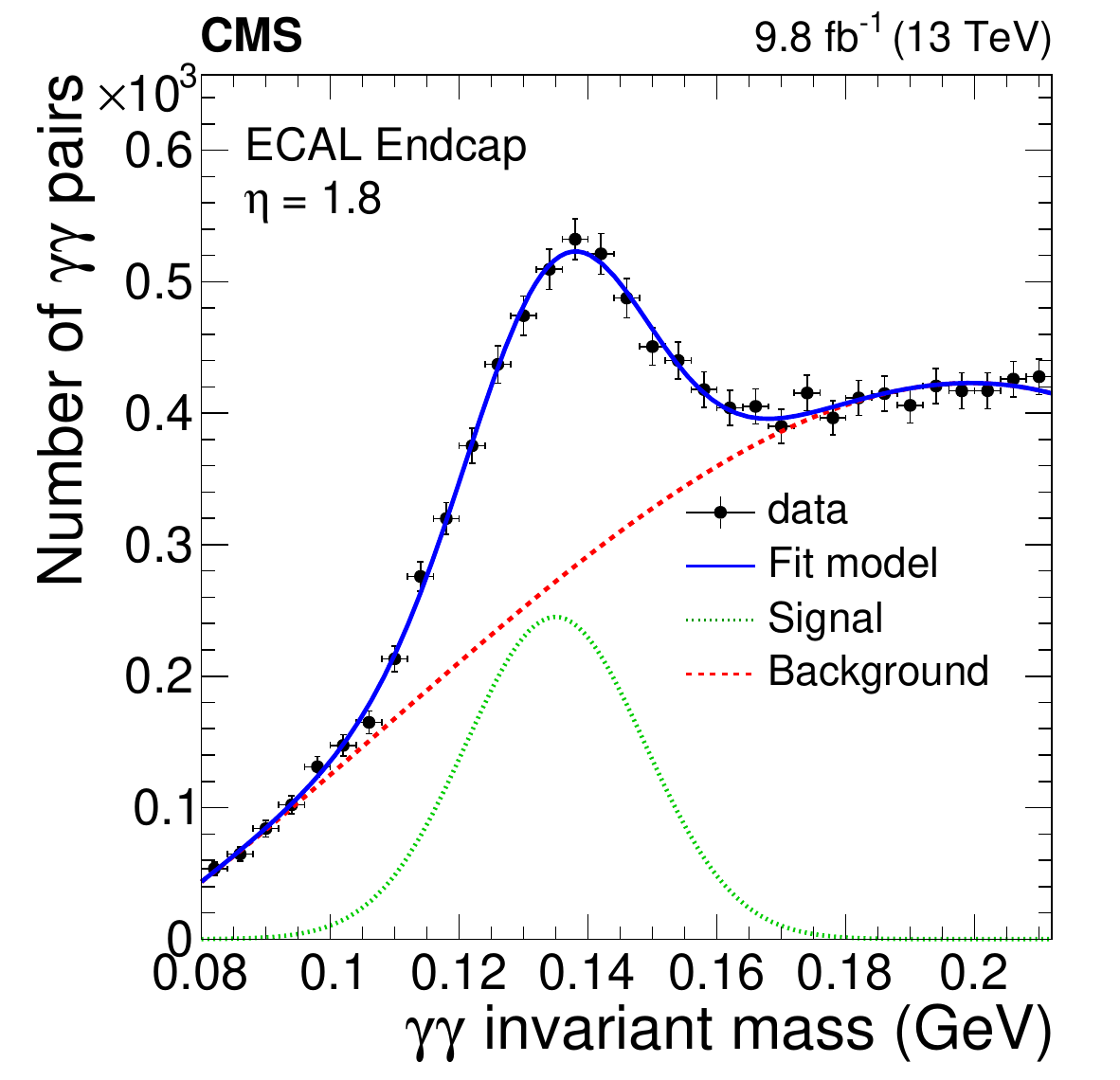}\label{fig:pi0MassEE}
\caption{
    The invariant mass distribution of photon pairs around the \Pgpz mass peak for one crystal in EB (left) and EE (right).
    Data (black points) are fitted with the sum (solid blue line) of signal (dashed green line) and background components (dotted red line), 
    as detailed in the text.
        The vertical bars on the points represent the statistical uncertainty.
}
\label{fig:pi0Mass}
\end{figure}

The IC constants are derived from a fit to the \mgg distribution measured in each channel.
The single-channel invariant mass distribution is obtained from all \Pgpz candidates for which one photon has
deposited a fraction of its energy in that crystal, with a weight based on the corresponding energy fraction.
Data are fitted with a sum of signal and background components modelled as a Gaussian function and a 
second-order Chebyshev polynomial, respectively. 
The peak of the \mgg distribution is obtained from the
mean parameter of the Gaussian function. Figure~\ref{fig:pi0Mass}
shows the fitting result for two representative channels, one in the EB and one in the EE.

The IC constant is computed as:
\begin{equation}
\label{eq:pi0_IC_formula}
\mathrm{IC}^{i} = \frac{1}{1+r^{i}} ~~~~~~ \text{with} ~~~~~~ r^{i} = \frac{1}{2} \left[ \left(\frac{m_{\Pgpz}^{i}}{m_{\Pgpz}^\text{true}}\right)-1 \right],
\end{equation}
where $m_{\Pgpz}^\text{true} = 0.1349\GeV$ is the world-average mass of the \Pgpz meson~\cite{PhysRevD.98.030001}, 
and $m_{\Pgpz}^{i}$ is the measured mass from the fit in the $i$th channel.
The formula in Eq.~(\ref{eq:pi0_IC_formula}) originates from a Taylor series expansion of the expression for the reconstructed invariant mass divided by the true mass,
neglecting second-order terms~\cite{Meridiani:934068}.
Since $m_{\Pgpz}^{i}$ also involves the energy deposited in the crystals surrounding the $i$th channel, the procedure in Eq.~(\ref{eq:pi0_IC_formula})
is iterated 
multiple times, where in each iteration the measured photon energy in each channel is corrected according to the IC constants obtained in the previous iterative step.
The convergence criterion is that the change in the IC constants from one iteration to the next is less than one tenth of the statistical uncertainty.

Because of the much lower energy of photons used in the calibration with respect to the typical energies of photons and electrons
used in physics analysis, 
the \Pgpz method is not used to derive the absolute scale of the ECAL as a function of $\eta$.

\subsection{The \texorpdfstring{$E/p$}{E/p} method}\label{sec:sub:Eop}
\begin{figure}[htbp]
\centering
\includegraphics[width=.6\textwidth]{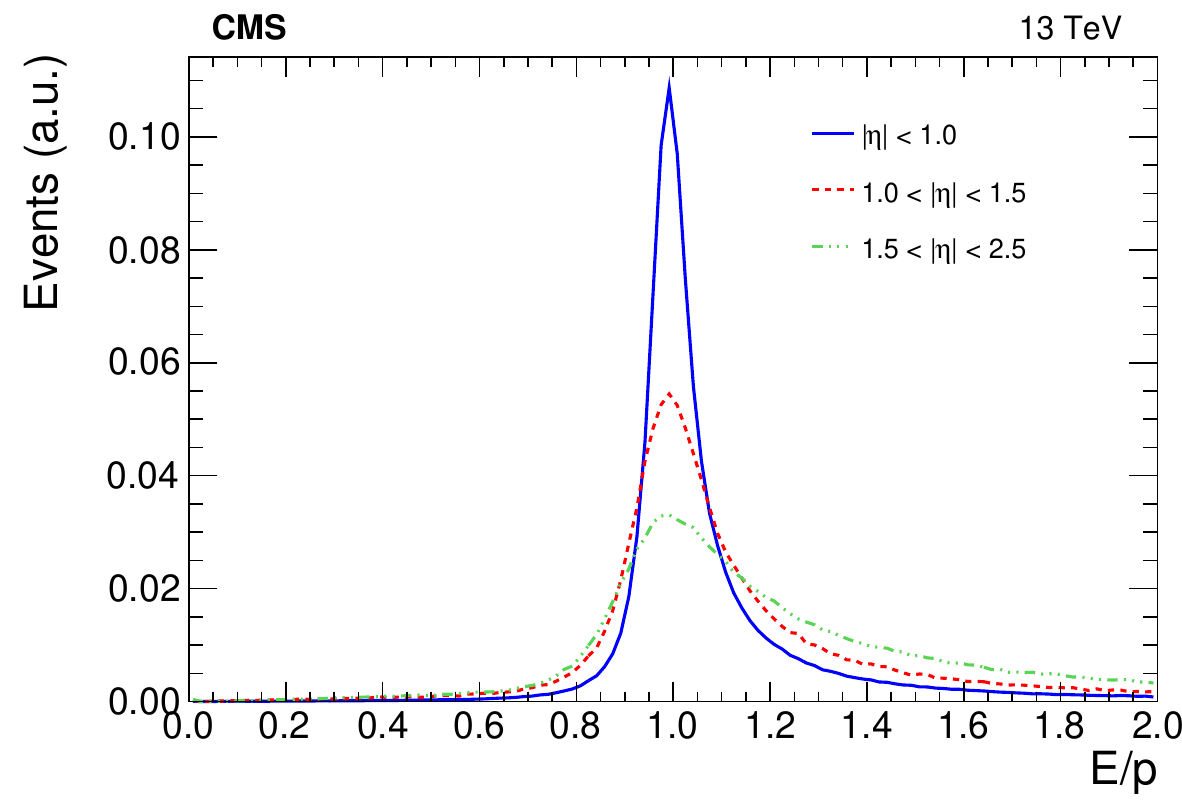}
\caption{
    The $E/p$ distributions measured from data in different intervals of $\eta$.
    Electrons from \PW and \PZ boson decays are selected, whose \ET ranges between 30 and 70\GeV.
    }
\label{fig:Eop_distribution}
\end{figure}

In the $E/p$ method, the calibration of each channel is obtained
exploiting the narrow distribution of the ratio between the reconstructed energy ($E$)
in the ECAL and the momentum ($p$) of 
electrons measured in the tracker. 
A set of selections is applied based on electron kinematics, 
identification, and isolation
in order to obtain a pure sample of 
electrons which mostly arise from
\PW and \PZ boson decays.
Events from \PW boson decays are selected requiring:
\begin{itemize}
\item  exactly one electron reconstructed within the tracker acceptance ($|\eta|<2.5$), 
       with transverse momentum \ptmiss greater than 30\GeV, and satisfying a tight identification criterion~\cite{EGM17001}
\item  missing transverse momentum greater than 25\GeV
\item  transverse mass, computed as
  \begin{equation}
    M_T = \sqrt{(2 \ptmiss \ET) (1-\cos \Delta\phi)}
  \end{equation}
  greater than 50\GeV, with $\Delta\phi$ the angle between the electron and the \ptmiss in the transverse plane, and $\ET$ the transverse energy of the electron.
\end{itemize}
Events from \PZ boson decays are selected requiring:
\begin{itemize}
  \item an electron-positron pair, with both the particles satisfying loose identification criteria. If more than two pairs pass the selection, the pair with the highest \pt is used,
  \item the dielectron mass greater than 55\GeV.
\end{itemize}
The algorithm used for the calibration is iterative: for each iteration,
the IC constant for the $i$th crystal, IC$^i$, is updated to constrain the average $E/p$ ratio to 1.
In particular, the IC constant for the $i$th crystal at the $N$th iteration is computed with the formula:
\begin{equation}
\label{eq:EopIC}
\mathrm{IC}^i_N = \mathrm{IC}^i_{N-1}\, \frac{\sum_{j=1}^{N_\Pe} w_{ij}\, f(E_j^{N-1}/p_j,\eta_j) \, (p_j/E_j^{N-1})}{\sum_{j=1}^{N_\Pe} w_{ij}\, f(E_j^{N-1}/p_j,\eta_j)},
\end{equation}
where the index $j$ runs over the total number of selected electrons, ${N_\Pe}$. 
Furthermore:
\begin{itemize}
  \item $w_{ij}$ is the fraction of the SC energy of the $j$th electron deposited in the $i$th crystal.
  \item $f(E_j^{N-1}/p_j,\eta_j)$ is a weight assigned to the $j$th electron representing the probability
        of measuring a certain value of the $E/p$ ratio at $\eta$.
\end{itemize}
For each iteration, the SC energy for the $j$th electron is recomputed as the sum of the energy deposited in each crystal,
weighted by the corresponding IC values obtained in the previous iteration:
\begin{equation}\label{eq:SC_Energy}
  E_j^{N-1} = \sum_{k \in SC} E_{kj} \, \mathrm{IC}^k_{N-1},
\end{equation}
where $E_{kj}$ is the reconstructed energy of the $k$th crystal contributing to the $j$th electron.
The event weight $f(E_j/p_j,\eta_j)$
is built using the $E/p$ distribution from data
in windows of $\eta$.
The variation of the $E/p$ distributions as a function of $\eta$
is shown in Fig.~\ref{fig:Eop_distribution}. These differences are
due to the variations in the material budget upstream of the ECAL.
At each iteration, the $E/p$ distributions are recomputed using the updated values of the SC energy~$E$ (Eq.~\ref{eq:SC_Energy}).
The left and right tails of the $E/p$ distribution extend to 0.5 and 2, respectively.
Because of the expected energy resolution of the ECAL for the selected electrons ($\Delta E/E \approx 10^{-2}$),
such a significant deviation from the expected value of 1 is ascribed to the momentum resolution of the tracker for electron tracks,
mainly due to bremsstrahlung. Therefore, the weight $f(E_j/p_j,\eta_j)$ in Eq.~(\ref{eq:EopIC})
is intended to reduce the impact of events with poor momentum reconstruction.
Studies have shown that the inclusion of events populating the tails of the $E/p$ distribution, even if with a lower $f(E_j/p_j,\eta_j)$ weight,
results in a worsening of the IC precision. A selection based on the $E/p$ values itself is therefore applied to each electron,
at each iteration of the algorithm.

A fundamental assumption of the $E/p$ method is that the momentum measured by the tracker
does not have any $\phi$-dependent bias. This assumption is not completely true because the momentum of electrons and positrons measured
by the tracker is affected by the presence of the tracker support structures.
This effect can be
corrected
by means of scale factors that are derived 
exploiting the $\phi$-symmetry of the  invariant mass for \zee events. 
These are computed using the tracker momentum for one of the two decay products in a given $\phi$ window
and the ECAL energy for the other, that could be in any region of the detector.
Due to the bending in the magnetic field, the effect of the tracker structures is different for electrons and positrons,
therefore two separate corrections are derived.
The modulation of the correction factor in $\phi$, up to 1\%, matches well with the position of the tracker support structures and services,
and the effect is confirmed by MC simulations.

The $E/p$ method is not used to derive the absolute scale of the ECAL as a function of $\eta$
because the dependence of the $E/p$ distribution along $\eta$ is difficult to model 
with the required accuracy, but it is used to derive relative ICs between crystals in the same $\eta$-ring. 

\subsection{Calibration with \texorpdfstring{\zee}{Zee} events}\label{sec:sub:calibrationZee}
The \PZ boson properties have been extremely well measured by the LEP experiments~\cite{PhysRevD.98.030001},
in particular the mass, which has a relative uncertainty of about $2 \times 10^{-5}$. 
This is used to calibrate the ECAL response to electrons using \zee decays,
assuming that the same  invariant mass should be observed for pairs of electrons in any 
region of the detector.
The \PZ boson natural width is usually nonnegligible with respect to the electron energy resolution.
To make the best use of the available data sample,
a method based on maximising the unbinned likelihood has been developed, 
as described in detail in Ref.~\cite{couderc:tel-02106984}.

The likelihood compares the lineshape of the invariant mass of electron-positron pairs ($m_{\ell\ell}$) after reconstruction to the expected lineshape. 
The energy scale and resolution of the two electrons are vectors of free parameters
for different regions of the detector, to be determined in the fit. 
In principle, the likelihood of $m_{\ell\ell}$  can be obtained from a 
simulation that includes a realistic description of the detector effects and background contributions. 
However, this approach requires complex modelling of the energy resolution and is not practical 
with the large number of free parameters to be determined. 
Instead, a simplified description of the $m_{\ell\ell}$  lineshape is used with the following assumptions:
\begin{itemize}
\item The background contamination in the dielectron sample is negligible.
\item The underlying dielectron mass distribution from \PZ boson decays is well represented by a classical Breit-Wigner function 
      and relativistic effects can be neglected.
      In addition any deviation from the classical Breit-Wigner function due to acceptance effects can be ignored.
\item The energy response function of ECAL is described by a Gaussian function and the tails can be neglected.
\end{itemize}

With these assumptions, the probability distribution of $m_{\ell\ell}$
can be computed from a Voigtian function~\cite{cite:voigtian:1975},
which is a Breit-Wigner distribution convolved with a Gaussian distribution.
The variable $m_{\ell\ell}$ is given by
\begin{equation}
 m_{\ell\ell} = \sqrt{2 E_1  E_2  (1-\cos(\theta))},
\end{equation}
where $E_1$ and $E_2$ are the energies of the two electrons and $\theta$ their angular separation.
The two parameters of the Gaussian smearing are then defined as:
\begin{equation}
 \mu = \sqrt{\mu_{1} \mu_{2}} ~~~~~~~~~~~~  \sigma = 0.5  \sqrt{\sigma_{1}^2 + \sigma_{2}^2}
\end{equation}
where $\mu_{i}$ and $\sigma_{i}$ are the energy scale and the resolution parameters
associated with the $i$th electron.
Since there are many free parameters in the fit, an approximate form of the Voigtian function is used~\cite{cite:voigt:approx}
that allows for an analytical computation of the gradients used in the likelihood maximisation.

This method is used for:
\begin{itemize}
 \item the absolute calibration and $\eta$-scale,
 \item the IC measurement along $\phi$, and
 \item the estimation of the energy resolution for electrons, for performance estimation and IC combination (described in Section~\ref{sec:combination}). 
\end{itemize}

The $\eta$-scale calibration involves equalizing the energy response for each $\eta$ ring.
This is achieved using \zee decays and the scale is determined by matching measurement to the simulation.
Electrons that are less affected by bremsstrahlung, and therefore have a lower dependence on the upstream material included in the simulation, 
are used. They are selected by means of a topological selection based on the $\RNINE$ variable, which is defined as
\begin{equation}
 \RNINE = \frac{ E_{3\times 3} }{E_\mathrm{SC}} , 
 \label{eq:cms:r9}
\end{equation}
where $E_{3\times 3}$ is the energy sum in the 9 crystals around and including the seed crystal, 
and $E_\mathrm{SC}$ is the SC energy.
Only electrons with large values of $\RNINE$ ($>0.94$) are used to compute the $\eta$ scale.
The selection on $\RNINE$ is slightly relaxed compared to the one used to assess the performance ($\RNINE>0.965$) to increase the number of events for the $\eta$-scale determination.
A fit is performed to this sample, with one free scale parameter per $\eta$-ring (the i-th ring energy scale, $S_{i\eta}$) and
20 free parameters for the resolution (2 bins in $\RNINE$ and 10 bins in $\abs{\eta}$). 
This amounts to $85 \times 2$ ($39 \times 2$) scale parameters for the EB (EE).
In the fit, the electron energy is rescaled according to the $S_{i\eta}$ value of the seed crystal in the SC,
the scale of the ring where the shower seed falls.
Since the tracker covers the EE only up to $\abs{\eta}<2.5$, 
the $\eta$ scale calibration is performed with reconstructed electrons only within the tracker.
For $\abs{\eta}>2.5$, pairs of electrons and SCs are used, where the electron is within the tracker coverage. 
The fit is performed on different samples: first for the EB-EB pairs; then the EB-EE~$+$~EE-EE pairs with the parameters for the EB fixed;
and finally for the SCs with $\abs{\eta}>2.5$ using the electron-SC pairs, keeping the $S_{i\eta}$ for $\abs{\eta}<2.5$ fixed (together with a free parameter for the resolution of the SC).
The same fitting procedure is applied to data and 
simulation
samples, and the  $\eta$ scale is defined as the ratio of the 
$S_{i\eta}$
parameters measured 
in data to those measured in simulation.
The $\eta$-scale is then applied as a scaling factor to the calibration constants applied to data. 

The second use of \zee events is for intercalibration purposes.
The \zee method is based solely on the electron energy from ECAL energy deposits and is therefore
negligibly affected by
uncertainties in the distribution of upstream detector material, which can
 affect the momentum measurement from the tracker.
Pileup biases are minimised in the energy reconstruction of electrons 
by means of a dedicated regression~\cite{EGM17001}.
Nevertheless, the \zee event count is small compared to the other methods,
which is partially compensated by the improved \zee energy resolution. 
Similarly to the other techniques described in earlier sections, this method is used to equalize the crystal energy response along~$\phi$. 
A free energy scale parameter is assumed for each crystal, and the energy of the electron is scaled
according to the scale parameter of the seed crystal in the electron SC,
while there are 20 additional parameters related to the resolution in the fit, as described previously.
The calibration is performed with 90\% of the available \zee events, 
 and the remaining 10\% are used for validation.
The absolute $\eta$-scale is applied
before performing the IC derivation,
so that all the $\eta$-rings have the same average energy.

The simultaneous calibration of the full EB is not 
computationally sustainable.
To reduce the number of free parameters, one scale parameter per crystal is defined in a 9 $\eta$-ring window,
while 360 scale parameters are used for the rest of the EB, grouping along $\eta$ the crystals outside the window.
To ensure that the ICs are measured with each SC fully contained in the window, 
only the scale parameters of the 5 rings in the centre of the window are used as the calibration constants.
This procedure is repeated in different $\eta$ regions to scan the whole EB.
This method is sensitive to gaps between the EB supermodules (which occur every 20 crystals in  $\phi$),
because the energy regression does not completely recover the energy lost in the gaps.
An empirical correction is derived 
by performing a preliminary fit of the whole EB, 
folding all the supermodules together 
(since the effect should be the same for all modules). Likewise, the $\eta>0$ (EB+) and $\eta<0$ (EB--) regions are folded together,
and only 4 $\eta$-regions are used.
Since after the folding each scale parameter corresponds to the average of the IC constants of about 775 crystals,
the scale parameters are expected to be uniform along $\phi$, and the observed nonuniformity (at the percent level) is applied
as a multiplicative factor to the ICs to correct for the $\phi$ modulation.

For the EE, the \zee fit is performed over the full EE+ ($\eta>0$) and EE-- ($\eta<0$) regions and no additional corrections are applied.

\subsubsection{The \texorpdfstring{\zee}{Zee} systematic studies}\label{app:Zee}
The \zee method has been used extensively
for the intercalibration,
to determine the absolute energy scale,
and to extract resolution parameters for the IC combination.
Here the biases in the method and how they are taken into account are discussed.

The scale and resolution parameters obtained from the fit roughly correspond to the mean and RMS 
of the energy response function over a truncated range. 
Since only the relative difference of the scale parameters
is used, either with respect to the simulations as in the $\eta$-scale determination, or between crystals as in the intercalibration,
the bias introduced by these assumptions should be small.
To evaluate this, the procedure was tested with 
simulated \zee events.
For this, the reconstructed energy was rescaled and the resolution of the electrons degraded
using a Gaussian distribution with parameters that depended on $\eta$.
It was confirmed that the coefficients are correctly retrieved by the \zee method. 
The energy scale parameters were fitted in the phase spaces $i\eta$ and $\RNINE$ 
and the resolution parameters in the phase spaces $\abs{\eta_\mathrm{SC}}$ and $\RNINE$,
where the variable $i\eta$ refers to the $\eta$-ring number of the SC seed crystal, $\eta_\mathrm{SC}$
is the energy log-weighted average of the pseudorapidity of the crystals in the SC, and $\RNINE$ is defined in Eq.~(\ref{eq:cms:r9}).
Electrons with a high value of $\RNINE$ usually lose a small fraction of their energy by bremsstrahlung,
and thus have a better energy resolution than low-$\RNINE$ ones. 
The $\RNINE$ dimension is split into 2 bins,
the $i\eta$ dimension contains 118 bins, and the $\eta_\mathrm{SC}$ dimension contains 25 bins,
which represents a total of 
236 scale and 50 resolution parameters.
The test has been performed on four different 
\zee samples:
\begin{itemize}
\item \textit{Original simulation.}   This sample is used to define the reference for the energy scale and resolution parameters.
\item \textit{Scale-only simulation.} Only the electron energy response is scaled, with a sinusoidal dependence on $\eta$, leaving the energy resolution unchanged.
\item \textit{Smear-only simulation.} The scale of the electron energy response is not modified but the energy resolution is degraded.
\item \textit{Scale+Smear simulation.} The energy response is scaled and the energy resolution is degraded.
\end{itemize}

\begin{figure}[h]
\centering
\includegraphics[width=0.49\textwidth]{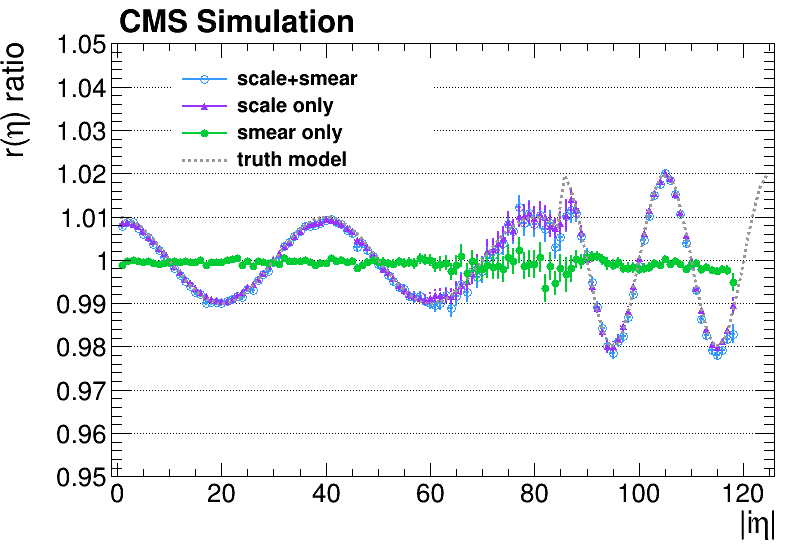}
\includegraphics[width=0.49\textwidth]{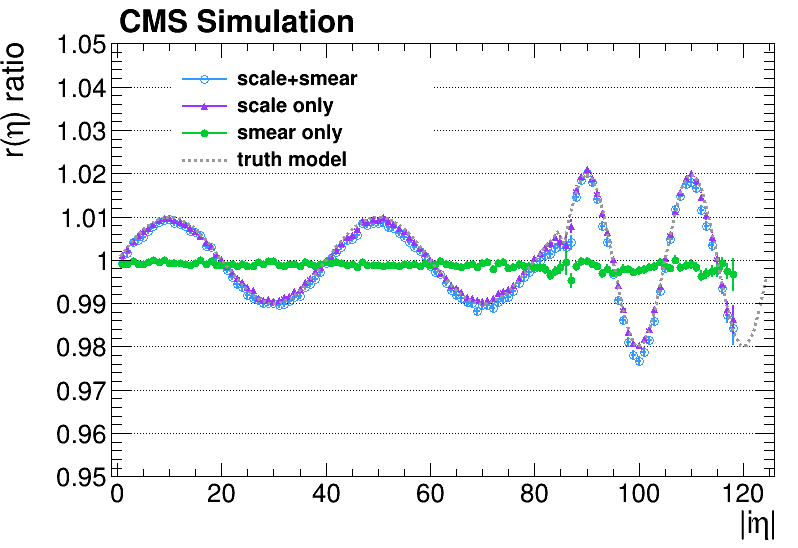}
\includegraphics[width=0.49\textwidth]{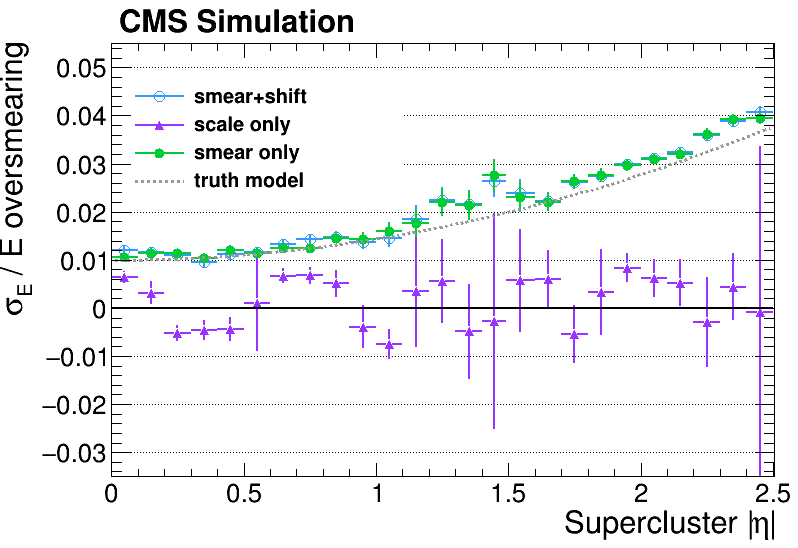}
\includegraphics[width=0.49\textwidth]{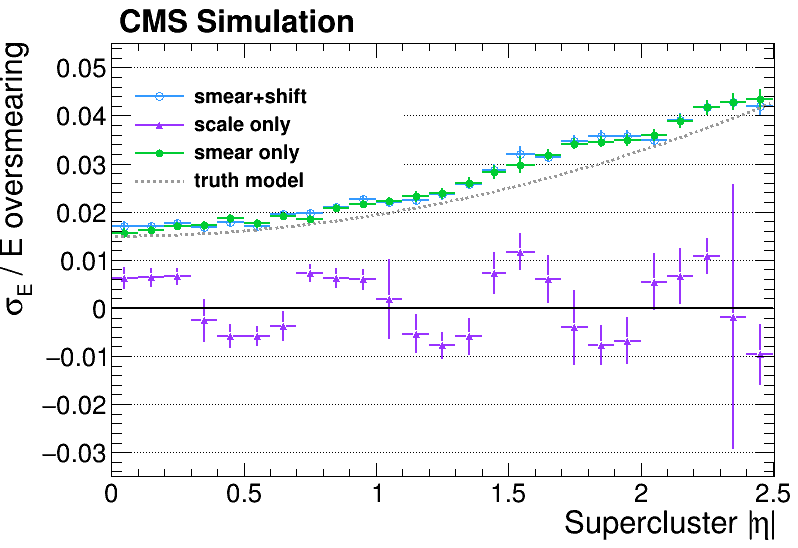}
\caption{
    The result of the \zee method validation fit to the four simulations described in the text. 
    The upper row shows the ratio to the original simulation.
    The lower row corresponds to the resolution quadratic difference between the two simulations.
    The grey dashed lines correspond to the input parameters used in the simulation, that should be retrieved by the fit. 
    The results for the low- and high-bremsstrahlung electrons are shown, respectively, on the left and on the right.
    The error bars shown correspond to the uncertainty, as determined from the fit.
    }
\label{fig:ijazz:validation}
\end{figure}

The scale and resolution parameters are fitted in the four aforementioned samples,
and the ratio with respect to the \textit{Original} simulation parameters 
are shown in Fig.~\ref{fig:ijazz:validation}. 
In the upper row of Fig.~\ref{fig:ijazz:validation}
the simulated energy scaling is shown in grey.
The scale is retrieved with an accuracy of better than 0.2\%,
even when an additional smearing is simulated, as shown by the green and light blue points.
This test is also used to evaluate the bias in the estimate of the resolution parameters, for which the results
of the fit are shown in 
the lower row of Fig.~\ref{fig:ijazz:validation}.
The resolution degradation applied to the \textit{Scale+Smear} and \textit{Smear} simulations 
should correspond to an oversmearing of the original resolution, 
\ie $\sigma_\text{Smear} = \sqrt{\smash[b]{\sigma_\text{Orig}^2+\sigma_\text{Degrad}^2}}$,
where $\sigma_\text{Smear}$ and $\sigma_\text{Orig}$ 
are the resolutions measured in the \textit{Smear} and in the \textit{Original} simulation, respectively,
and $\sigma_\text{Degrad}$ is the smearing degradation 
applied on top of the \textit{Original} simulation (grey dashed line in the figure). 
In general, the fit converges to the expectation, but a small 
overestimation
of the order of 10\% arises in EE, 
especially for high-bremsstrahlung electrons ($\RNINE<0.94$). 
The resolution parameters are used to determine the accuracy of the different intercalibration methods and
in the procedure this effect is taken into account.
Those values are also used to measure the electron resolution in Section~\ref{subsec:performance:energy}
and the bias has a negligible impact on the values reported.

\section{The ES calibration}\label{sec:sub:es-calibration}
Charged particles with momenta close to minimum ionizing (MIPs) are used to calibrate the ES.
An accuracy of 5\% of the channel-by-channel calibration is needed,
corresponding to a contribution of about 0.25\%
to the overall EE+ES energy resolution for 
electrons or photons,
since only a few percent of the electron or photon energy is deposited in the ES.
For each channel, the energy distribution is fit with a Landau function to model the signal
convolved with a Gaussian function to model the intrinsic noise. The energy per MIP calibration
factor is obtained from the most probable value of the Landau function after the fit.
The sources of response variation (sensor-to-sensor and channel-to-channel)
are the sensor thickness seen by the incident particles, which depends on the angle of incidence,
the gain of the front-end electronics chain,
and the charge collection efficiency which varies with radiation damage.

\section{Timing calibration}\label{sec:sub:timingcalib}
The timing of the digitized signal pulse in the EB and EE is measured using the ratio method \cite{Chatrchyan:2009aj},
which determines the pulse timing
relative to the time of the maximum amplitude from a reference template, exploiting the ratio between consecutive samples, $A(T)/A(T + 25\unit{ns})$.
The final timing measurement is computed by applying corrections and calibrations to the ratio method output:
\begin{equation}
\mathrm{T} = \mathrm{T}_\text{Max} +\mathrm{T}_\text{Correction} +\mathrm{T}_{IC}+\mathrm{T}_\text{Offset}
\label{PhiSymCorr:time}
\end{equation}
where
\begin{itemize}
\item $\mathrm{T}_\text{Max}$: output of the ratio method;
\item $\mathrm{T}_\text{Correction}$: effective correction, as a function of the pulse amplitude, optimized with 
simulations;
\item $\mathrm{T}_{IC}$: timing calibrations, measured for each crystal;
\item $\mathrm{T}_\text{Offset}$: global offset, not crystal dependent, applied to match data and 
simulations.
\end{itemize}
The crystal pulse shapes evolve with time, due to the changes in transparency under irradiation,
which modifies the propagation time distribution of photons from scintillation.
This affects the timing measurement obtained from the ratio method that assumes a fixed pulse shape.
To cope with this effect, which amounts to about 2\unit{ns} over one year, 
the timing calibration is frequently updated, typically on a daily basis.

This calibration also reduces the effect of the variations in the clock distribution
between different regions of the ECAL and different CMS runs.
Timing calibrations are computed by extracting the average of the timing distribution for each crystal,
using events from the $\phi$-symmetry calibration stream.
Additional energy selections are applied in addition to the online trigger selections: a constant threshold of 2\GeV in the EB and an $\eta$-dependent threshold,
from 4.6 up to 8.8\GeV, in the EE. These thresholds are optimized to remove pulses that are smaller than 10 times the expected RMS of the noise.
In addition, selections on the quality of the reconstruction are applied, 
requiring the estimated uncertainty of the timing measurement to be less than 3\unit{ns}
and a selection based on the $\chi^{2}$ of the pulse fit with the multifit method.
The statistical precision of the timing calibration is about 1\%.
The performance of the timing measurement is discussed in Section~\ref{subsec:performance:timing}.

\cleardoublepage \section{The CMS Collaboration \label{app:collab}}\begin{sloppypar}\hyphenpenalty=5000\widowpenalty=500\clubpenalty=5000
\cmsinstitute{Yerevan Physics Institute, Yerevan, Armenia}
{\tolerance=6000
A.~Hayrapetyan, A.~Tumasyan\cmsAuthorMark{1}\cmsorcid{0009-0000-0684-6742}
\par}
\cmsinstitute{Institut f\"{u}r Hochenergiephysik, Vienna, Austria}
{\tolerance=6000
W.~Adam\cmsorcid{0000-0001-9099-4341}, J.W.~Andrejkovic, T.~Bergauer\cmsorcid{0000-0002-5786-0293}, S.~Chatterjee\cmsorcid{0000-0003-2660-0349}, K.~Damanakis\cmsorcid{0000-0001-5389-2872}, M.~Dragicevic\cmsorcid{0000-0003-1967-6783}, A.~Escalante~Del~Valle\cmsorcid{0000-0002-9702-6359}, P.S.~Hussain\cmsorcid{0000-0002-4825-5278}, M.~Jeitler\cmsAuthorMark{2}\cmsorcid{0000-0002-5141-9560}, N.~Krammer\cmsorcid{0000-0002-0548-0985}, D.~Liko\cmsorcid{0000-0002-3380-473X}, I.~Mikulec\cmsorcid{0000-0003-0385-2746}, J.~Schieck\cmsAuthorMark{2}\cmsorcid{0000-0002-1058-8093}, R.~Sch\"{o}fbeck\cmsorcid{0000-0002-2332-8784}, D.~Schwarz\cmsorcid{0000-0002-3821-7331}, M.~Sonawane\cmsorcid{0000-0003-0510-7010}, S.~Templ\cmsorcid{0000-0003-3137-5692}, W.~Waltenberger\cmsorcid{0000-0002-6215-7228}, C.-E.~Wulz\cmsAuthorMark{2}\cmsorcid{0000-0001-9226-5812}
% ...
\par}
\vskip\cmsinstskip
\dag:~Deceased\\
$^{1}$Also at Yerevan State University, Yerevan, Armenia\\
$^{2}$Also at TU Wien, Vienna, Austria\\
% ...
\end{sloppypar}
\end{document}